\documentclass[apj,revtex4,numberedappendix]{emulateapj}
\usepackage{graphics}
\usepackage{subfigure}
\usepackage{hyperref}
\usepackage{graphicx}
\usepackage{pdflscape}
\usepackage{rotating}
\usepackage{color}
\usepackage[dvipsnames]{xcolor}
 
\setcitestyle{notesep={ }}

\usepackage{amsmath}
\usepackage{natbib}
\usepackage{float}
\newcommand{\CH}[1]{\colhead{#1}}
\newcommand{\A}{$\alpha$}
\newcommand{\B}{$\beta$}
\newcommand{\D}{$\delta$}
\newcommand{\G}{$\gamma$}

\newcommand\ii{{\sc ii}}
\newcommand\iii{{\sc iii}}
\newcommand\iv{{\sc iv}}
\newcommand\W{{$\lambda$}}
\newcommand\ND{\nodata}

\newcommand{\mc}[1]{\multicolumn{2}{c}{#1}}

\begin{document}

\shortauthors{Berg et al.}
\title{The Chemical Evolution of Carbon, Nitrogen, and Oxygen in Metal-Poor Dwarf Galaxies \footnote{
Based on observations made with the NASA/ESA Hubble Space Telescope,
obtained from the Data Archive at the Space Telescope Science Institute, which
is operated by the Association of Universities for Research in Astronomy, Inc.,
under NASA contract NAS 5-26555.
These observations are associated with program \#14628.}}

\author{Danielle A. Berg\altaffilmark{1}, 
	    Dawn K. Erb\altaffilmark{2},
	    Richard B.C. Henry\altaffilmark{3},
	    Evan D. Skillman\altaffilmark{4}, 
	    Kristen B.W. McQuinn\altaffilmark{5}
	    }
\altaffiltext{1}{Department of Astronomy, The Ohio State University, 140 W. 18th Avenue, Columbus, OH 43202, USA; berg.249@osu.edu}
\altaffiltext{2}{Center for Gravitation, Cosmology and Astrophysics, Department of Physics, University of Wisconsin Milwaukee, 3135 N Maryland Ave., Milwaukee, WI 53211, USA; erbd@uwm.edu}
\altaffiltext{3}{H. L. Dodge Department of Physics and Astronomy, University of Oklahoma, Normal, OK 73019, USA ; rhenry@ou.edu}
\altaffiltext{4}{Minnesota Institute for Astrophysics, University of Minnesota, 116 Church St. SE, Minneapolis, MN 55455, USA;  skillman@astro.umn.edu}
\altaffiltext{5}{Department of Astronomy, The University of Texas at Austin, 2515 Speedway, Stop C1400, Austin, TX 78712, USA; kmquinn@astro.as.utexas.edu}

\begin{abstract}
Ultraviolet nebular emission lines are important for understanding the time 
evolution and nucleosynthetic origins of their associated elements, 
but the underlying trends of their relative abundances are unclear. 
We present UV spectroscopy of 20 nearby low-metallicity, high-ionization dwarf 
galaxies obtained using the Hubble Space Telescope.  
Building upon previous studies, we analyze the C/O relationship for a combined 
sample of 40 galaxies with significant detections of the UV O$^{+2}$/C$^{+2}$ 
collisionally-excited lines and direct-method oxygen abundance measurements. 
Using new analytic carbon ionization correction factor relationships,
we confirm the flat trend in C/O versus O/H observed for local metal-poor galaxies. 
We find an average ${\rm log(C/O)}=-0.71$ with an intrinsic dispersion of $\sigma=0.17$ dex.  
The C/N ratio also appears to be constant at ${\rm log(C/N)}=0.75$, plus significant scatter 
($\sigma=0.20$ dex), with the result that carbon and nitrogen show similar evolutionary trends. 
This large and real scatter in C/O over a large range in O/H implies that measuring the UV 
C and O emission lines alone does not provide a reliable indicator of the O/H abundance. 
By modeling the chemical evolution of C, N, and O of individual targets, we find that the C/O 
ratio is very sensitive to both the detailed star formation history and to supernova feedback. 
Longer burst durations and lower star formation efficiencies correspond to low C/O ratios, 
while the escape of oxygen atoms in supernovae winds produces decreased 
effective oxygen yields and larger C/O ratios. 
Further, a declining C/O relationship is seen with increasing baryonic mass due to increasing effective oxygen yields.
\end{abstract}



\section{INTRODUCTION}\label{sec1}


Carbon, Nitrogen, and Oxygen are thought to originate primarily from stars of different mass ranges; 
O is synthesized mostly in massive stars (MSs; $M > 10 M_{\odot}$), while C and N are produced in 
both MSs and intermediate-mass stars (IMS). 
Theoretically, only primary (metallicity independent) nucleosynthetic processes are known to produce 
C such that we expect the C/O ratio to be constant under the assumption of a universal initial mass function. 
However, the first prominent analysis of nebular C/O in galaxies of low O/H came from the study of 
\citet{garnett95}, who used \textit{Hubble Space Telescope} (HST) \textit{Faint Object Spectrograph} 
(FOS) observations to suggest that C/O continuously increases with O/H in metal-poor \ion{H}{2} regions.
Recently, \citet[][\llap, hereafter B16]{berg16} presented HST 
\textit{Cosmic Origins Spectrograph} (COS) 
UV spectroscopy of 12 nearby, low-metallicity, high-ionization dwarf galaxies,
finding that the relationship between C/O and O/H is nominally consistent with the increasing trend 
reported by \citet{garnett95}, but with 3 significant outliers. 
Considering these outliers, a flat relationship with large dispersion is also viable.
This empirical trend of C/O $\propto$ O/H suggests secondary\footnotemark[1] 
(metallicity dependent) C production is prominent
and the large scatter in C/O could be due to the delayed C released from IMSs.
Additionally, B16 found C/N exhibited a relatively constant trend over a large range in O/H, 
indicating that C may be produced by similar nucleosynthetic mechanisms as N.
Therefore, the C abundance may be affected by pseudo-secondary C enrichment such as the 
metallicity driven winds of MSs \citep{henry00}. 

At higher metallicities (12 + log(O/H) $> 8.0$), nebular C/O abundances can be determined from optical 
recombination lines \citep[e.g.,][]{peimbert05,garcia-rojas07,lopez-sanchez07,esteban14}. 
These abundances extend the C/O relationship to larger values of O/H and show C/O increasing 
with increasing O/H.  Thus, the overall trend of C 
production appears to be bi-modal, where primary production of C dominates at low metallicity, but 
pseudo-secondary production becomes prominent at higher metallicities. 

 \footnotetext[1]{If a particular isotope is produced from the original H and He in a star, 
the production is said to be ``primary" and its abundance relative to other heavy elements remains constant.
If instead the isotope is a daughter element of heavier elements initially present in the star,
the production is called ``secondary" and is linearly dependent on the initial abundance
of the parent heavier elements (metallicity). }

Realistically, chemical enrichment does not occur in a closed box, and 
the relative abundance trends of galaxies may also be significantly shaped by the
selective loss of heavy elements in star formation driven outflows 
\citep[e.g.,][]{matteucci85,heckman90,veilleux05}. 
This is particularly relevant for the low-mass end of the mass-metallicity relationship
\citep[e.g.,][]{tremonti04}.
Owing to their smaller gravitational potentials, galactic outflows in low-mass galaxies are expected to 
more efficiently remove their metals \citep[e.g.,][]{dekel86,dalcanton07,peeples11}, and this
has been demonstrated by the trend of increasing effective yield with increasing galaxy mass
\citep{garnett02}.
Further, \citet{chisholm18b} recently reported the first observational evidence for a sample of
low-mass galaxies to have larger metal-loading factors than massive galaxies, indicating that
their galactic outflows remove metals very efficiently.
However, while the escaping metals have been measured for several atoms including oxygen
(e.g., \ion{O}{1} in the warm outflow phase and \ion{O}{6} in the hot wind),
no studies to date have measured the mass outflow of carbon due to the tendency of the
C \W1334 absorption feature (a common tracer of galactic outflows) to be saturated.
While the relative escape of C and O in galactic outflows is unknown empirically, 
theoretically, supernova-driven outflows could allow for the preferential loss of O 
on short time scales prior to C production in IMSs. \looseness=-2

Even so, the primary source and significance of the scatter in C/O at low metallicity is unclear, in part,
due to the paucity of C/O detections in metal-poor \ion{H}{2} regions.
We, therefore, build upon the results of B16, using HST spectroscopic measurements of the UV
\ion{O}{3}] $\lambda\lambda$1660,1666 and \ion{C}{3}] $\lambda\lambda$1907,1909
collisionally-excited lines (CELs) in dwarf galaxies in the range of $7.6 \lesssim$ 12 + log(O/H) $\lesssim 8.0$, 
where the C/O relationship is poorly constrained and many suitable targets are available. 
These data provide a statistically robust sample of secure C/O abundances in 19 metal-poor dwarf galaxies 
from which we report on the relative variation of C with respect to O, and
discuss the nucleosynthetic origin and time evolution of C. \looseness=-2

As rest-frame UV emission-line spectra are increasingly being used to determine the physical 
properties of high-z galaxies, these data also provide templates for interpreting the distant universe.
In addition to the \ion{O}{3}] and \ion{C}{3}] emission observed in the UV spectra of our sample,
very high ionization \ion{C}{4} and \ion{He}{2} emission lines are detected.
These features require especially hard ionizing radiation fields 
($>$ 47.9 and 54.4 eV for \ion{C}{4} and \ion{He}{2}, respectively), where models of massive stars predict few photons.
However, recent studies of $z\sim5-7$ galaxies have revealed prominent high-ionization nebular UV emission lines 
(i.e., \ion{O}{3}], \ion{C}{3}], \ion{C}{4}, \ion{He}{2}) with extremely large EWs 
indicating that extreme radiation fields may also characterize 
reionization-era systems \citep{stark15, stark16, mainali17}. 
Indeed, strong high-ionization nebular emission appears to be more common in high redshift galaxies 
\citep[e.g., $z\sim5-7$:][]{smit14}.
Further, high-ionization optical lines have been linked to the leakage of Lyman continuum 
(LyC) photons (necessary for reionization) both theoretically \citep[e.g.,][]{jaskot13,nakajima13} and 
observationally in $z\sim0$ systems \citep[e.g.,][]{chisholm18a,izotov18}. 
Therefore, the high-ionization UV emission lines of our sample, especially those with large EWs, 
could hold important clues about the sources of ionizing photons during the Epoch of Reionization. 

We describe our sample in Section~\ref{sec:sample} and details of the UV HST 
and complementary optical observations in Section~\ref{sec:observations}.
Incorporating other recent studies in the literature, 
we define a high-quality, comprehensive C/O sample in Section~\ref{sec:expanded}.
We derive the nebular properties and chemical abundances in a uniform manner
 as described in Section~\ref{sec:abundances}.
In order to properly estimate the relative abundances,
we use a grid of {\sc cloudy} photoionization models to derive new analytic fits to 
the carbon ionization correction factors in Section~4.4.
Trends of the resulting relative abundances are discussed in Section~5.
Constrained by these data, we develop single-burst chemical evolution models in Section~6
and use them to interpret the scatter in the C/O and C/N relationships in Section~6.4,
and the evolution of C/O in Section~6.5.
Finally, we summarize our results in Section~\ref{sec:summary}.
For convenience, we include details of our chemical evolution code in Appendix~A
and tables of our emission-line and abundance measurements in Appendix~B.


\begin{deluxetable*}{lccccccccc}
\tablewidth{0pt}
\tablecaption{Bright, Compact, Nearby Dwarf Sample}
\tablehead{
\CH{1   	}  & \CH{2 } 	  & \CH{3 } 		& \CH{4 }    & \CH{5} 		& \CH{6} 		& \CH{7}	 	 & \CH{8} 				& \CH{9}	 			& \CH{10} \\ \hline
\CH{Target} & \CH{R.A.}	  & \CH{Dec.}		& \CH{$z$} & \CH{$m_{FUV}$} & \CH{m$_r$}	& \CH{12+} 	 & \CH{log $M_{\star,tot}$}	& \CH{log SFR$_{tot}$ }	& \CH{log sSFR} 	\\
\CH{   	} & \CH{(J2000)} & \CH{(J2000)}	& \CH{}	 & \CH{(mag)}		& \CH{(mag)}	& \CH{log(O/H)} & \CH{(M$_\odot$)}	 	& \CH{(M$_\odot$/yr)}	& \CH{(yr$^{-1}$)} }
\startdata
J223831		& 22:38:31.11	& +14:00:28.29		& 0.021	& 18.86		& 19.22	& 7.63	& 6.72	    & $-0.77$		& $-7.49$	\\	
J141851		& 14:18:51.12	& +21:02:39.84		& 0.009	& \ND		& 18.46	& 7.64	& 6.63    	& $-1.16$     	& $-7.79$	\\	
J120202		& 12:02:02.49	& +54:15:51.05		& 0.012	& \ND		& 19.29	& 7.65	& \ND	    & \ND		    & \ND	    \\	
J121402		& 12:14:02.40	& +53:45:17.28		& 0.003	& 18.40		& 18.75	& 7.72	& 6.02	    & $-1.94$		& $-7.97$	\\
J084236		& 08:42:36.48	& +10:33:14.04		& 0.010	& 19.21		& 19.09	& 7.74	& 7.01     	& $-1.19$     	& $-8.20$	\\	
J171236		& 17:12:36.72	& +32:16:33.60		& 0.012	& \ND		& 18.89	& 7.77	& 7.05    	& $-1.06$		& $-8.12$ 	\\	
J113116		& 11:31:16.32	& +57:03:58.68		& 0.006	& 18.78		& 19.31	& 7.81	& 6.51     	& $-1.69$     	& $-8.21$	\\	
J133126		& 13:31:26.88	& +41:51:48.24		& 0.012	& 18.04		& 18.28	& 7.83	& 7.16    	& $-0.88$     	& $-8.05$	\\	
J132853		& 13:28:53.04	& +15:59:34.44		& 0.023	& 19.03		& 19.12	& 7.87	& 7.18    	& $-0.68$     	& $-7.86$	\\	
J095430		& 09:54:30.48	& +09:52:12.11		& 0.005	& 18.96		& 19.06	& 7.87	& 6.53     	& $-1.61$     	& $-8.14$	\\	
J132347		& 13:23:47.52	& $-$01:32:51.94 	& 0.022	& 19.22		& 19.24	& 7.87	& 7.04    	& $-0.73$		& $-7.77$ 	\\	
J094718		& 09:47:18.24	& +41:38:16.44		& 0.005	& 19.37		& 19.12	& 7.88	& 6.44     	& $-1.79$     	& $-8.24$	\\	
J150934		& 15:09:34.08	& +37:31:46.20		& 0.033	& 18.68		& 18.91	& 7.92	& 7.78     	& $0.33$     	& $-7.45$	\\
J100348		& 10:03:48.72	& +45:04:57.72		& 0.009	& 19.11		& 18.53	& 7.95	& 7.03     	& $-1.17$		& $-8.21$	\\
J025346		& 02:53:46.70	& $-$07:23:43.98 	& 0.004	& 17.71		& 18.81	& 7.97	& 6.06     	& $-2.01$     	& $-8.07$	\\
J015809		& 01:58:09.38	& $-$00:06:37.23 	& 0.012	& \ND		& 19.50	& 7.97	& 6.15    	& $-1.41$     	& $-9.31$	\\
J104654		& 10:46:54.00	& +13:46:45.84		& 0.011	& 18.55		& 18.74	& 8.00	& 6.81     	& $-1.15$     	& $-7.96$	\\
J093006		& 09:30:06.48	& +60:26:53.52		& 0.014	& 16.82		& 17.78	& 8.00	& 7.32    	& $-0.75$		& $-8.07$   \\
J092055		& 09:20:55.92	& +52:34:07.32		& 0.008	& 17.48		& 18.18	& 8.02	& 7.52    	& $-0.74$     	& $-8.26$	\\
J084956		& 08:49:56.16	& +10:43:08.76		& 0.014	& 18.12		& 17.90	& 8.03	& 7.67    	& $-0.46$     	& $-8.13$
\enddata 
\tablecomments{Selected target sample listed in order of increasing oxygen abundance. All objects are bright, compact,
    nearby dwarf galaxies with low metallicities as measured by their ground-based optical spectra. 
    The first four columns give the target name used in this work, location, and redshift.
    Columns 5 and 6 give magnitudes from GALEX and the SDSS, respectively.     
    Column 7 lists the oxygen abundances of our sample taken from the literature.
    Columns 8$-$10 list the median total stellar masses, star formation rates, and specific star formation rates from the MPA-JHU DR8 database.}
\label{tbl1}
\end{deluxetable*}


\section{HIGH IONIZATION DWARF GALAXY SAMPLE} \label{sec:sample}


The main goal of this study is to obtain new gas-phase UV observations of 
C$^{+2}$ and O$^{+2}$ in an expanded sample of 20 dwarf galaxies to aid 
our understanding of the C/O versus O/H relationship and its scatter.
We demonstrated in B16 that using the \ion{C}{3}] $\lambda\lambda1907,1909$/\ion{O}{3}] 
$\lambda1666$ line ratio is a robust way to investigate the C/O ratio 
and minimize observational uncertainties.
In particular, this method benefits from the fact that C/O exhibits minimal uncertainty due to reddening,
as the interstellar extinction curve is nearly flat over the wavelength range of interest ($1600-2000$ \AA)
and the \ion{O}{3}] and \ion{C}{3}] lines have similar excitation and ionization
potentials such that their ratio has little dependence on the physical conditions of 
the gas (i.e., nebular $T_e$ and ionization structure).

In order to fill in the C/O relationship with O/H in the sparsely measured 
metal-poor regime, we need objects with large equivalent widths (EWs) of high-ionization 
emission lines and low metallicity in the range of $7.6 \lesssim$ 12 + log(O/H) $\lesssim 8.0$. 
High-ionization \ion{H}{2} regions are needed given the energies required to ionize
C$^{+}$ and O$^{+}$ are 24.8 eV and 35.1 eV respectively.
High nebular electron temperatures ($T_e$) in low-metallicity 
environments allow the collisionally excited C and O transitions of interest to be observed
despite their large excitation energies (6-8 eV).

Using the Sloan Digital Sky Survey Data Release 12\footnotemark[2] 
\citep[SDSS-III DR12;][]{eisenstein11,alam15}, and following the suggestions of B16, 
we select targets with the following updated criteria:

\begin{enumerate}
\item $7.6 \lesssim$ 12 + log(O/H) $\lesssim 8.0$: This fills in the sparsely 
populated region of the low-metallicity C/O relationship.
\item $z < 0.04$: The G140L grating is the only grating on COS that allows simultaneous observations 
of \ion{O}{3}] \W\W1660,1666 and \ion{C}{3}] \W\W1907,1909 in nearby galaxies. 
Limited wavelength coverage and rapidly declining redward throughput ($>2000$ \AA) requires targets with redshifts of $z < 0.04$.
\item $D_{25} \lesssim 5$\arcsec: Through a visual inspection of SDSS photometry \citep{gunn98} 
using the Catalog Archive Server (CAS) database, we selected candidate targets which have compact 
morphologies in the sense that the diameter of their optical light profiles $\lesssim 5$\arcsec. 
In B16, the selected targets were all more compact in the UV than in the optical.
The compactness of our sample is demonstrated by the near-UV acquisition images shown in Figure~1,
and allows for maximum flux through the 2.\arcsec5 COS aperture.
Similarly, the SDSS spectra were taken through a 3\arcsec\ fiber, 
capturing roughly the same light profile as the COS aperture.
\item $m_{FUV} \lesssim 19.5$ AB: Galaxy Evolution Explorer (GALEX) photometry \citep[GR6;][]{bianchi14} 
detections are required in order to ensure targets are bright enough in the FUV to enable continuum detections with COS.
Note that the FUV magnitudes we inferred for three of our targets are based on their spectral energy distribution 
fits to their SDSS photometry and similarities to the other targets in our sample. 
\item EW(\W5007) $> 500$ \AA: To improve upon previous studies which largely 
lack significant detections of \ion{O}{3}] \W1666, we selected galaxies with large 
[\ion{O}{3}] \W5007 EWs. B16 found targets with 
EW(\W5007) $> 500$ \AA\ had significant \ion{O}{3}] \W1666 detections.
\item Predicted F(\ion{O}{3}] \W1666) $> 10^{-15}$ erg s$^{-1}$ cm$^{-2}$ and S/N $> 5$: 
B16 found that large predicted \ion{O}{3}] \W1666 fluxes were needed to overcome the low sensitivity of UV detectors.
The expected \ion{O}{3}] \W1666 fluxes were determined from the optical spectra using the [\ion{O}{3}] \W5007 fluxes and 
the theoretical line emissivities at the measured $T_e$ and $n_e$ of each galaxy.
Additionally, the GALEX FUV fluxes within the 2.\arcsec5 HST/COS aperture were used alongside the optical [\ion{O}{3}] line 
fluxes to select targets for which the COS exposure time calculator predicted S/N $> 5$ in the \ion{O}{3}] \W1666 line, 
which is the weakest of the desired UV lines.
\end{enumerate}


\begin{figure*}[h]
   \begin{center}
   \plotone{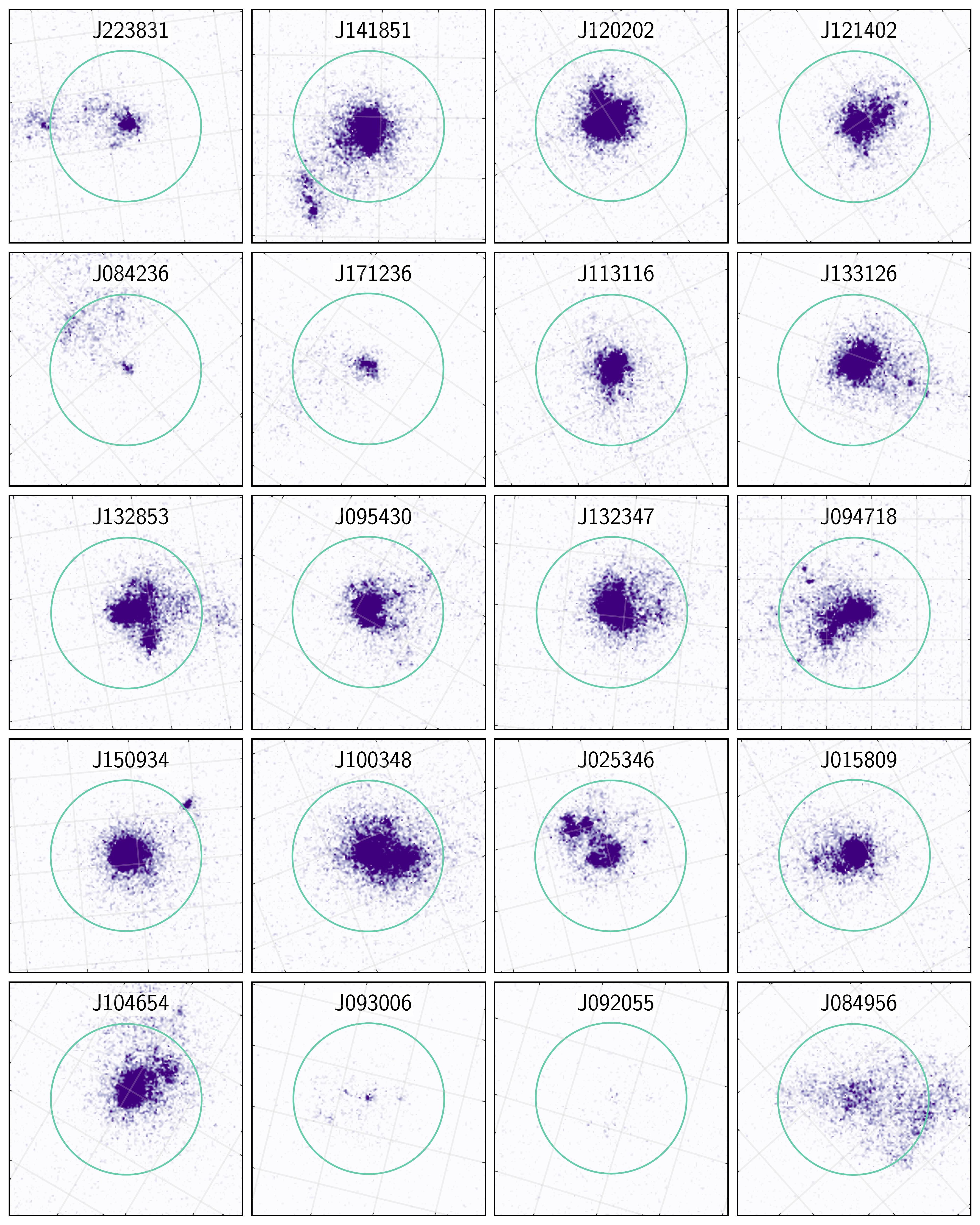}
	\caption{
	COS acquisition images of all 20 targets. 
	 The compact morphologies of our target allow for maximum flux through the 2.\arcsec5 COS aperture,
	 which is shown as a green circle.
	 The observations resulted in 3$\sigma$ C/O detections for 19 of these galaxies 
	 (all but J084956).
	 Note that MirrorB was used to safely acquire the FUV-brightest galaxies in our sample,
	 resulting in the weak acquisition detections of J093006 and J092055 shown here.}
   \end{center} 
\label{fig1}
\end{figure*}     


\footnotetext[2]{\url{http://www.sdss.org/dr12/}} 

Our final sample contains 20 nearby (0.003 $< z <$ 0.033), UV-bright ($m_{FUV} \leq 19.5$ AB), 
compact ($D_{25} < 5\arcsec$), low-metallicity (12+log(O/H) $\lesssim 8.0$) dwarf galaxies.
Basic properties of our final sample are listed in Table~\ref{tbl1}.
Median total stellar masses \citep[following][]{kauffmann03b}, average total 
star formation rates \citep[SFRs; based on][]{brinchmann04}, and
average total specific SFRs (sSFRs) are taken from the MPA-JHU catalogue\footnotemark[3].
Given our selection of nearby, compact, bright targets, our sample has
very low masses and high sSFRs. 
Figure~1 displays our sample targets.

\footnotetext[3]{Data catalogues are available from \url{http://wwwmpa-garching.mpg.de/SDSS/}.
The Max Plank institute for Astrophysics/John Hopkins University(MPA/JHU) SDSS data base was produced by a collaboration 
of researchers(currently or formerly) from the MPA and the JHU. The team is made up of Stephane Charlot (IAP), 
Guinevere Kauffmann and Simon White (MPA),Tim Heckman (JHU), Christy Tremonti (U. Wisconsin-Madison $-$ formerly JHU) 
and Jarle Brinchmann (Leiden University $-$ formerly MPA).} 

\section{SPECTROSCOPIC OBSERVATIONS AND DATA REDUCTION}\label{sec:observations}


\subsection{HST/COS FUV Spectra} \label{sec:hst}

Following the observational strategy laid out in B16, we efficiently and simultaneously observed 
the C and O CELs in the UV with a single orbit of COS/HST for each of our targets (HST-GO-14628).
Since all sources have excellent input coordinates from the SDSS, which are generally accurate to 0.1\arcsec, 
we acquire our targets using the ACQ/IMAGE mode with the PSA aperture and MirrorA\footnotemark[4] for the 
COS/NUV configuration. 
The acquisition images in Figure~1 confirm that the COS 2.5\arcsec\ aperture was well aligned 
with the UV peak of our targets and captures the majority of the UV emission.
COS FUV observations were taken in the TIME-TAG mode using the 2.5\arcsec\ 
PSA aperture and the G140L grating at a central wavelength of 1280 \AA.
In this configuration, segment A has an observed wavelength range of 1282$-$2000 \AA\footnotemark[5],
allowing the simultaneous observation of the \ion{C}{4} \W\W1548,1550, 
\ion{He}{2} \W1640, \ion{O}{3}] \W\W1660,1666, \ion{N}{3}] \W1750, 
\ion{Si}{3}] \W\W1883,1892, and \ion{C}{3}] $\lambda\lambda1907,1909$ emission lines. 
We used the FP-POS=ALL setting, which takes 4 images offset from one 
another in the dispersion direction, increasing the cumulative S/N and 
mitigating the effects of fixed pattern noise. 
The 4 positions allow a flat to be created and cosmic rays to be eliminated.
Each target was observed for the maximum time allotted in a single orbit as 
determined by the object orbit visibility.
All data from GO-14628 were processed with CALCOS version 3.2.1\footnotemark[6].

\footnotetext[4]{MirrorB was used to safely acquire our brightest FUV galaxies: J093006 and J092055.}
\footnotetext[5]{The G140L grating on COS is characterized as having wavelength coverage out to 2405 \AA.
However, our experience with this setup indicates a range of usefulness out to only 2000 \AA.}
\footnotetext[6]{\url{http://www.stsci.edu/hst/cos/pipeline/CALCOSReleaseNotes/notes/}}

In order to gain signal-to-noise we chose to bin the spectra in the dispersion direction.
The COS has a resolution of R = 2,000 for a perfect point source, which corresponds 
to six detector pixels in the dispersion direction. 
We re-binned the data by a factor of six 
to reflect this, improving S/N without losing information.
For the G140L grating, six pixels (80.3 m\AA/pix) span a resolution element 
of roughly 0.55 \AA\ at $\lambda1660$.
By measuring individual airglow emission lines in our spectra, we found a typical
FWHM $\approx$ 3 \AA, allowing us to re-bin our spectra by the six pixels of a 
resolution element while maintaining six resolution elements per FWHM.

\subsection{SDSS Optical Spectra} \label{sec:sdss}

Each of the targets in our sample has been previously observed as part of the SDSS DR12.
We used the publicly available SDSS data \citep{york00}, 
which have been reduced with the SDSS pipeline \citep{bolton12}.
Preliminary emission line fluxes from the MPA-JHU data catalog were used
to select these targets such that they had significant [\ion{O}{3}] $\lambda4363$ 
auroral line detections.
However, to ensure uniformity, we have re-measured the SDSS emission lines, as
described below, and used the most recent atomic data for the subsequent analysis.


\begin{figure*}
\plotone{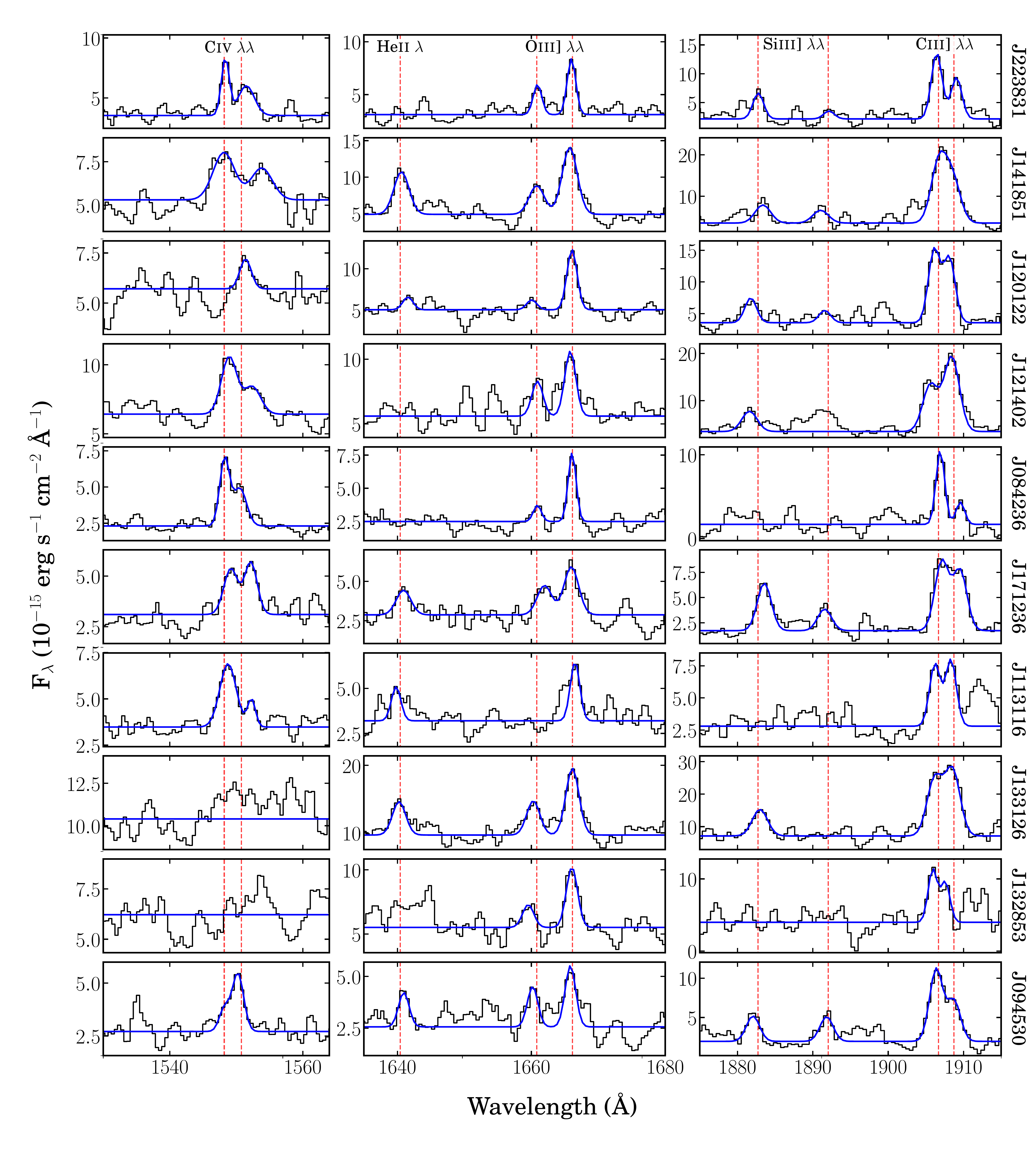}
\caption{HST/COS rest-frame emission line spectra for 10 of the galaxies in our sample with significant C and O detections.
The blue lines represent the best fit to the emission lines as described in Section~\ref{sec:lines}.
Rest-frame vacuum wavelengths of the nebular emission lines are indicated by red, vertical 
dashed lines. 
Note that the spectra have been binned by 6 pixels and box-car smoothed by 3.}
\label{fig2}
\end{figure*}

\begin{figure*}
\plotone{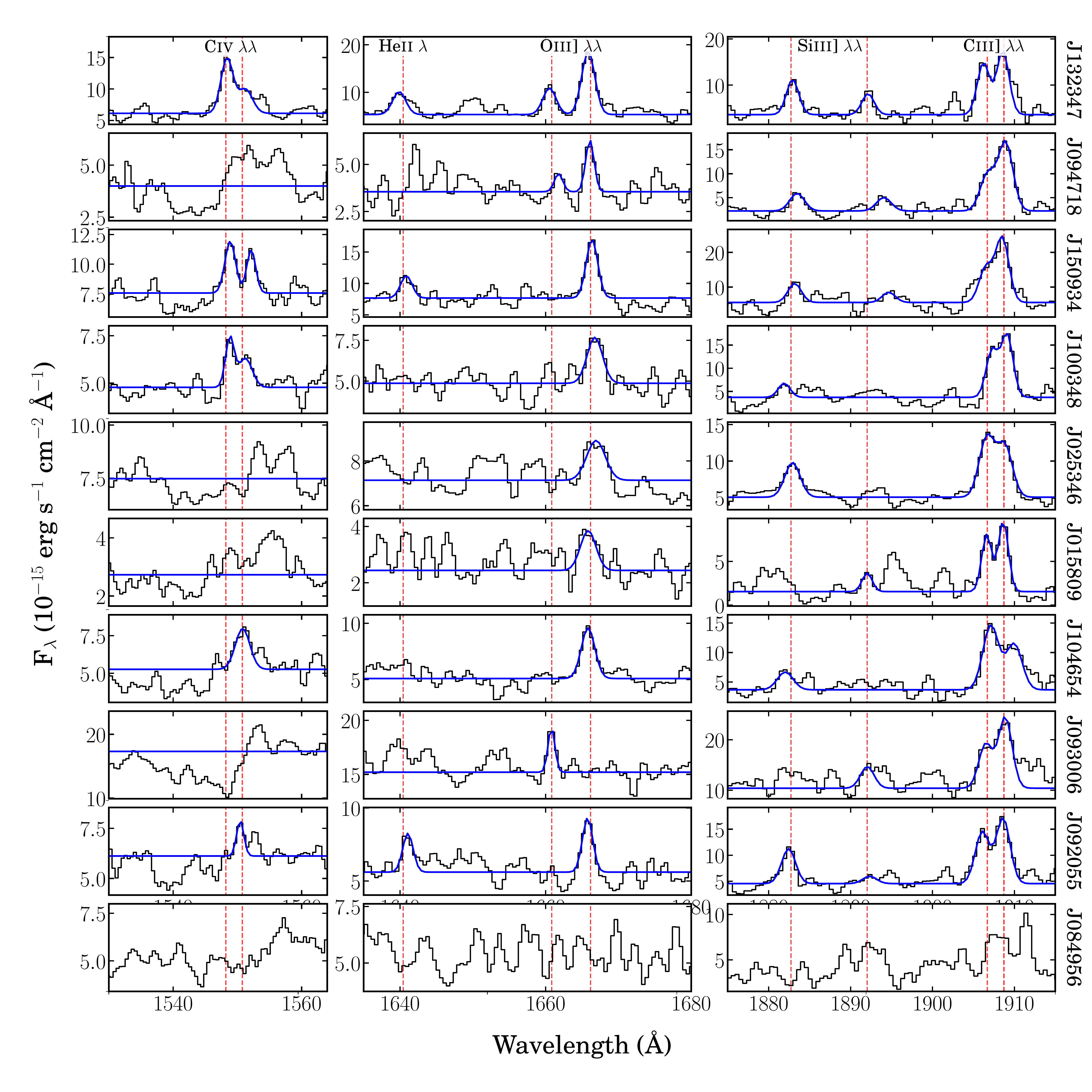}
\caption{HST/COS rest-frame emission line spectra for the remaining 10 galaxies in our sample, 
9 of which have significant C and O detections (all but J084956).
The blue lines represent the best fit to the emission lines as described in Section~\ref{sec:lines}.
Rest-frame vacuum wavelengths of the nebular emission lines are indicated by red, vertical 
dashed lines. 
The spectra have been binned by 6 pixels and box-car smoothed by 3.}
\label{fig3}
\end{figure*}


\subsection{Nebular Emission Line Measurements} \label{sec:lines}

The rest-frame (corrected by SDSS redshift) emission line features of the 
UV HST/COS spectra are plotted in Figures~2 and 3.
The emission line strengths for both the UV and optical spectra were measured 
using the {\tt SPLOT} routine within IRAF\footnotemark[7]. 
Groups of nearby lines were fit simultaneously, constrained by a single Gaussian 
FWHM and a single line center offset from the vacuum wavelengths (i.e., redshift).

Emission lines in the optical SDSS spectra were fit in the same manner.
However, in the cases where Balmer absorption was also clearly visible, 
the bluer Balmer lines (H$\delta$ and H$\gamma$) were fit simultaneously with 
multiple components such that the absorption was fit by a broad, negative Lorentzian 
profile and the emission was fit by a narrow, positive Gaussian profile. 
Note that our sample is composed of high ionization star-formation regions that 
display only weak Balmer absorption, consistent with the hard radiation 
fields from main sequence O stars, such that the absorption component
is negligible for the stronger H$\beta$ and H$\alpha$ emission lines.  
To ensure that noise spikes are not fit, only emission lines with a 
strength of 3$\sigma$ or greater are used in the subsequent abundance analysis.

The high-ionization \ion{C}{4} \W\W1548,1550 and \ion{He}{2} \W1640 emission lines were detected in several of our targets.
Since these features can result from a combination of nebular and stellar features, and, in the case of \ion{C}{4},
can be affected by interstellar absorption, special care must be taken in their measurement and interpretation.
The UV \ion{He}{2} emission features appear narrow and so were fit with Gaussian
profile widths constrained to the other nebular lines.
On the other hand, many of the \ion{C}{4} profiles are clearly broadened and so were fit with a free width parameter.
For this reason, the \ion{C}{4} fits do not necessarily measure solely the nebular emission.
Further, we do not attempt to fit the \ion{C}{4} emission in galaxies with low S/N or very complicated profiles.
We compare the \ion{C}{4} measurements to \ion{C}{3} line strengths in Section~5.1,
but reserve a more detailed analysis for a future paper.

\footnotetext[7]{IRAF is distributed by the National Optical Astronomical Observatories.}

The errors of the flux measurements were approximated using
\begin{equation}
	\sigma_{\lambda} \approx \sqrt{(2 \times \sqrt{N} \times \mbox{RMS})^2 + (0.01 \times F_{\lambda})^2},
	\label{eq1}
\end{equation}
where N is the number of pixels spanning the Gaussian profile fit to the narrow emission lines. 
The root mean squared (RMS) noise in the continuum was taken to be the average of the RMS on each side of an emission line. 
The two terms in Equation~\ref{eq1} approximate the errors from continuum subtraction and flux calibration.
For weak lines, such as the UV CELs, the RMS term determines the approximate uncertainty. 
In the case of strong Balmer, [\ion{O}{3}], and other lines, 
the error is dominated by the inherent uncertainty in the flux calibration and
accounted for by adding the 1\% uncertainty of standard star calibrations in CALSPEC \citep{bohlin10}.
We note that this assumption applies only to relative line fluxes, as the absolute flux calibration for 
both COS and SDSS spectra can have much larger uncertainties.
Additionally, B16 tested the relative flux calibration of the SDSS spectrum of one of their targets
compared to an optical spectrum observed at the MMT and found the spectrophotometry to be consistent. 

The COS and SDSS spectra were corrected for the Galactic extinction from \citet{schlafly11}
using the reddening law of \citet{cardelli89}, parametrized by $A_{V}=3.1\times E(B-V)$.
The spectra were then de-reddened for dust extinction using a correction 
based on the H\A/H\B, H\G/H\B, and H\D/H\B\ Balmer line ratios.
The \citet{calzetti00} reddening law, parametrized by $A_{V}=4.05\times E(B-V)$ was used
for the UV emission lines (\W\ $< 2000$ \AA) and the \citet{cardelli89} reddening 
law was used for the optical emission lines (3727 \AA\ $<$ \W\ $<$ 10,000 \AA).
An initial estimate of the electron temperature was determined 
from the ratio of the [\ion{O}{3}] $\lambda4363$ auroral line to the 
[\ion{O}{3}] $\lambda\lambda4959,5007$ nebular lines and used
to determine the theoretical Balmer ratios in solving for the reddening.
The final reddening estimate is an error weighted average of the 
individual reddening values determined from the H$\alpha$/H$\beta$, 
H$\gamma$/H$\beta$, and H$\delta$/H$\beta$ ratios. 
All of our targets have low extinction in the range of $E(B-V)$ of $\sim0.03-0.16$.

Figures~2 and 3 show the Gaussian fits to the 
\ion{C}{4}, \ion{He}{2}, \ion{O}{3}], \ion{Si}{3}], and \ion{C}{3}] 
emission lines in the COS spectra for each of the targets in our sample.
Nineteen of our 20 targets (all but J084956) have significant C and O detections, 
demonstrating the efficacy of our target selection criteria.
The reddening corrected line intensities measured for our C/O targets are reported in 
Tables $7-9$ in the appendix.
An anomaly was noted in the spectrum of J093006 (see Figure~3), where the
\ion{O}{3}] \W1661 line is detected but the \W1666 line (the stronger of the doublet) is absent.
This is discussed further in section 4.4.


\section{Chemical Abundances} \label{sec:abundances}


\begin{deluxetable*}{rll}
\tabletypesize{\scriptsize}
\setlength{\tabcolsep}{3pt} 
\tablewidth{0pt}
\tablecaption{Sources of Atomic Data}
\tablehead{
Ion			& Radiative Transition Probabilities	& Collision Strengths }
\startdata
{C$^{+2}$}	& {\citet{wiese96}}			& {\citet{berrington85}}	\\
{C$^{+3}$}	& {\citet{wiese96}}			& {\citet{aggarwal04}}	\\
{N$^{+}$}		& {\citet{fft04}$^{\star}$}		& {\citet{tayal11}}		\\
{O$^{+}$}		& {\citet{fft04}$^{\star\dagger}$}	& {\citet{kisielius09}}		\\
{O$^{+2}$}	& {\citet{fft04}$^{\star\dagger}$}	& {\citet{aggarwal99}}	\\
{Ne$^{+2}$}	& {\citet{fft04}$^{\star}$}		& {\citet{mclaughlin00}} 	\\
{S$^{+}$}		& {\citet{podobedova09}}		& {\citet{tayal10}} 		\\
{S$^{+2}$}	& {\citet{podobedova09}}		& {\citet{hudson12}}		\\
{Ar$^{+2}$}	& {\citet{mendoza83}}		& {\citet{munoz-burgos09}} \\
{Ar$^{+3}$}	& {\citet{mendoza82}}		& {\citet{ramsbottom97}} 	
\enddata
\tablecomments{
The atomic data used with the P{\sc y}N{\sc eb} package to calculate ionic abundances. 
Note that the O$^{+2}$ collision strengths from \citet{aggarwal99} are calculated from a 6-level atom 
approximation, which is required to analyze our \ion{O}{3}] \W\W1661,1666 measurements.  \\
$^{\star}$ Agrees with updated values from \citet{tayal11}. \\
$^{\dagger}$ Equivalent to \citet{tachiev02}, as recommended by \citet{stasinska12}.}
\end{deluxetable*}		


\subsection{Expanding The Optimum C/O Sample} \label{sec:expanded}

Until recently, observations of collisionally excited C and O emission existed for only a small number of galaxies.
In B16 we assembled the most comprehensive picture to date of C/O determinations
by combining all published observations with 3$\sigma$ detections of the UV \ion{O}{3}] $\lambda1666$ 
and \ion{C}{3}] $\lambda\lambda1907,1909$ emission lines.
The resulting combined optimal sample consisted of 12 $z\sim0$ objects 
\citep[7 new observations and 5 from past studies; see][for more details]{berg16}.
Recent studies of UV emission lines in nearby galaxies allow us to expand the optimal C/O sample,
adding 19 detections from this work, 6 detections from 
\citet[][\llap, using \W1661 in place of \W1666 for one target]{senchyna17}, 
and 3 detections from \cite{pena-guerrero17}, for a total sample of 40 targets.
For convenience, we provide the EW measurements of the UV emission lines for the 
expanded sample (when available) in Table~10 in the appendix.

Building on the B16 sample, we compute the chemical abundances for our sample, 
as well as the supplemented literature sources, in a uniform, consistent manner.
We use the P{\sc y}N{\sc eb} package in {\sc python} \citep{luridiana12, luridiana15}, assuming a 
five-level atom model \citep{derobertis87} and default atomic data, for most of the calculations in this section. 
For \ion{O}{3}] \W\W1661,1666 we use \citet{aggarwal99} who calculate the collision strengths for the necessary 6-level atom.
The adopted atomic data is listed in Table~2.
Nebular physical conditions and the O/H, N/O, S/O, Ne/O, and Ar/O abundances are calculated from the optical spectra,
whereas the C/O abundances are determined from the UV O$^{+2}$ and C$^{+2}$ CELs.


\subsection{Temperature and Density} \label{sec:temden}

The derivation of the nebular physical conditions in our sample is identical to that of B16.
For convenience, we briefly summarize the main points below.
All of the galaxies in our sample have SDSS spectra with significant [\ion{O}{3}] \W4363
detections that allow us to directly calculate the electron temperature, $T_e$, of the 
high-ionization gas via the \ion{O}{3}] I(\W\W4959,5007)/I(\W4363) ratio.\footnotemark[8]
Temperatures of the low and intermediate ionization zones are then determined using the 
photoionization model relationships from \citet{garnett92}:
\begin{align}
       \mbox{T[S~\iii]} & =  0.83\times \mbox{T[O~\iii]} + 1700\mbox{ K} \label{eqn:G92-1} \\
       \mbox{T[N~\ii]} & =  0.70\times \mbox{T[O~\iii]} + 3000\mbox{ K,} \label{eqn:G92-2}
\end{align}
The [\ion{S}{2}] $\lambda\lambda$6717,6731 ratio is used to determine the electron densities.
In cases where the [\ion{S}{2}] ratio is consistent with the low density upper limit,
we assume $n_e = 100$ cm$^{-3}$.
The electron temperature and density determinations are listed in Tables~$11-13$ in the appendix.

\footnotetext[8]{We note that the electron temperature can also be determined from the 
[\ion{O}{3}] $\lambda5007$/\ion{O}{3}]$\lambda1666$ ratio, as is commonly done 
in high redshift targets where the intrinsically faint optical auroral line is often undetected.
However, for nearby targets, this requires combining space- and ground-based observations,
potentially introducing flux matching issues and mismatched aperture effects.
As a result of this mismatch, and the large dependence on reddening over this 
wavelength range, B16 found the [\ion{O}{3}] \W5007/\ion{O}{3}] \W1666 electron 
temperatures to be unreliable.}


\begin{figure*} 
\begin{center}
	\includegraphics[scale=0.3,trim=0mm 0mm 0mm 0mm,clip]{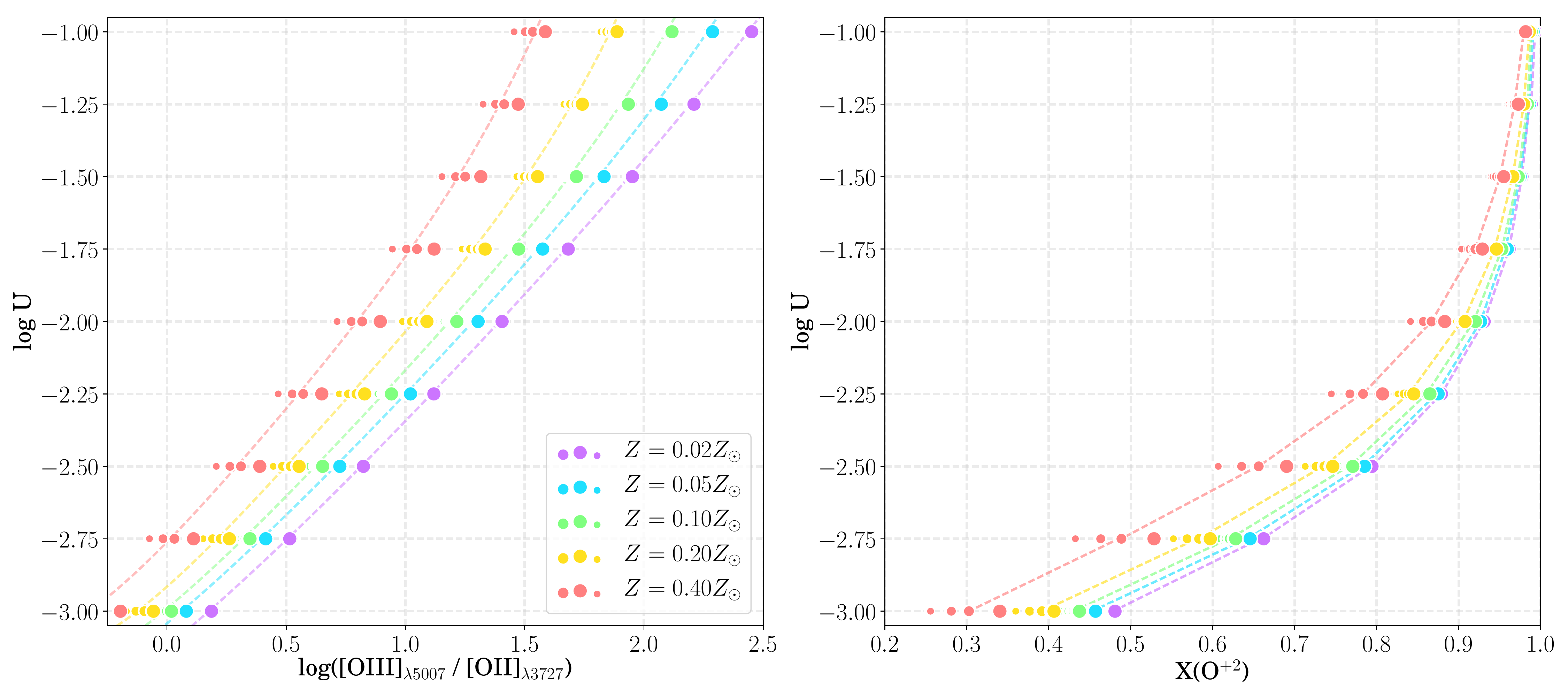}
\caption{
Ionization parameter, log $U$, as a function of ({\it left}) log([\ion{O}{3}] \W5007/[\ion{O}{2}] \W3727) 
and ({\it right}) O$^{+2}$ ionization fraction from {\sc cloudy} models. 
Model metallicity is indicated by point color and burst age from $t=10^{6.0}-10^{7.0}$ yrs
is denoted by increasing point size. 
Each metallicity model is fit with a polynomial of the shape:
log $U = c_3\cdot{x^2} + c_2\cdot{x} + c_1$, where 
$x = \rm{log} O_{32} = $ log([\ion{O}{3}] \W5007/[\ion{O}{2}] \W3727).
The coefficients parametrizing these fits are given in Table~\ref{tbl:coeff}.}
\label{fig4}
\end{center}
\end{figure*}


\subsection{Ionic And Total Abundances} \label{sec:ionic}

Ionic abundances relative to hydrogen are calculated using:
\begin{equation}
	{\frac{N(X^{i})}{N(H^{+})}\ } = {\frac{I_{\lambda(i)}}{I_{H\beta}}\ } {\frac{j_{H\beta}}{j_{\lambda(i)}}\ },
	\label{eq:Nfrac}
\end{equation}
where the emissivity coefficients, $j_{\lambda(i)}$, are functions of both 
temperature and density.

Photoionization models show contributions
from O$^{+3}$ (requiring an ionization potential of 54.9 eV) 
are typically negligible in \ion{H}{2} regions (also see Figure~5).
Further, we do not detect any \ion{O}{4} \W\W1401,1405 emission,
and so the total oxygen abundances (O/H) are calculated from the simple 
sum of O$^{+}$/H$^{+}$ and O$^{+2}$/H$^{+}$. 
We note, however, the presence of \ion{He}{2} emission in some of our sample, which has a similar ionization potential as \ion{O}{4}.
Given the difficulties to reproduce the observed \ion{He}{2} emission using photoionization models 
in recent studies \citep[e.g.,][]{kehrig15,kehrig18,berg18a}, the O$^{+3}$ contribution
may also be under-predicted.  On the other hand, because of the relatively high emissivity of
the \ion{He}{2} $\lambda$4686 line, and the relative weakness of this line when present,
this under-prediction cannot be very large.
Traditionally, O$^{+}$/H$^{+}$ is determined from the optical [\ion{O}{2}] \W3727 blended line. 
Due to the limited SDSS blue wavelength coverage and the
low redshifts in our sample, [\ion{O}{2}] \W3727 is not detected in our targets. 
Alternatively, O$^{+}$/H$^{+}$ is determined from the 
optical red [\ion{O}{2}] \W\W7320,7330 doublet \citep[e.g.,][]{kniazev04,berg16}.

Other abundance determinations require ionization correction 
factors (ICF) to account for unobserved ionic species. 
For nitrogen, we adopt the convention of N/O = N$^{+}$/O$^{+}$ 
\citep{peimbert67}, which is valid at a precision of about 10\% \citep{nava06}.
Collisionally excited emission lines for $\alpha$-elements are also observed
in many of the SDSS spectra.
For Ne we employ the ICF of \citet{peimbert69}, which has been confirmed by 
\citet{crockett06}: ICF(Ne) = (O$^+$ + O$^{+2}$)/ O$^{+2}$.
However, unlike the simple ionization structure of nitrogen or neon, 
for sulfur, both S$^{+2}$ and S$^{+3}$ lie in the O$^{+2}$ zone, and
for argon, Ar$^{+2}$ spans both the O$^+$ and O$^{+2}$ zones.
To correct for the unobserved S and Ar states, we employ the ICFs from \citet{thuan95}: 

\footnotesize{
\begin{align}
        \rm{ICF(S)}  & =  \frac{\rm{S}}{\mbox{S}^{+} + \rm{S}^{+2}} \nonumber \\
        			   & = \big[0.013 + x\{5.10 + x[-12.78 + x(14.77 - 6.11x)]\}\big]^{-1},  \\
        \rm{ICF(Ar)} & =  \frac{\rm{Ar}}{\mbox{Ar}^{+2} + \rm{Ar}^{+3}} \nonumber \\
        			   & = \big[0.99+ x\{0.091 + x[-1.14 + 0.077x]\}\big]^{-1},  \\
			   & =  \frac{\rm{Ar}}{\mbox{Ar}^{+2}} \nonumber \\
        			   & = \big[0.15+ x(2.39 - 2.64x)\big]^{-1}, 
\end{align}}
\normalsize 
where $x =$ O$^{+}$/O and Equation~(7) is used when [\ion{Ar}{4}] \W4740 is not observed.
Ionic and total O, N, S, Ne, and Ar abundances determined from the optical spectra are listed for 
our targets in Tables~$11-13$ in the appendix.

\subsection{The C/O Abundance Ratio} \label{sec:co}

In the simplest case, C/O can be determined from the C$^{+2}$/O$^{+2}$ ratio alone.
Since O$^{+2}$ has a higher ionization potential than C$^{+2}$ (54.9 eV versus 47.9 eV, respectively), 
regions ionized by a hard ionizing spectrum may have a significant amount of carbon in the C$^{+3}$ form, 
causing the C$^{+2}$/O$^{+2}$ ionic abundance ratio to underestimate the true C/O abundance. 
The metallicity dependence of the stellar continua, the stellar mass-$T_{\rm{eff}}$
relation \citep{maeder90}, and stellar mass \citep{terlevich85} will also systematically 
affect the relative ionization fractions of these species.  
To correct for this effect, we apply a C ICF:
\begin{align}
	{\frac{\mbox{C}}{\mbox{O} }} & = {\frac{\mbox{C}^{+2}}{\mbox{O}^{+2}}\ }\times \Bigg[{\frac{X(\mbox{C}^{+2})}{X(\mbox{O}^{+2})}}\Bigg]^{-1} \nonumber \\
			     			    & = {\frac{\mbox{C}^{+2}}{\mbox{O}^{+2}}\ }\times{\mbox{C ICF}},	
\end{align}
where X(C$^{+2}$) and X(O$^{+2}$) are the C$^{+2}$ and O$^{+2}$ volume fractions, respectively.


\begin{figure}
\begin{center}
	\includegraphics[scale=0.35,trim=0mm 0mm 0mm 0mm,clip]{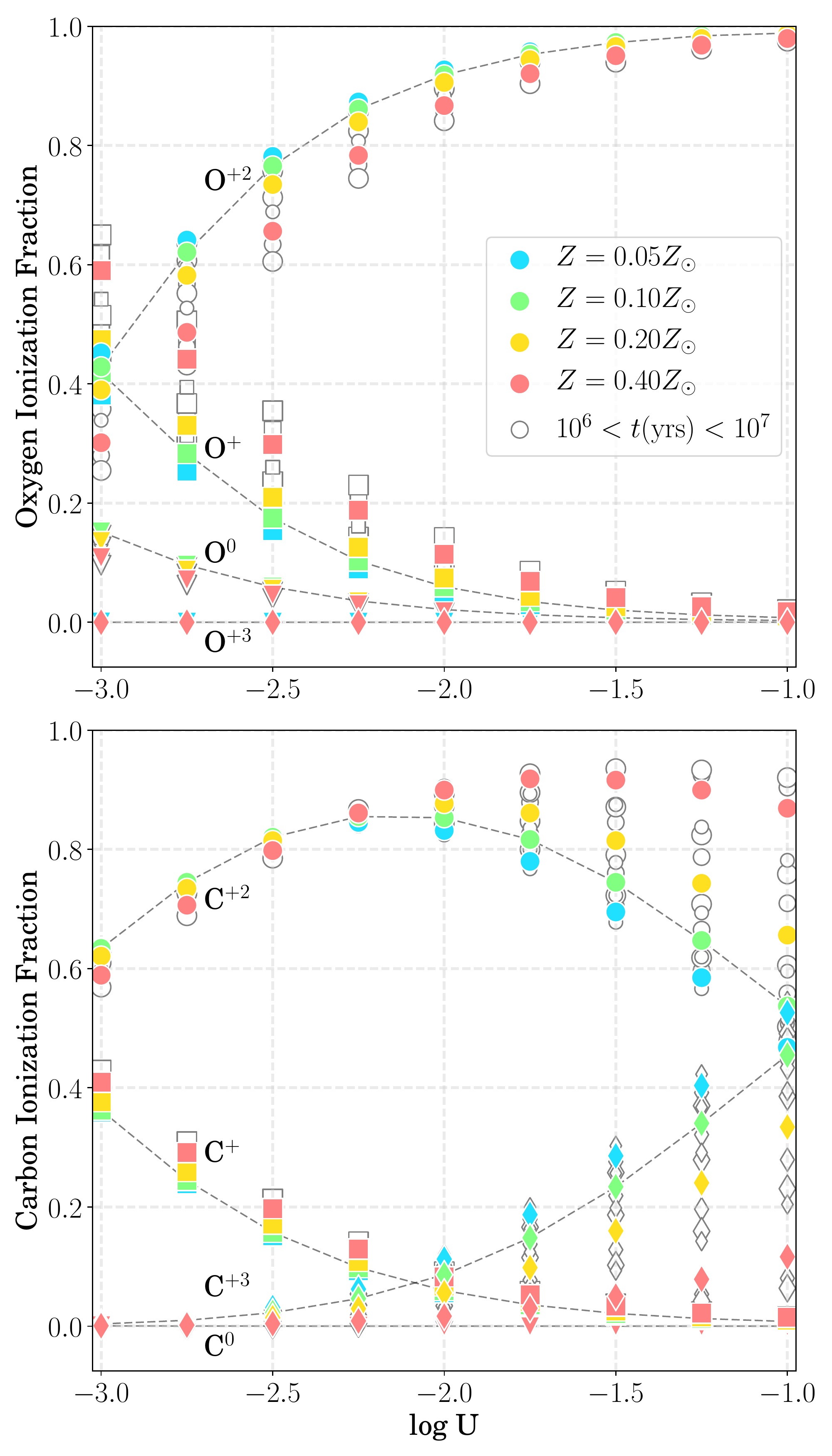}
\caption{The ionization fractions of O and C species as a function of ionization parameter.
Open symbols designate deviations due to burst age, with size increasing from $t=10^{6.0}$ to $t=10^{7.0}$ years.
Solid symbols are color coded by the gas-phase oxygen abundance at $t=10^{6.7}$ yrs.
In each panel the different ionic species are designated by triangles, circles, squares, and diamonds in order of increasing ionization. 
Dashed gray lines connect the $Z=0.1\times Z_{\odot}$ models, or 12+log(O/H) = 7.7,
which is a typical nebular abundance for our sample.}
\label{fig5}
\end{center}
\end{figure}


\begin{figure}
\begin{center}
	\includegraphics[scale=0.3175,trim=0mm 0mm 228mm 0mm,clip]{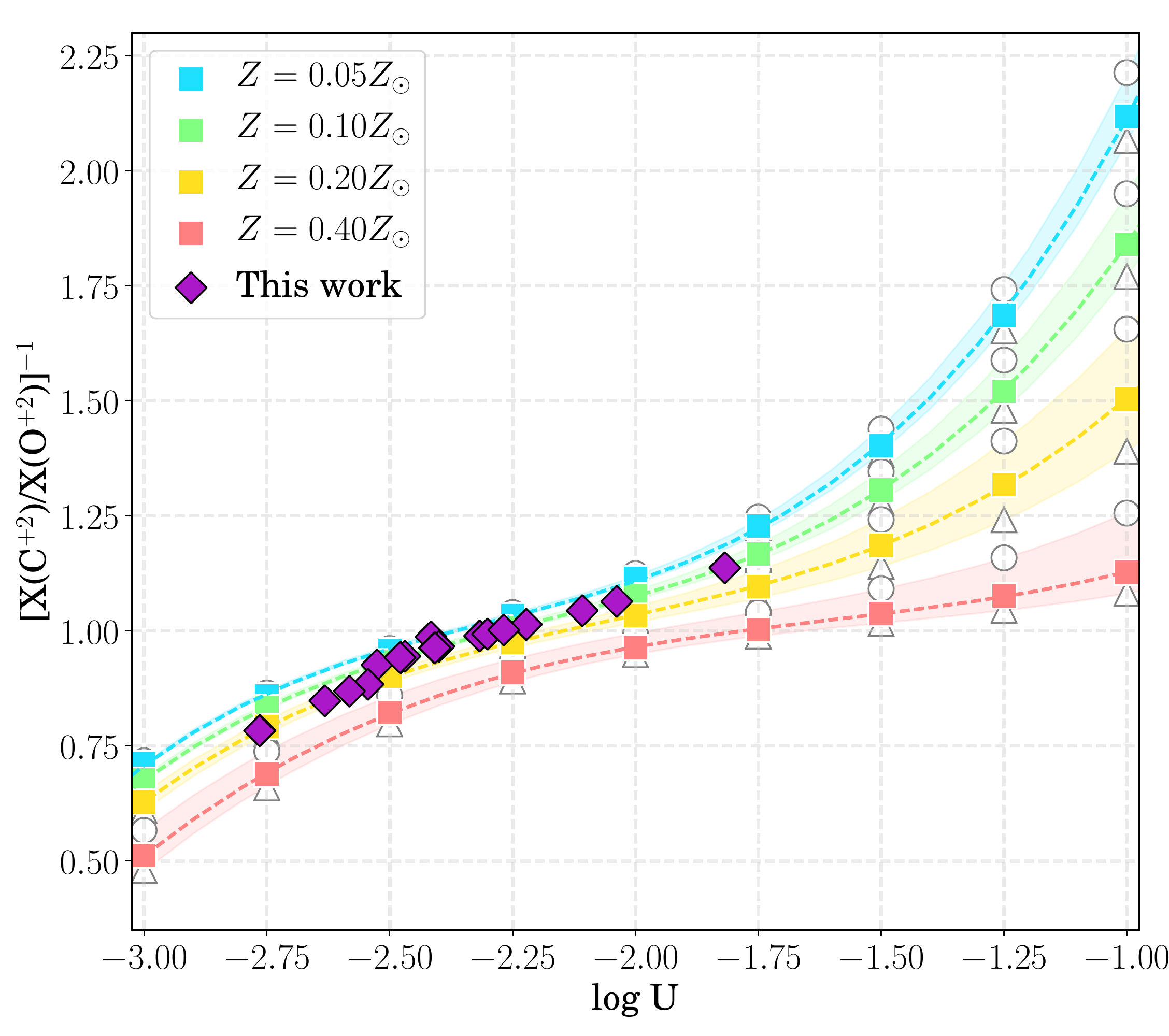}
	\includegraphics[scale=0.3175,trim=0mm 0mm 228mm 0mm,clip]{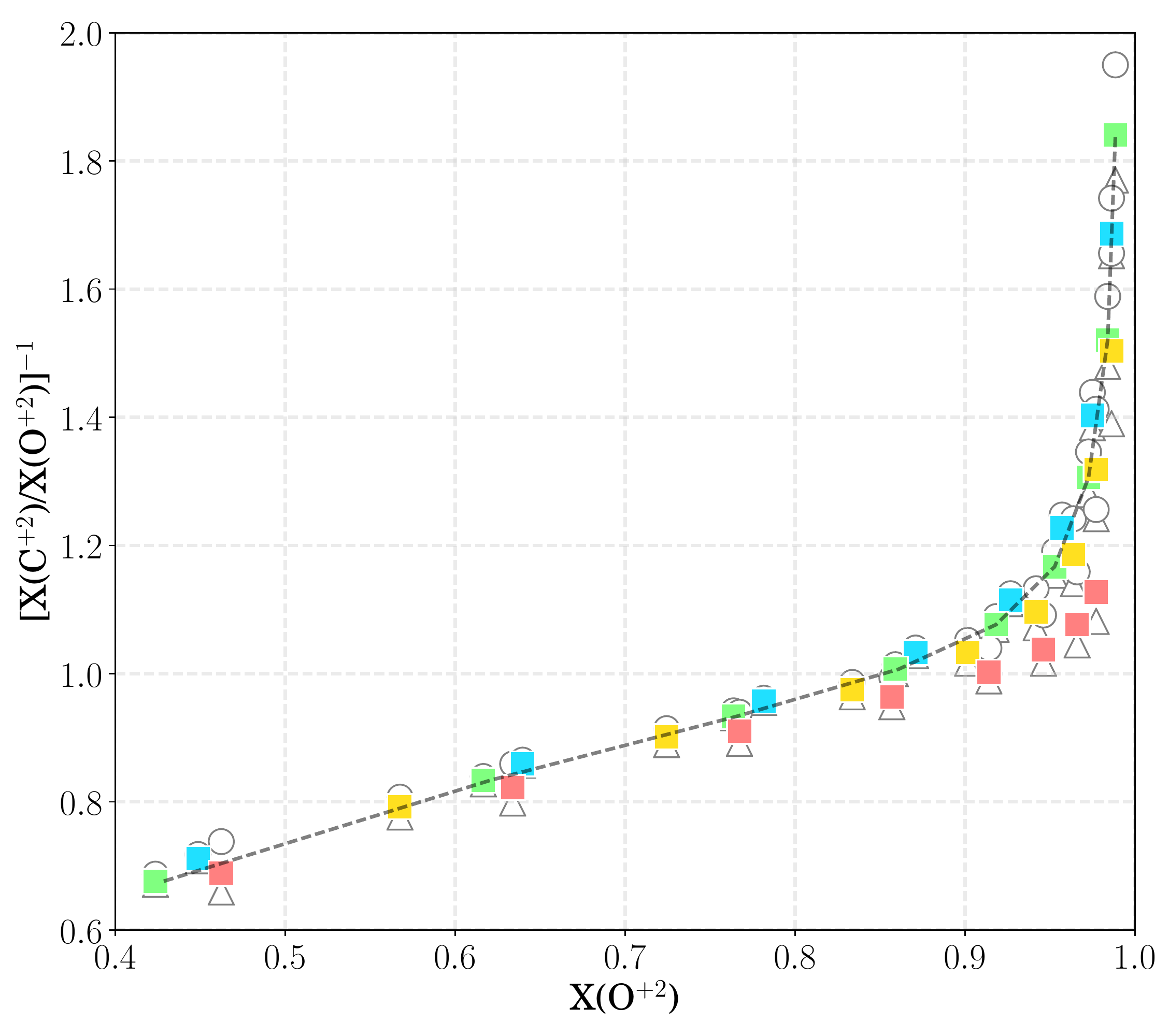}
\caption{
{\sc cloudy} models of the C ionization correction factor versus 
ionization parameter in the top panel and O$^{+2}$ ionization fraction in the bottom panel.
Symbols are color coded by the gas-phase oxygen abundance.
Open symbols demonstrate the effects of burst age, where circles are for a burst age of 
$t=10^{6.0}$ yrs and triangles for $t=10^{7.0}$ yrs.
In the top plot, the best polynomial fit for each metallicity model is plotted as a dashed line,
where the shaded bands span the difference in burst age.
Polynomial fit coefficients are given in Table~\ref{tbl:coeff}.
Far less scatter is seen when the C ICF is plotted versus O$^{+2}$ ionization fraction,
and so a single fit is sufficient.}
\label{fig6}
\end{center}
\end{figure}


We estimate the ICF as a function of the ionization parameter using {\sc cloudy} 17.00 \citep{ferland13}.
In B16 we used Starburst99 models \citep{leitherer99}, with and without rotation, for a region of continuous star formation.
Since the targets in our sample are capable of producing very high ionization emission lines
(see Figures~2 and 3),
we ran new {\sc cloudy} models using ``Binary Population and Spectral Synthesis'' \citep[BPASSv2.14;][]{eldridge16, stanway16}
burst models for the input ionizing radiation field.
For these models we intentionally covered parameter space appropriate to our sample, including an
age range of $10^{6.0}-10^{7.0}$ yrs for our young bursts, a range in ionization parameter of $-3.0 <$ log $U <-1.0$,
with matching stellar and nebular metallicities  (Z$_{\star}$ = Z$_{neb} =0.001,0.002,0.004,0.008 = 0.05,0.10,0.20,0.40$ Z$_{\odot}$) 
that more than cover our observed gas-phase abundances ($7.5<$ 12+log(O/H) $< 8.0$ or 0.065 Z$_{\odot} < $ Z$_{neb} <0.20$ Z$_{\odot}$). 
The {\sc gass10} solar abundance ratios within {\sc cloudy} were used to initialize the relative gas-phase abundances. 
These abundances were then scaled to cover the desired range in nebular metallicity, 
and relative C, N, and Si abundances ($0.25 <$ (X/O)/(X/O)$_{\odot} < 0.75$).
The ranges in relative N/O, C/O, and Si/O abundances were motivated by the
observed values for nearby metal-poor dwarf galaxies \citep[e.g.,][]{garnett99,berg12, berg16}. \looseness=-2


\begin{deluxetable}{crrrr}
\tablecaption{Coefficients for Ionization Correction Model Fits}
\tablehead{
\multicolumn{1}{c}{} 	& \multicolumn{4}{c}{$Z(Z_{\odot})$} \\ 
\cline{2-5}
\CH{$y = f(x)$} 			& \CH{0.05}	& \CH{0.10} 		& \CH{0.20} 	& \CH{0.40}}
\startdata	
\multicolumn{1}{l}{\bf{log $U$:}} & & & & \\
\multicolumn{1}{l}{$x =$ log $O_{32}$} & & & & \\
{$c_3$ ..............}		& 0.0887		&  0.1203			& 0.1668		& 0.2231	\\
{$c_2$ ..............}		& 0.6941		& 0.6858			& 0.6985		& 0.7774	\\
{$c_1$ ..............}		& $-3.0390$	& $-2.9830$		& $-2.9050$	& $-2.7610$ \\
\vspace{-1ex} \\
\multicolumn{1}{l}{\bf{C ICF:}} & & & & \\
\multicolumn{1}{l}{$x =$ log $U$} & & & & \\
{$c_4$ ..............}		& 0.3427		& 0.2807  			& 0.2038  		& 0.1224  \\	
{$c_3$ ..............}		& 2.3588    	& 1.8670			& 1.2553  		& 0.5887  \\	
{$c_2$ ..............}		& 5.6825    	& 4.4009  			& 2.8099  		& 1.0722  \\	
{$c_1$ ..............}		& 5.7803 		& 4.6543			& 3.2636		& 1.7340	\\
\vspace{-2ex} 
\enddata
\tablecomments{
{\sc cloudy} photoionization model fits of the form 
$f(x)=c_4\cdot{x^3}+c_3\cdot{x^2}+c_2\cdot{x}+c_1$.
The model grids and polynomial fits are shown in Figures~4, 5, and 6.
 }
\label{tbl:coeff}
\end{deluxetable}


These {\sc cloudy} models allow us to plot the ionization parameter, log $U$, 
as a function of both log([\ion{O}{3}] \W5007/[\ion{O}{2}] \W3727) and 
O$^{+2}$ ionization fraction in Figure~4.
The points are color coded by model stellar/nebular metallicity, showing the dependence of the 
[\ion{O}{3}] \W5007/[\ion{O}{2}] \W3727 line ratio on both metallicity and ionization.
The age of the burst has a small effect, where $t=10^{6.0}-10^{7.0}$ yrs
are denoted by increasing point size. 
We fit each of the metallicity models with a polynomial of the shape:
log $U = c_3\cdot{x^2} + c_2\cdot{x} + c_1$, where 
$x = \rm{log} O_{32} = $ log([\ion{O}{3}] \W5007/[\ion{O}{2}] \W3727).
The coefficients for these fits are listed in Table~3.


\begin{figure*}
	\includegraphics[scale=0.34,angle=0,trim=0mm 0mm 0mm 0mm]{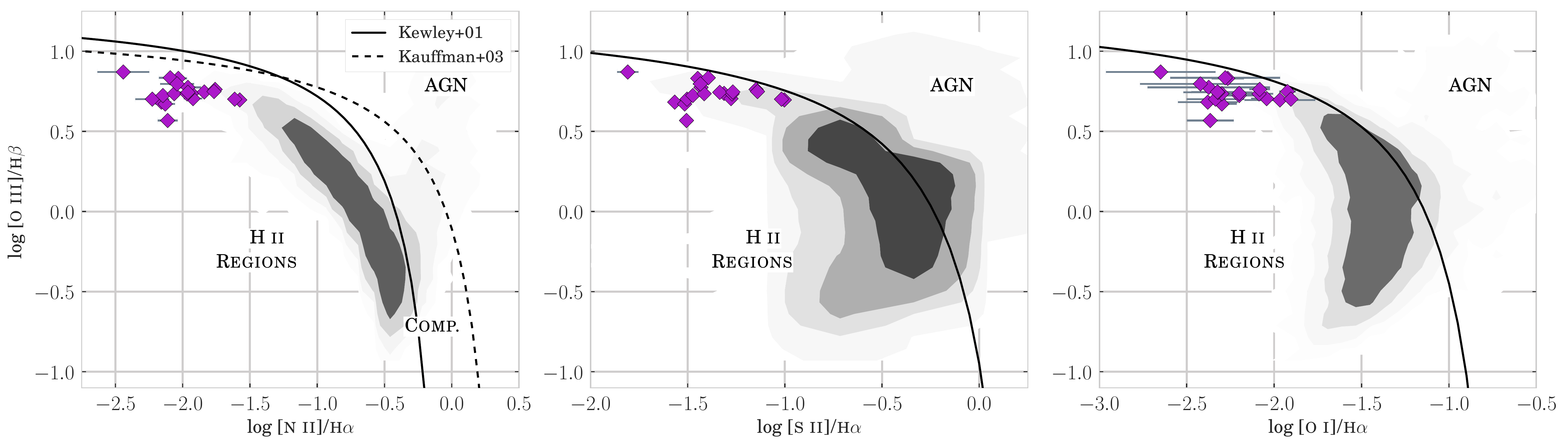}
\caption{Optical emission line ratios for the dwarf galaxies in our sample, as measured from their SDSS spectra. 
A low-mass subset ($M_{\star} < 10^9 M_{\odot}$) of the SDSS from the MPA-JHU database is plotted 
in gray as a comparison sample. The contours represent the $1-$,$2-$,$3-$, and $4-\sigma$ levels of
the probability density function (PDF).
Defined by the theoretical starburst \citep[solid line;][]{kewley01} and AGN \citep[dashed line;][]{kauffmann03} limits,
the parameter space is divided into three sections: photoionized \ion{H}{2} regions (below the solid line), 
composite regions (between the line), and AGN (above the dashed line).
In all three panels the optical line ratios of our sample lie below the theoretical starburst limits (solid line),
distinguishing them as photoionized \ion{H}{2} regions.}
\label{fig7} 
\end{figure*}


The ionization fractions of C and O species as a function of ionization parameter, metallicity,
and burst age are shown in Figure~5.
For the most part, the resulting trends are very similar to those in B16 for the Starburst99 models with and without rotation.
One exception is the increased dependence of C$^+$ and C$^{+2}$ on metallicity at high ionization parameter (log $U > -2.0$).
However, this is not a significant concern for the current sample given its low metallicity.

From the ratio of the modeled C$^{+2}$ and O$^{+2}$ ionization fractions,
the C/O ICF is plotted in Figure~6.
The light color shading depicts the minimal variation in the C/O ICF with burst age,
centered on models with an age of $t=10^{6.7}$ yrs (colored lines) and extending from $t=10^{6.0}-10^{7.0}$ yrs.
However, the ICF is sensitive to metallicity, and has a smoothly varying dependence on
excitation \citep[as highlighted by][]{garnett95}, 
so we fit each of the four metallicity models with a polynomial of the shape:
ICF(C) = $c_4\cdot{x^3} + c_3\cdot{x^2} + c_2\cdot{x} + c_1$,
where $x =$ log$U$ and the $c$ coefficients are listed in Table~3.

Using the analytic fits determined here (see Table~3),
the measured log $O_{32}$ values for our sample were used to estimated  
ionization parameters.
These ionization parameters were subsequently used to determine the C ICF,
and then correct the C/O abundances.
The resulting C ICFs for our sample are plotted in the top panel of Figure~6, showing
only a small correction for the majority of our sample that lies between log$U=-2.5$ to $-2.0$.
We estimate the uncertainty in the ICF as the scatter amongst the different
models considered (relative abundances and burst age) at a given O$^{+2}$ volume fraction. 
Ionization parameters, C ICFs, ionic and total C abundances, as well as the corrected C/O ratio, 
are provided in Tables~$11-13$ in the appendix.

As mentioned in section 3.3, we noted an anomaly in the spectrum of J093006 (see Figure~3), where the
\ion{O}{3}] \W1661 line is detected but the \W1666 line (the stronger of the doublet) is absent.
This effect is also seen in the UV spectrum of SB191 in \citet{senchyna17},
but with an apparent absorption feature replacing the expected \ion{O}{3}] \W1666 emission.
Investigating SBS191 further, we find that the galaxy sits at an unfortunate redshift ($z\sim0.003$) 
such that the \ion{O}{3}] \W1666 emission is coincident with Milky Way \ion{Al}{2} \W1671 absorption.
Interestingly, the two largest C/O outliers in B16 (J082555 and J120122) also have 
redshifts of $z\sim0.003$ that make their \ion{O}{3}] detections susceptible to MW contamination.
This is not the case, however, for J093006 or J084956: the two galaxies in our sample without
\ion{O}{3}] \W1666 detections which both have a redshift of $z\sim0.014$.
For targets with significant \ion{O}{3}] \W1661 but lacking or weaker-than-expected \ion{O}{3}] \W1666, 
we have recalculated their C/O abundances using only the \ion{O}{3}] \W1661 detections.

Note that the C and O abundances presented here have not been corrected 
for the fraction of atoms embedded in dust.
\citet{peimbert10} have estimated that the depletion of O ranges between 
roughly 0.08$-$0.12 dex, and has a positive correlation with O/H abundance.
C is also expected to be depleted in dust, 
mainly in polycyclic aromatic hydrocarbons and graphite.
The estimates of the amount of C locked up in dust grains in the local 
interstellar medium shows a relatively large variation depending on the 
abundance determination methods applied \citep[see, e.g.,][]{jenkins14}.
For the low abundance targets presented here, and their corresponding 
small extinctions, the depletion onto dust grains is likely small,
and therefore no correction is applied.


\section{PROPERTIES OF HIGH-IONIZATION DWARF GALAXIES} \label{sec:prop}


\subsection{Diagnostic Diagrams}
Many of the targets in our extended sample exhibit strong high-ionization emission lines. 
In Figures~\ref{fig2} and \ref{fig3} we see that significant \ion{C}{4} \W\W1548,1550 and 
\ion{He}{2} \W1640 emission is not uncommon for the high-ionization sample presented here.
The presence of collisionally excited \ion{C}{4} and \ion{He}{2} from recombination indicates 
very hard radiation fields are needed to reach their ionization potentials of 47.9 eV and 54.4 eV respectively.
Such emission line features are more commonly observed in high energy objects, such as active galactic nuclei (AGN). 
While previous AGN studies have shown that large \ion{C}{4}/\ion{C}{3}] ratios ($\sim7.5$) are typical of 
narrow-line AGN \citep[e.g.,][]{alexandroff13}, B16 found that \ion{C}{4}/\ion{C}{3}] $< 1.0$ can be found
in high-ionization star-forming galaxies.
Similarly, all of our targets have \ion{C}{4}/\ion{C}{3}] flux ratios that are $< 1.0$. \looseness=-2

Using the standard BPT optical emission-line diagnostics \citep{baldwin81},
we show in Figure~7 that the targets in our sample have properties which 
are consistent with photoionized \ion{H}{2} regions.
Line measurements for the current sample are plotted as purple diamonds in comparison to 
the gray locus of low-mass ($M_{\star} < 10^9 M_{\odot}$) SDSS dwarf star-forming galaxies 
taken from the MPA-JHU database. 
The solid lines represent the theoretical starburst limits from \citet{kewley01},
and the dashed line is the AGN boundary derived by \citet{kauffmann03}. 
Based on these plots, our sample exhibits the expected optical properties of typical photoionized regions. 
In the first two panels of Figure~7 our sample is located on the low-end tail of the 
log([\ion{N}{2}] \W6584/H\A) and log([\ion{S}{2}] \W\W6717,6731/H\A) distributions,
as is expected for metal-poor galaxies.
In the last panel, the low log([\ion{O}{1}] \W6300/H\A) ratios of our sample indicate that shock
excitation has a negligible contribution to the ionization budget.


\begin{figure}
\begin{center}
	\includegraphics[scale=0.315,angle=0,trim=5mm 50mm 30mm 65mm,clip]{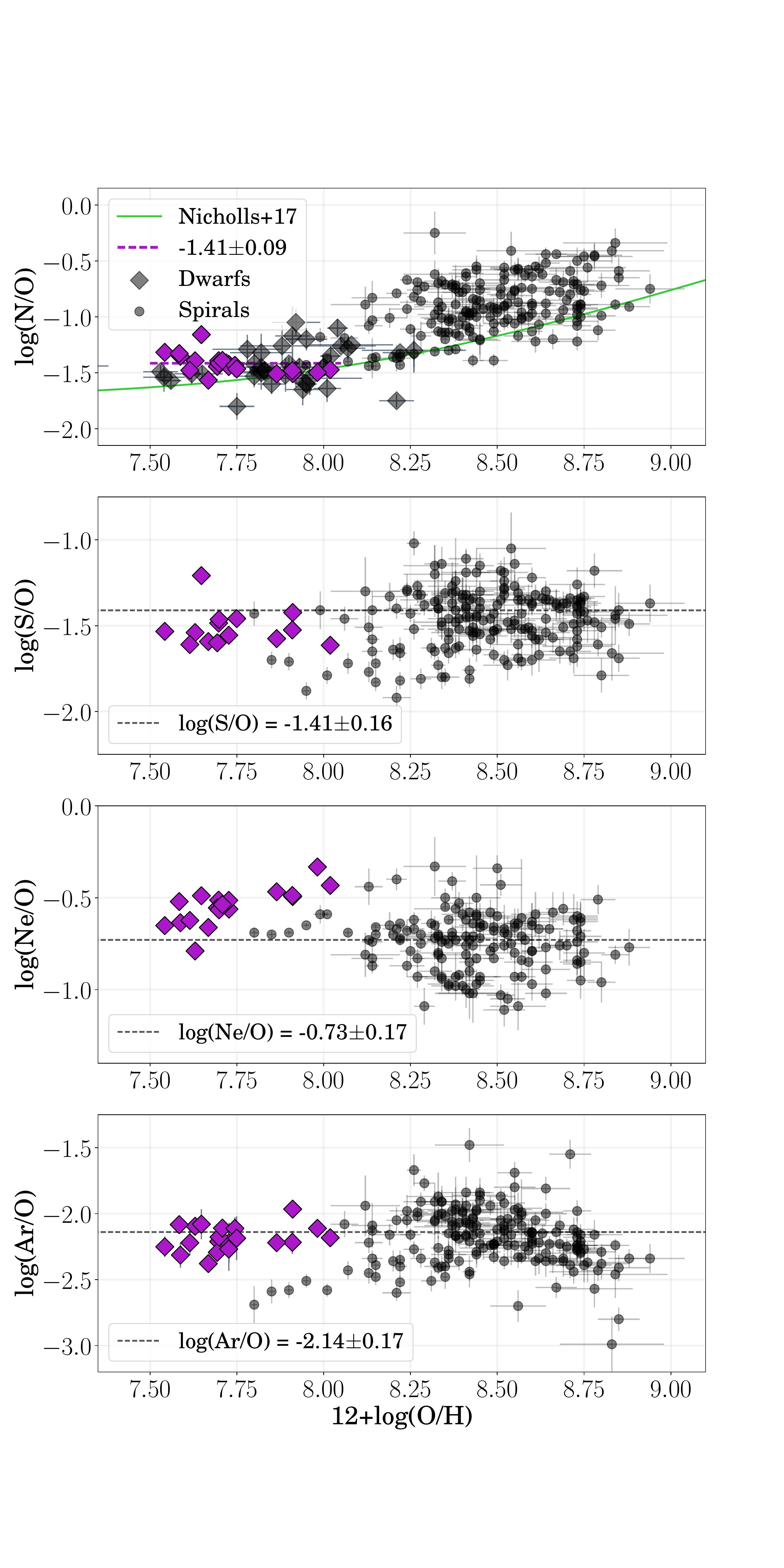}
\caption{
Relative enrichment of the N and \A-elements for our sample (purple diamonds). 
To extend the oxygen abundance range, we plot direct abundances from \ion{H}{2} regions of spiral galaxies 
taken from the CHAOS survey \citep{berg15,croxall15,croxall16,berg19} as gray circles.
We also plot additional dwarf galaxy N/O abundances from \citet{vanzee06a} and \citet{berg12} as gray diamonds.
The purple dashed line is the N/O weighted-mean for our sample and
the gray dashed lines in the bottom three panels denote the weighted-mean values of the CHAOS observations.}
\label{fig8} 
\end{center}
\end{figure}


\subsection{Relative N, S, Ne, and Ar Abundances}
We plot the relative abundance ratios derived from the optical spectra for N, S, Ne, and Ar
in Figure~\ref{fig8}.
For comparison, we plot direct abundances from \ion{H}{2} regions of spiral galaxies 
taken from the CHAOS survey \citep{berg15,croxall15,croxall16,berg19} as gray circles,
which extends the oxygen abundance range covered by CEL measurements. 
For nitrogen, we plot additional dwarf galaxies from \citet{vanzee06a} and \citet{berg12},
and it is shown that our sample lies to the left of the characteristic knee in the N/O versus O/H relationship 
around 12+log(O/H)$\sim 8.0$ \citep[e.g.,][]{henry00,vanzee06a,berg12},
defined by the transition from primary nitrogen enrichment to primary $+$ secondary nitrogen contributions.
As a visual aid, we plot the empirical fit to stellar observations by \citet{nicholls17} as a a solid-green line.
For our sample, we find a weighted-average of log(N/O) $= -1.41\pm0.09$ (purple dashed-line).
This is consistent with the value of log(N/O) $= -1.46\pm0.14$ found by 
\citet{izotov99} for their metal-poor blue compact galaxies with oxygen abundances of 
12$+$log(O/H) $> 7.60$. \looseness = -2

Since \A-elements are predominantly produced on relatively short timescales by 
type II supernovae (SNe; massive stars) explosions,
we expect the \A-element ratios in the bottom three panels of Figure~\ref{fig8} 
to be constant.
Indeed, our S/O, Ne/O, and Ar/O values are consistent with the dispersion of the CHAOS 
observations and their weighted-mean values, denoted by the gray dashed lines.


\begin{figure*}	
\begin{center}
	\includegraphics[scale=0.3,angle=0,trim=0mm 0mm 0mm 0mm]{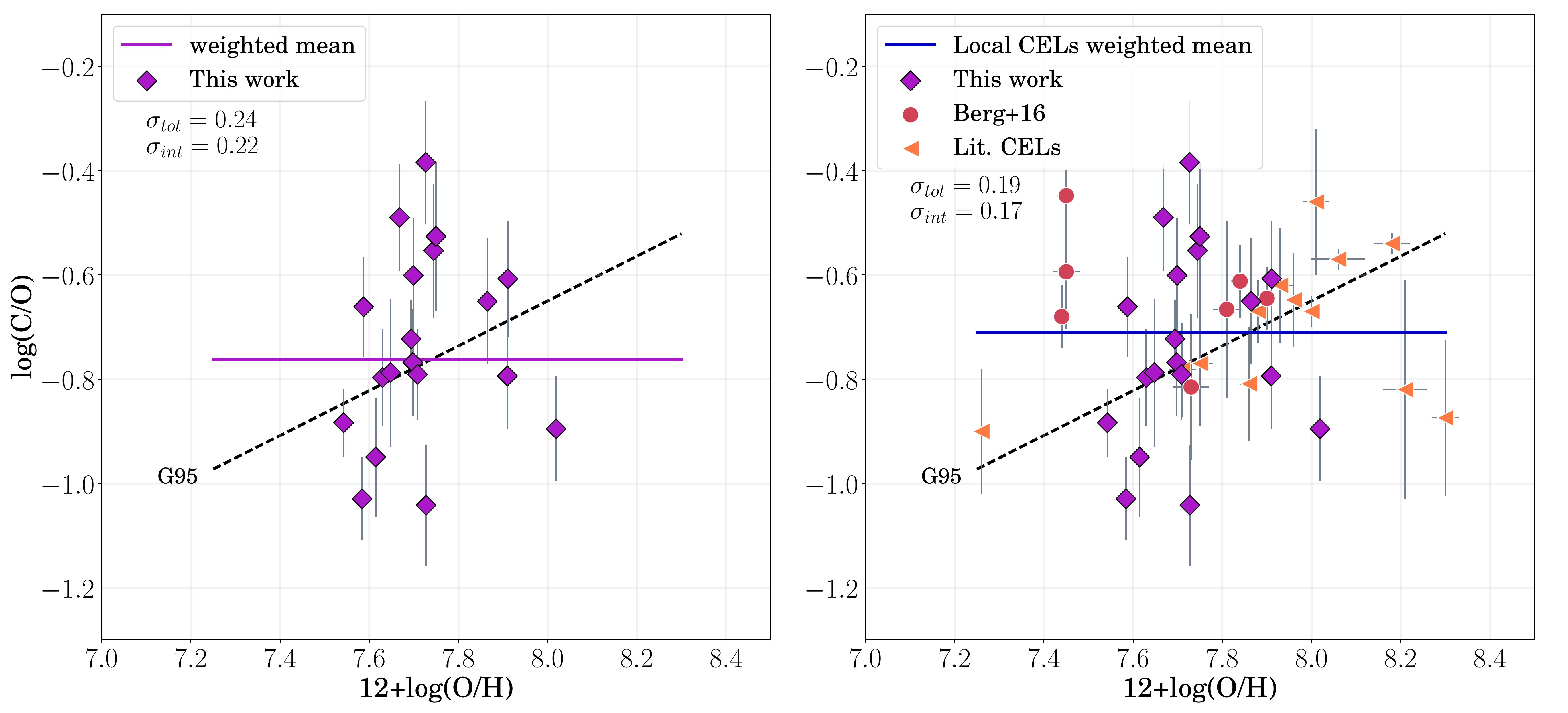} 
\caption{
{\it Left:} Carbon to oxygen ratio vs. oxygen abundance for the 19 star forming galaxies presented in this work.
The data are compared to the weighted mean of the data (solid purple line) and the
increasing linear relationship of Garnett et al. (1995: G95, dashed line).
No clear trend emerges, but the significant total scatter ($\sigma_{tot}$) is largely intrinsic ($\sigma_{int}$),
and not due to observational uncertainties. 
{\it Right:} The C/O and O/H abundances for our sample (purple diamonds) compared to the 
\citet{berg16} sample (red circles) and other CEL measurements from the literature \citep[orange triangles;][]{senchyna17,pena-guerrero17}.
Again, the combined sample of 40 local galaxies with CEL C/O measurements can be fit with a
flat trend, albeit a large dispersion. }
\end{center}
\end{figure*}

\begin{figure*}	
\begin{center}
\begin{tabular}{cc}
	\includegraphics[scale=0.31,angle=0,trim=0mm 0mm 0mm 0mm]{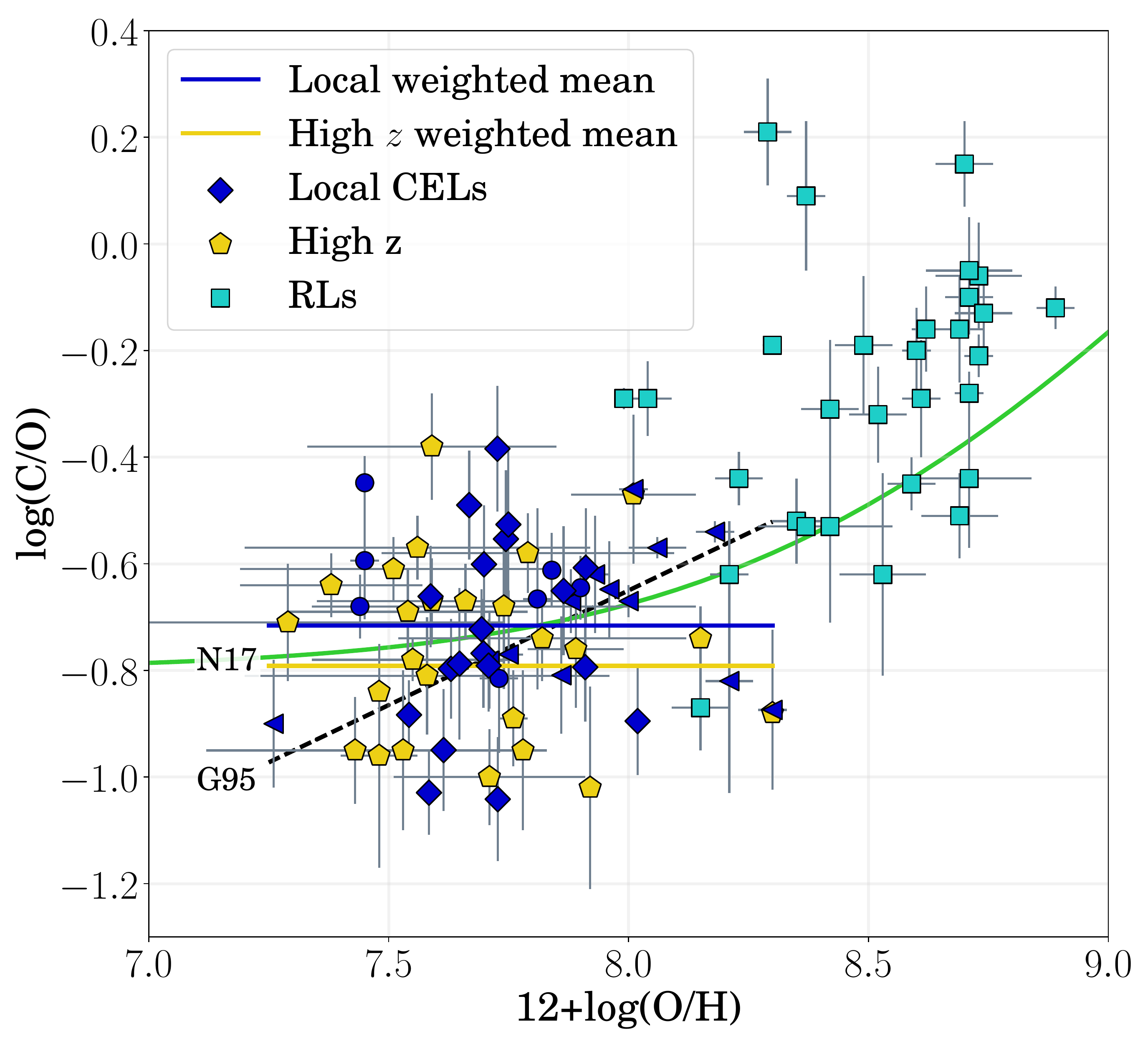}  &
	\includegraphics[scale=0.31,angle=0,trim=0mm 0mm 0mm 0mm]{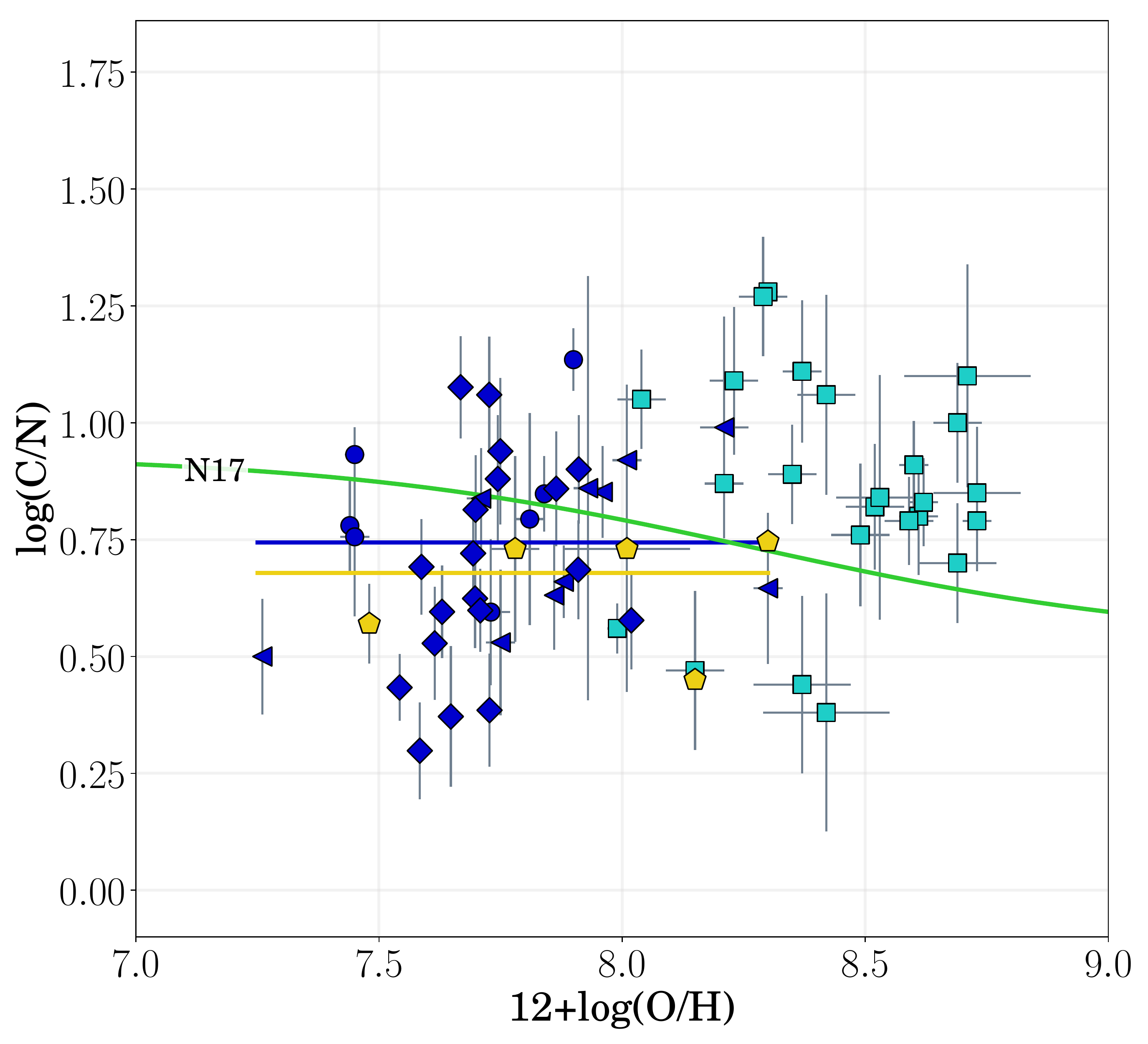}\label{fig10}
\end{tabular}
\caption{
{\it Left:} Carbon to oxygen ratio vs. oxygen abundance for star forming galaxies.
Local galaxies with CEL C/O measurements are plotted as blue points, where the symbol
types correspond to those in Figure~9.
Targets with C/O detections at intermediate redshifts ($z\sim2-3$) are shown as yellow pentagons.
Located at larger oxygen abundances, teal squares represent star forming galaxies with recombination 
line abundance determinations \citep{esteban02,pilyugin05,garcia-rojas07,lopez-sanchez07,esteban09,esteban14}.
The dashed line is the least-squares fit from \citet{garnett95}.
{\it Right:} Carbon to nitrogen abundance vs. oxygen abundance.
C/N appears to be relatively constant across oxygen abundance, suggesting carbon may
follow nitrogen in originating from primary (secondary) production at low (high) values of O/H,
as suggested by previous studies \citep{garnett99,esteban14,berg16}. 
However, there is significant scatter ($\sigma=0.20$ dex) present. 
In both plots the solid blue line is the weighted mean of the significant CEL C/O detections at $z\sim0$ (blue symbols),
the solid yellow line is the weighted mean of the high-$z$ ($2\lesssim z \lesssim 3$) detections,
and the solid green line is the empirical stellar curve from \citet{nicholls17}.}
\end{center}
\end{figure*}


\subsection{Relative C/O and C/N Abundances}
We present the C/O versus O/H relationship for our sample of 19 star forming galaxies in the left panel of Figure~9.
The large dispersion in C/O is striking, especially given the small range in O/H.
We determine the weighted mean and the scatter of the C/O abundances using 
the {\sc mpfitexy} IDL code, which uses a least-squares fitting algorithm that 
allows for errors in both the x and y coordinates. 
The code reports the total dispersion as the RMS distance of the data from the model,
as well as what fraction of the dispersion is observational versus intrinsic to the phenomenon being measured.
We find a mean log(C/O)$=-0.76\pm0.06$ with a total dispersion of $\sigma_{tot}=0.24$ dex
and an intrinsic dispersion of $\sigma_{int}=0.22$ dex.
This suggests that while no clear trend is apparent from our sample
(flat vs. increasing trend), there is significant and real scatter that is not due to observational uncertainties.  \looseness=-2

Next we compare our C/O relative abundances to other measurements in the literature.
In the right hand panel of Figure~9, we also plot the 12 C/O detections from B16 as red circles and 
the 9 additional significant detections from \citet{senchyna17} and \citet{pena-guerrero17} 
as orange triangles to form the most comprehensive sample of local UV CEL C/O detections to date.
The B16 study analyzed the 12 available UV CEL C/O detections at the time,
finding no clear trend evident in C/O versus O/H for 12+log(O/H) $< 8.0$, 
but noted a general trend of increasing C/O with O/H when recombination line observations 
at higher values of O/H were included.
Using our expanded sample of 40 C/O detections, we confirm the results of B16:
when only the CEL C/O detections are considered, the trend in C/O versus O/H
appears to be flat, with a weighted average of log(C/O) $=-0.71$ and large dispersion of
$\sigma=0.17$ dex. \looseness=-2

In Figure 10, the trend is extended to higher oxygen abundances by incorporating C/O determinations 
from optical recombination lines 
\citep[teal squares:][]{esteban02, esteban09, esteban14, pilyugin05, garcia-rojas07, lopez-sanchez07}.
When the CEL and RL data are combined, the C/O trend appears to be consistent with
the original \citet{garnett95} relationship, albeit with significant scatter.

Carbon and oxygen have also been measured from the UV CELs for a handful of $z\sim2-3$ galaxies 
\citep{pettini00, fosbury03, erb10, christensen12, stark14, bayliss14, james14, vanzella16, steidel16, amorin17, rigby17,berg18a}.
These data are plotted as yellow pentagons in Figure~10, and their weighted mean is shown as a solid yellow line.
Interestingly, C/O values for the $z\sim2-3$ galaxies are somewhat lower on average 
(high-$z$ sample: log(C/O)$_{{\rm h}z} = -0.79$) 
than the nearby dwarf galaxies of similar metallicity, but with similar dispersion ($\sigma_{{\rm h}z} = 0.17$). 
\citet{berg18a} suggest that these different redshift populations may be distinguished 
by their ages as higher C/O values could be due to a delayed 
carbon contribution from low- and intermediate-mass stars. 
Given this offset and the real dispersion present in the observed trend, 
we suggest caution when using UV spectra to interpret properties of the distant universe; specifically,   
measurements of the UV C and O emission lines alone should not be used to predict O/H abundance.

In the right panel of Figure~10 we plot C/N versus O/H for the entire dataset. 
Here, we find an average log(C/N) $=0.75$ for the $z\sim0$ CEL data (blue line), 
with a large dispersion that is similar to that of C/O: $\sigma = 0.20$.
Again, the average value for the high-redshift sample is smaller than the local sample,
with an average log(C/N)$_{{\rm h}z} = 0.68$.
The dispersion for the high-redshift sample is significantly smaller ($\sigma_{{\rm h}z} = 0.09$), 
but is based on a small sample of the only five points that have both UV C/O and optical N/O detections.

The absence of a trend in C/N abundance for \ion{H}{2} regions has been reported previously 
using both UV CELs \citep{garnett99, berg16} and RLs \citep{esteban14}.
In comparison, the flat C/N trend measured here is shifted lower than, but in agreement with, 
the nearly constant C/N ratio found by B16 (log(C/N)$\sim0.9$). 
The implication is that carbon is predominantly produced by nucleosynthetic mechanisms which
are similar to those of nitrogen. 
In this scenario, primary carbon production (a flat trend) dominates at low metallicity, 
but metallicity-dependent production (quasi-secondary production / an increasing trend) 
becomes prominent at higher metallicities. \looseness=-2

Further support for a bi-modal C/O relationship can be found in the analytic reduced$-\chi^2$ fits
to stellar abundance data by \citet{nicholls17}.
This fit is shown as a solid green curve in the lefthand panel of Figure~10, 
where the plateau for 12+log(O/H) $< 8.0$ corresponds to primary C production with log(C/O) $= -0.8$,
and the increasing trend at high metallicities is due to the onset of a pseudo-secondary C contribution.
By visual inspection, the empirical stellar trend follows the combined CEL + RL data nicely.
The empirical C/O and N/O trends fit by \citet{nicholls17} also allow us to predict the stellar C/N trend.
We combine their analytic C/O and N/O trends and plot the resulting stellar C/N relationship versus
oxygen abundance as a solid green line in the righthand panel of Figure~10.
The stellar trend suggests that the pseudo-secondary contribution to C could be less metallicity-sensitive
than secondary nitrogen production, as indicated by the shallow negative correlation between C/N 
and O/H.

Our data confirm the similar behavior between C and N production observed from stars. 
Both C and N production appear to be metallicity-dependent, however, the presence of significant scatter 
could indicate that they are synthesized in stars of different average masses.
Then C and N would be returned to the interstellar medium on different timescales. 
Therefore, the dispersion in C/N of our sample may be the result of taking a snapshot of 
many galaxies at different times since their most recent onset of star formation.




\section{THE CHEMICAL EVOLUTION OF C, N, AND O}\label{sec:models}

\subsection{Chemical Evolution Modeling}

Chemical evolution models can be used to investigate the observed abundance trends in this work,
as has been done previously with samples of dwarf irregular galaxies 
\citep[e.g.,][]{matteucci83,matteucci85,bradamante98,yin11}.
By comparing the relative abundances of a number of chemical species with different 
nucleosynthetic production mechanisms, it is possible to investigate and constrain
chemical evolution scenarios. 
For example, \citet{yin11} were able to reproduce the observed spread in the C/O
versus O/H abundance plane by combining simulations of individual bursting systems in 
which the timing, number, and burst duration, as well as metal-enhanced outflow rates, 
varied from model to model.
Through this work, they found that, in addition to variations in the above parameters, 
metal-enhanced winds were necessary to fully explain the observed range in abundance ratios.

Here we investigate the relative abundance of C, N, and O in a similar exercise.
Our greatly enhanced number of C/O observations relative to that used by \citet{yin11}
gives a much better characterization of the behavior in the C/O versus O/H diagnostic diagram.
In order to test how well the C/O vs.\ O/H relationship for our sample
could be explained by variations in a bursty star formation history 
(SFH; i.e., varying the number of bursts, their timing, and their duration) and 
the fraction of newly synthesized oxygen retained by the galaxy (the effective yield, $Y_{\rm{eff}}$(O))
or lost through outflow, we ran a set of numerical models using the {\sc opendisk} chemical evolution code
\citep[for details, see Appendix~\ref{A1} and][]{henry00}.   
The code treats the galaxy as a single, well-mixed zone
and follows the formalism for chemical evolution described in \citet{tinsley80}.


\begin{figure*}
\begin{centering}
   \includegraphics[scale=0.475]{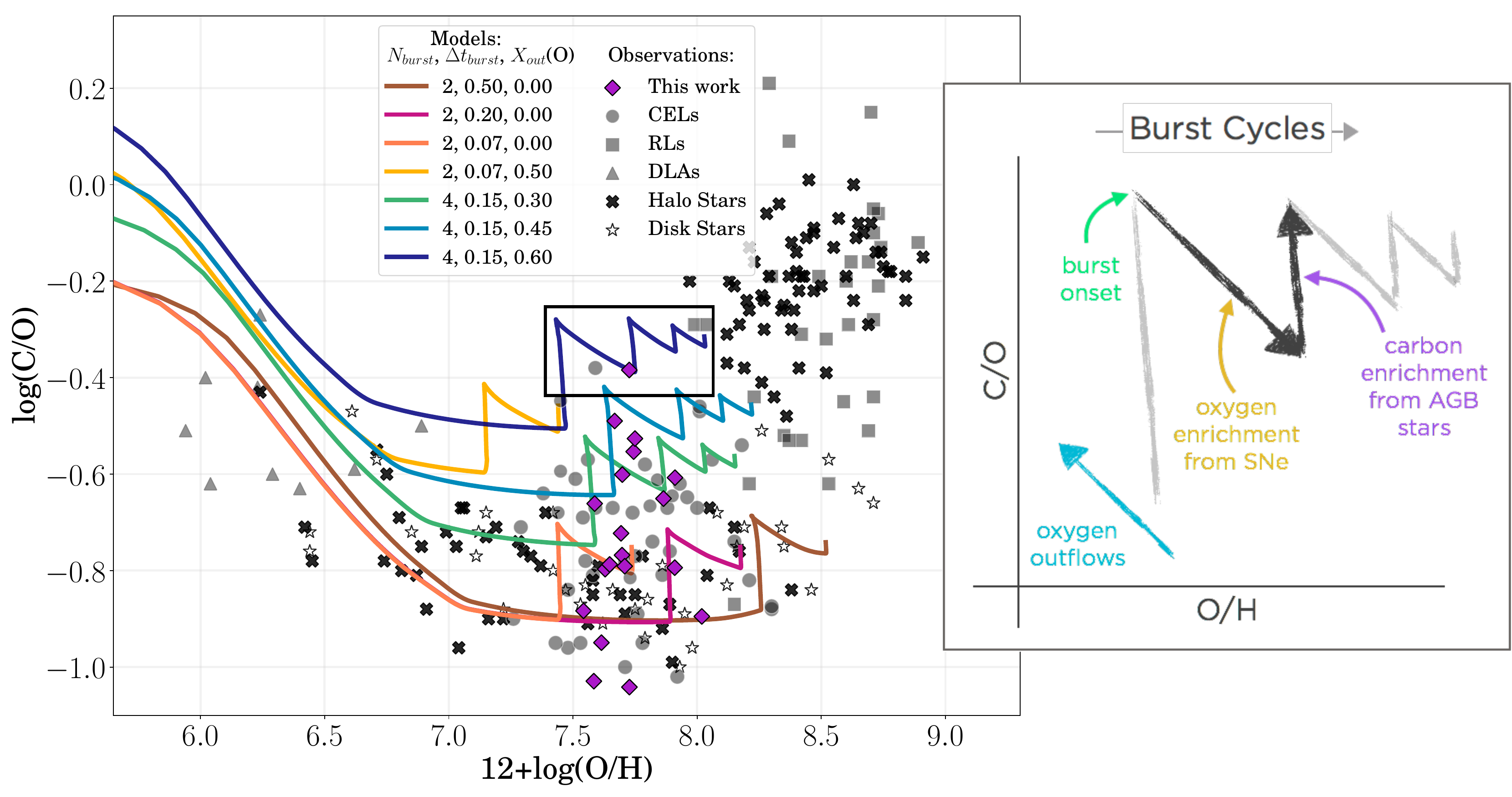} \label{fig11}
   \caption{Comparing chemical evolution models with the observed log(C/O) versus 12+log(O/H) relationship. 
   The objects from this paper are shown as purple diamonds with respect to the
   collisionally-excited and recombination line measurements from Figure~10 (gray circles and squares, respectively).
   Additionally, we plot damped Ly$\alpha$ emitters from \citet{cooke17} (gray triangles), as well as Milky Way
   halo stars (crosses) and disk stars (stars) from \citet{gustafsson99}, \citet{akerman04}, \citet{fabbian09},
   and \citet{nissen14}.
   The solid lines show model results (see details in Section~6). 
   The properties of the models are provided in the legend, where the numbers refer, in order, to the number of bursts ($N_{burst}$), 
   the burst duration ($\Delta t_{burst}$), and the fraction of new oxygen exiting the galaxy via outflows ($X_{\rm out}(O)$). 
   Note that these representative models nicely span the observed vertical spread in C/O.
   The panel to the right shows the sawtooth pattern for the navy blue model track and the possible physical sources modifying the C and O abundances.
   In this picture, a single burst cycle is indicated by the single black sawtooth, as its progression with time advances from left to right.
   Just a few million years after the onset of star formation, oxygen is produced by type II SNe causing the C/O to decrease as O/H increases.
   Later, after the SNe production of O has ceased, C is returned by AGB stars producing the vertical climb in C/O.
   Also note the blue arrow in the lower left corner demonstrating the effect of preferential outflows of oxygen.}
    \end{centering}
\end{figure*}

To set the stage, we first created a set of 
representative models which demonstrate
the effects of various parameters in the C/O versus O/H diagnostic diagram.
The parameters of these models are listed in Table~4.  We have varied the number of 
bursts of star formation ($N_{burst}$), the durations of the bursts ($\Delta t_{burst}$),
when the bursts occur ($t_{burst}$) (effectively the star formation history), and 
the fraction of oxygen that is ejected during a burst of star formation ($X_{\rm out}$(O)).
 
Figure~11 shows our results for the models listed in Table~4 in the C/O versus O/H
diagnostic diagram against a background of abundance observations of several object 
types identified in the figure legend. 
Each solid line represents a model SFH with the number of bursts ($N_{burst}$), 
their duration ($\Delta t_{burst}$) in Gyr, 
and the ejected oxygen fraction ($X_{\rm{out}}$(O)) given in the legend. 
For these bursty models, the SFR prior to the first burst and between subsequent bursts is quiescent
(zero), while infall continues throughout. 
For example, the nave blue track corresponds to a model with four bursts each with a duration of 
0.15~Gyr, and where 60\% of the total oxygen production is ejected into the circumgalactic medium.
 
All of the model tracks portrayed in Figure~11 have a characteristic shape.
Starting from the lowest metallicities, the models show a declining slide in C/O, 
during which the interstellar medium is enriched with newly synthesized oxygen from the first SNII
that in turn drive the C/O down with the build up of oxygen.
The track then halts at a particular value of O/H and the C/O ratio increases at constant O/H 
as oxygen production is exhausted but carbon is ejected by the slower-evolving stars.
Finally, C/O peaks as the burst terminates. 
The resulting sawtooth shape is associated with a burst episode, where
a succession of bursts creates the overall zigzag shape of the curves.
The number of cycles is determined by the number of bursts and 
each burst results into a downward-rightward translation.
The burst duration plays a key role: a longer burst episode produces 
a larger rightward motion as more oxygen is created. 
Note that the shorter vertical motion in later bursts is due to a logarithmic effect, where 
the relative chemical enrichment of each episode becomes less as the overall metallicity increases. 
Finally, the vertical location of a track is directly related to the amount of oxygen outflow in the model
(as can be seen by comparing the green, blue, and navy models which only vary in ($X_{\rm out}$(O)). 


\begin{deluxetable}{lcccr}
\tablecaption{Chemical Evolution Model Track Parameters}
\tablehead{
\CH{Track} 
& \CH{$X_{\rm out}$(O)} 		
& \CH{$N_{burst}$}	
& \CH{$\Delta t_{burst}$} 		
& \CH{$t_{burst}$} }
\startdata	
\color{RawSienna} {Brown} \color{black}	& 0.00	& 2	& 0.50	& 9, \ \ \ \ \ 12.5	\\
\color{RedViolet} Pink \color{black}		& 0.00	& 2	& 0.20	& 11, 12.5	\\
\color{Orange} Orange \color{black}		& 0.00	& 2	& 0.07	& 11, 12.5	\\
\color{Dandelion} Gold \color{black}		& 0.50	& 2	& 0.07	& 11, 12.5	\\
\color{Green} Green \color{black}		& 0.30	& 4	& 0.15	& 7, 9, 11, 12.5	\\
\color{Cerulean} Blue \color{black}		& 0.45	& 4	& 0.15	& 7, 9, 11, 12.5	\\
\color{Blue} Navy \color{black}			& 0.60	& 4	& 0.15	& 7, 9, 11, 12.5	
\enddata
\tablecomments{
Parameters describing the chemical evolution model tracks depicted in Figure~11.
The first column lists the line color plotted in Figure~11.
Column 2 is the fraction of oxygen that is ejected during a burst of star formation ($X_{\rm out}$(O)).
Columns $3-5$ describe the star formation history of the track, where column 3
lists the number of bursts that occur ($N_{burst}$), each lasting for the duration listed 
in Column 4 ($\Delta t_{burst}$, in Gyr), and occurring at the midpoint times 
($t_{burst}$, in Gyr) given in Column 5. }
\label{tbl:tracks}
\end{deluxetable}


The above exercise clearly demonstrates that the spread observed in the C/O-O/H plane 
for our sample of metal-poor dwarf galaxies in Figure~11 can largely be reproduced 
by differences in SFH and the amount of oxygen ejected. 
However, as one attempts to model individual objects, there will clearly be a uniqueness 
problem in the absence of a galaxy's detailed SFH such that more than one set of burst and outflow parameters 
can be used to model each object (i.e., the different models can overlap in the diagnostic diagram). 
In order to move forward in our investigation of the chemical evolution of bursting systems, 
we decided to chose a subset of 10 of our sample objects and compute a heuristic single-burst 
model of each. 
The result is a set of simple toy models that demonstrate how chemical abundances can be 
affected by a single burst of star formation that is characterized by a set star formation efficiency,
burst duration, and oxygen outflow.
We describe our sample of 10 objects and their corresponding models below.


\begin{deluxetable*}{lcccccccccc}
\tablewidth{0pt}
\tablecaption{Observed and Derived Quantities for the Model Sample}
\tablehead{
\CH{1} & \CH{2} & \CH{3} & \CH{4} & \CH{5} & \CH{6} & \CH{7} & \CH{8} & \CH{9} & \CH{10} & \CH{11} \\ \hline
\CH{Target}    						&  
\CH{$z$ } 							&  
\CH{$L_D$ }						&  
\CH{$A_{fib}$ }						&  
\CH{log $M_{\star,fib}$ } 				&  
\CH{$\Sigma_\star$ } 				&  
\CH{log SFR$_{fib}$ }				&  
\CH{$\Sigma_{\rm{SFR}}$ } 			&  
\CH{$\Sigma_{\rm{gas}}$ }			&  
\CH{$\mu_{\rm{gas}}$ }				&  
\CH{log $M_{T}$ } 				             
\\	
\CH{} 				    			&  
\CH{} 							&  
\CH{(Mpc) }						&  
\CH{(kpc$^{-2}$) } 					&  
\CH{($M_\odot$) }	 				&  
\CH{($M_\odot$ pc$^{-2}$) } 			&  
\CH{($M_\odot$ yr$^{-1}$) }			&  
\CH{($M_\odot$ yr$^{-1}$ kpc$^{-2}$) } 	&  
\CH{($M_\odot$ pc$^{-2}$) }	 		&  
\CH{}							& 
\CH{($M_\odot$) }	 				}  
\startdata
J223831		& 0.021	& 91.39		& 1.39	& 7.02	& 7.57	& $-0.54$     	& 0.21	& 166.25		& 0.96	& 8.42	\\
J141851		& 0.009	& 38.81		& 0.25	& 6.47  	& 11.82  	& $-1.02$     	& 0.38   	& 259.90    	& 0.96 	& 7.87	\\
J121402		& 0.003	& 12.88		& 0.03	& 6.30  	& 72.18   	& $-2.13$     	& 0.27   	& 201.19    	& 0.76 	& 6.92	\\
J171236		& 0.012	& 51.87		& 0.45	& 6.75  	& 12.49	& $-1.16$     	& 0.16   	& 136.60    	& 0.93 	& 7.90	\\
J113116		& 0.006	& 25.82		& 0.11	& 6.24  	& 15.81	& $-1.98$     	& 0.09   	&  95.05    	& 0.87 	& 7.13	\\
J133126		& 0.012	& 51.87		& 0.45	& 6.86 	& 16.06	& $-0.94$     	& 0.26   	& 194.25    	& 0.93 	& 8.01	\\
J132347		& 0.022	& 95.81		& 1.53	& 7.09 	& 8.09	& $-0.55$     	& 0.19   	& 155.23    	& 0.96 	& 8.49	\\
J094718		& 0.005	& 21.50		& 0.08	& 6.21  	& 21.21	& $-1.91$     	& 0.16   	& 138.87    	& 0.88 	& 7.13	\\ 
J025346		& 0.004	& 17.18		& 0.05	& 6.26  	& 37.28	& $-1.96$     	& 0.22   	& 175.37    	& 0.84 	& 7.06	\\
J084956		& 0.014	& 60.61		& 0.61	& 7.22  	& 27.41	& $-0.51$     	& 0.50   	& 314.21    	& 0.93	& 8.37	
\enddata 
\tablecomments{
Observed and derived sample properties used as inputs to the chemical evolution models described in \S6.3.
Column 3 lists the luminosity distance in Mpc as determined from the SDSS redshift in Column 2,
and Column 4 lists the projected SDSS fiber area, assuming a 3\arcsec\ fiber.
Columns 5 and 7 list the median fiber stellar masses and SFRs from the MPA-JHU DR8,
scaled by a factor of 1.5 to convert from a Kroupa to a Salpeter IMF.
The fiber masses and SFRs were used in order to simplify the surface density calculations,
listed in Columns 6 and 8.
The gas surface density in Column 9 was then determined using the Schmidt law of \citet{kennicutt98}.
Finally, the gas fraction, $\mu_{\rm{gas}} = \Sigma_{\rm{gas}} / (\Sigma_{\rm{gas}} + \Sigma_{\star})$, in Column 10
is used to estimate the total baryonic mass in Column 11.}
\label{tbl:mod}
\end{deluxetable*}


\begin{deluxetable*}{lcccccccccc}
\tablewidth{0pt}
\tablecaption{Chemical Evolution Model Derived Properties}
\tablehead{
\CH{1} & \CH{2} & \CH{3} & \CH{4} & \CH{5} & \CH{6} & \CH{7} & \CH{8} & \CH{9} & \CH{10} & \CH{11} \\ \hline
\CH{Target}    			&  
\CH{log $M_{\star}$ } 	&  
\CH{log SFR}			&  
\CH{$\mu_{\rm{gas}}$ } 	&  
\CH{12+log(O/H) }	 	&  
\CH{log(C/O) }			&  
\CH{log(N/O) }			&  
\CH{log SFE}			&  
\CH{$Y_{\rm{eff.}}(O)$ }	&  
\CH{$\Delta t_{burst}$}	&  
\CH{log $M_T$ } 	         \\  
\CH{ } 			   	&  
\CH{Mod. O/M}	 		&  
\CH{\ Mod. O/M}	 	&  
\CH{Mod. O/M}	 		&  
\CH{Mod. O/M}	 		&  
\CH{\ Mod. O/M}	 	&  
\CH{\ Mod. O/M}	 	&  
\CH{(Gyr $^{-1}$)}		&  
\CH{}				&  
\CH{(Gyr)}  	 		& 
\CH{($M_\odot$) } }	 	   
\startdata
J223831		& 7.04$\ \ $0.96  	& $-1.1\ $ 3.60    & 0.96$\ \ $1.00	& 7.54$\ \ $1.12	& $-0.65\ \ $0.98	& $-1.45\ \ $1.26 	& $-3.82$	& 0.55	& 0.15	& 8.44\ \  \\
J141851		& 6.40$\ \ $1.18	& $-2.0\ $ 9.55    & 0.97$\ \ $0.99	& 7.60$\ \ $0.87  	& $-0.79\ \ $0.81  	& $-1.37\ \ $1.12	& $-3.92$	& 0.77 	& 0.25	& 7.92 \\
J121402		& 5.90$\ \ $2.51	& $-2.0\ $ 0.74    & 0.90$\ \ $0.84	& 7.69$\ \ $0.95   	& $-0.46\ \ $0.93    	& $-1.54\ \ $0.93	& $-2.64$	& 0.35	& 0.09	& 6.90 \\
J171236		& 6.60$\ \ $1.41	& $-1.7\ $ 3.47    & 0.95$\ \ $0.98	& 7.73$\ \ $0.93   	& $-0.75\ \ $0.95    	& $-1.36\ \ $0.93	& $-3.70$	& 0.70	& 0.22	& 7.90 \\
J113116		& 5.76$\ \ $3.05	& $-2.7\ $ 5.04    & 0.96$\ \ $0.91	& 7.72$\ \ $0.85   	& $-0.77\ \ $0.95     	& $-1.19\ \ $1.07	& $-3.70$	& 0.75	& 0.40	& 7.16 \\
J133126		& 6.76$\ \ $1.25	& $-1.4\ $ 2.87    & 0.94$\ \ $0.99	& 7.71$\ \ $0.95   	& $-0.69\ \ $0.93    	& $-1.47\ \ $1.07	& $-3.52$	& 0.60	& 0.15	& 7.98 \\
J132347		& 6.94$\ \ $1.42	& $-1.5\ $ 9.44    & 0.97$\ \ $0.99	& 7.64$\ \ $0.87   	& $-0.90\ \ $0.74    	& $-1.33\ \ $1.00	& $-4.46$	& 1.00	& 0.40	& 8.46 \\
J094718		& 6.20$\ \ $1.03	& $-1.7\ $ 0.62    & 0.88$\ \ $1.00	& 7.74$\ \ $0.98   	& $-0.39\ \ $1.02    	& $-1.46\ \ $1.05	& $-2.60$	& 0.30	& 0.09	& 7.12 \\ 
J025346		& 6.18$\ \ $1.21	& $-2.0\ $ 1.09    & 0.86$\ \ $0.98	& 7.98$\ \ $0.85   	& $-0.59\ \ $0.95    	& $-1.54\ \ $1.07	& $-2.60$	& 0.48	& 0.11	& 7.03 \\
J084956		& 7.26$\ \ $0.92	& $-0.9\ $ 2.54    & 0.92$\ \ $1.01	& 7.98$\ \ $1.00   	& $-0.82\ \ $0.93    	& $-1.54\ \ $1.10	& $-3.59$	& 0.82	& 0.17	& 8.36
\enddata 
\tablecomments{
Target parameters derived from chemical evolution models.
Columns $2-4$ list the modeled value (Mod.) and observed-to-modeled (O/M) ratio of the stellar mass, SFR, and gas fraction.
Similarly, the modeled relative chemical abundances and their comparison to the observed values are given in Columns $5-7$.
The star formation efficiency (SFE), effective oxygen yield ($Y_{\rm{eff}}(O)=1-X_{\rm out}(O)$), and burst duration ($\Delta t_{burst}$) 
associated with the model best fits are listed in Columns $8-10$.
The units of mass and SFR are $M_\odot$ and $M_\odot$~yr$^{-1}$, respectively.
The final column lists the total baryonic masses of the models.}
\label{tbl:modout} 
\end{deluxetable*}


\subsection{Properties of the Representative Sample}
A significant strength of the present study is that the galaxies in our sample have a large 
number of observed properties that can be used to place tight constraints on our models.
In turn, this allows us to determine which modes of evolution are necessary to reproduce 
the properties of interest to this study, i.e., the C/O, O/H, and N/O abundances.
In order to compose chemical evolution models, we need a number of characteristics 
for each of the galaxies. 
The chemical abundances have been derived from their UV and optical spectra 
(see Tables~$11-13$ in Appendix~\ref{A2}).
Additional galaxy properties can be determined from the SDSS photometry.
Following the method laid out in \citet{tremonti04}, we calculate the gas fraction for the sample galaxies:
$\mu_{\rm{gas}} = \Sigma_{\rm{gas}} / (\Sigma_{\rm{gas}} + \Sigma_{\star})$,
where the gas surface density is determined by inverting the Schmidt law of \citet{kennicutt98}:
$\Sigma_{\rm{gas}} (M_{\odot}$ pc$^{-2}) = (\Sigma_{\rm{SFR}}/1.6\times10^{-4})^{5/7}$.
Given the compactness of our bright targets (see Figure~1), 
we use the median fiber SFRs and stellar masses from the MPA-JHU DR8, along with 
the SDSS fiber area, to calculate the $\Sigma_{\rm{SFR}}$ and $\Sigma_{\star}$, respectively.
Note that the SDSS MPA-JHU derived quantities assume a \citet{kroupa01} initial mass function (IMF), whereas the models
used in this work adopt a \citet{salpeter55} IMF.
We therefore multiply the SDSS stellar masses and SFRs by a factor of $1.5$ to scale
from a Kroupa to a Salpeter IMF.
The gas fractions and other observation-derived quantities used to constrain our models 
are given in Table~5.


\begin{figure*}
\begin{center}
\begin{tabular}{ccc}
	\includegraphics[scale=0.21,angle=0,trim=20mm 0mm 0mm 0mm]{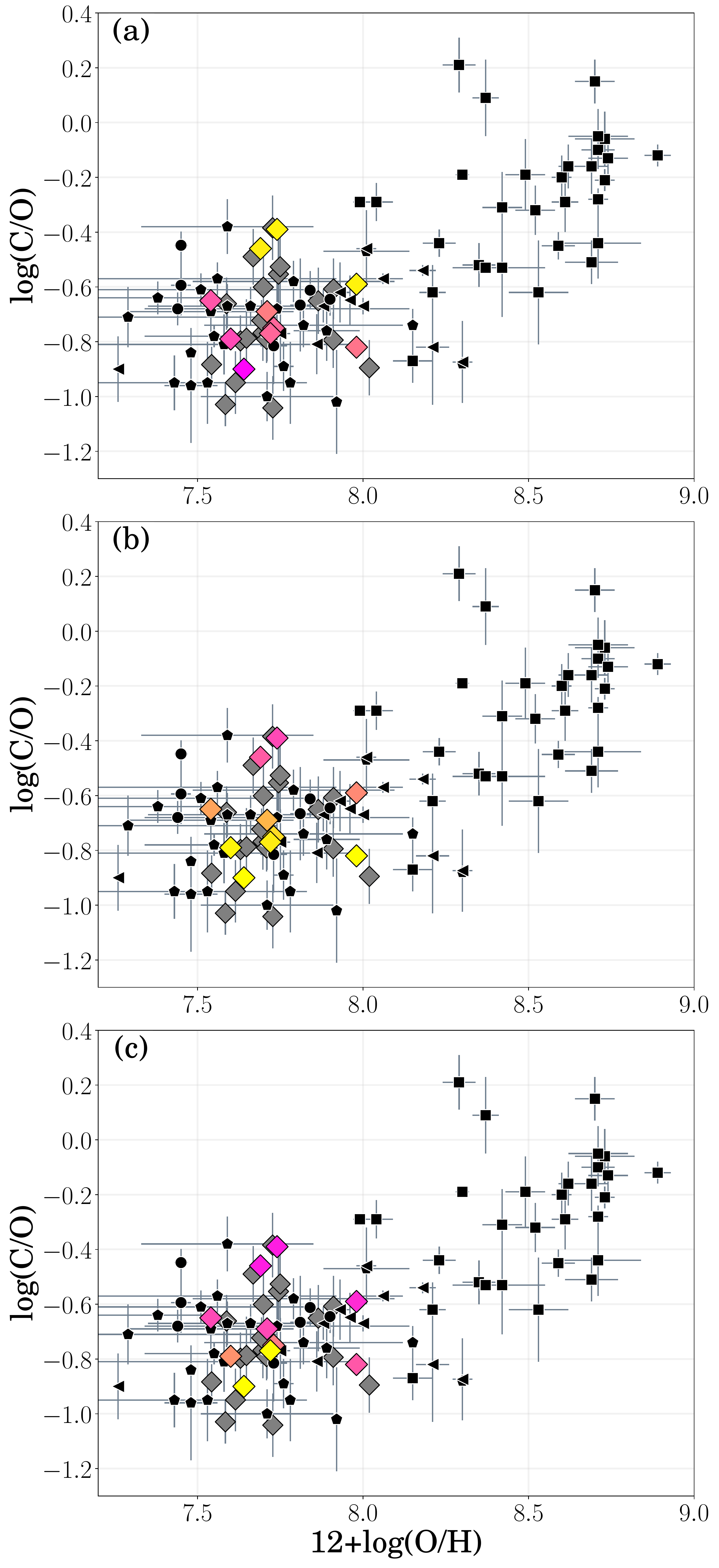}\label{fig12} &
	\includegraphics[scale=0.21,angle=0,trim=20mm 0mm 0mm 0mm]{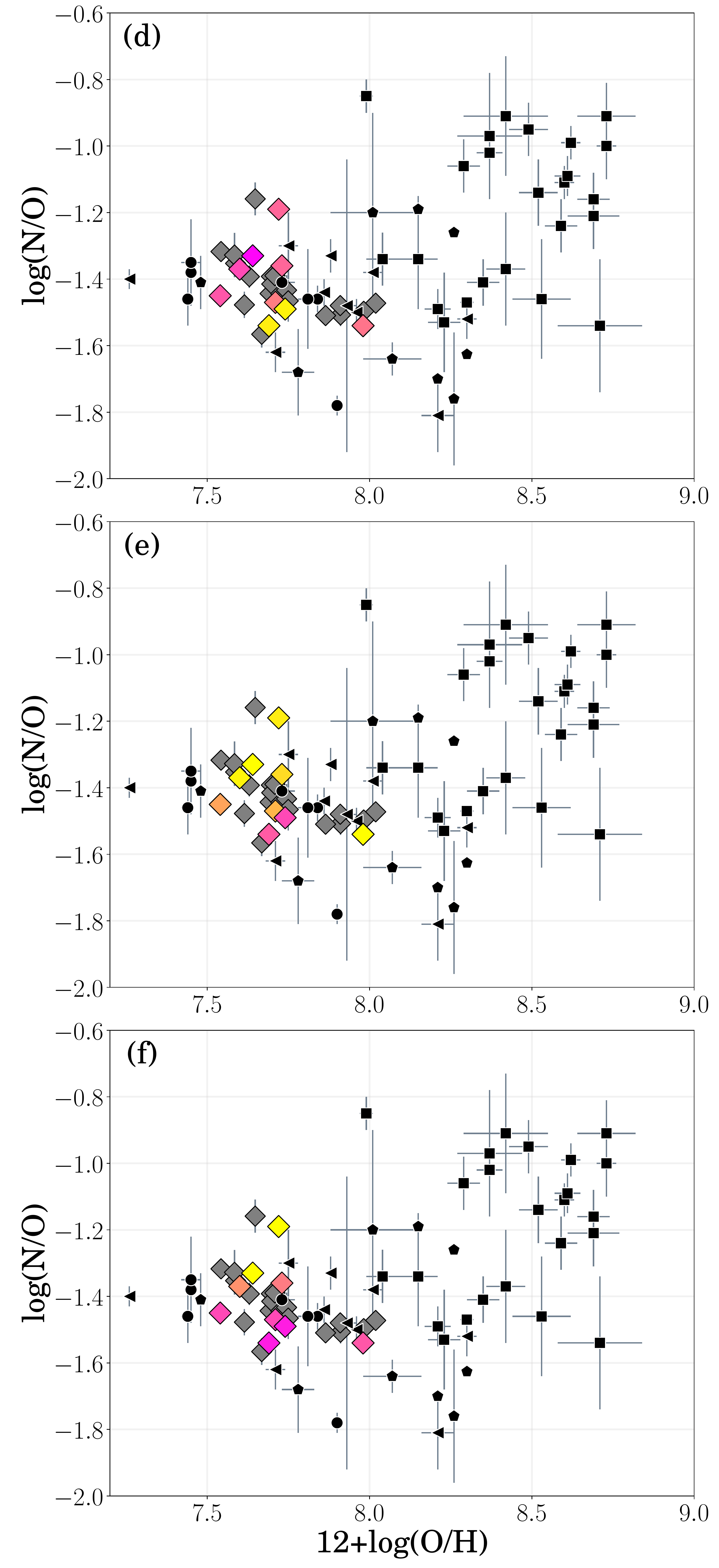} &
	\includegraphics[scale=0.21,angle=0,trim=20mm 0mm 0mm 0mm]{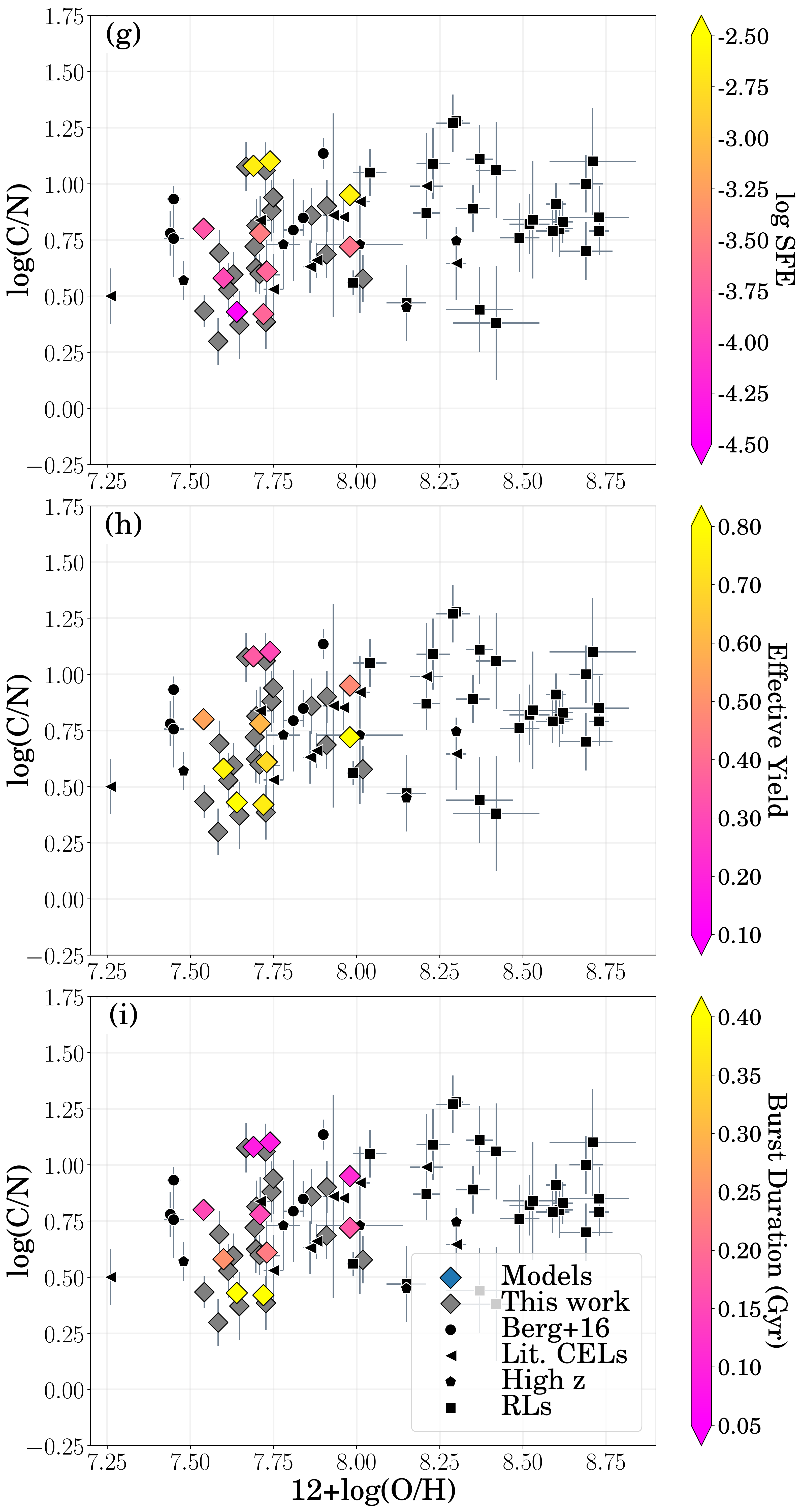}
\end{tabular}
\caption{
\textit{Left:} Carbon to oxygen abundance vs. oxygen abundance,
\textit{Middle:} nitrogen to oxygen abundance vs oxygen abundance, and 
\textit{Right:} carbon to nitrogen abundance vs. oxygen abundance.
Black plot symbols are the same observational data as Figure~10.
Colored points show the chemical evolution model results 
for a sub-sample of our targets
(described in Section~6 and listed in Table~6),
where the color-coding indicates 
star-formation efficiency (SFE; top row),
effective oxygen abundance yield ($Y_{\rm{eff}}(O)$; middle row), and 
burst duration ($\Delta t_{burst}$; bottom row).
The observed C/O ratio is clearly sensitive to all three model parameters.}
\end{center}
\end{figure*}


\subsection{Single-Burst Models}
We now fit a chemical evolution end point to each of the galaxies in the representative sample.
While each object in our sample is currently forming stars, we know relatively little about its star formation history, 
(e.g., the number of previous bursts it has experienced or each burst's duration or intensity). 
Therefore, we adopt a one-burst, one-zone model for simplicity. 
The burst intensity is a delta function centered at a galaxy age of 12.5~Gyr and a {\it burst duration},
$\Delta t_{burst}$, which is treated as a free parameter. 
Prior to the burst, the model galaxy forms by mass infall at a rate which decreases exponentially with a 
characteristic time of 5~Gyr, a current time of 13.0~Gyr, and with the time increment for our models set at 0.001~Gyr. 
Star formation begins as the burst begins, at one-half the duration time before the central time,
and then ceases after the specified duration time, $\Delta t_{burst}$. 
The rate of star formation is prescribed by a Schmidt--Kennicutt law with an index of 1.4 \citep{kennicutt98} and 
modulated by the {star formation efficiency} (SFE), our second free parameter. 
The stellar mass distribution is populated by a Salpeter IMF \citep{salpeter55} with an index of $-1.35$. 

Nucleosynthetic products are presumed to be expelled into the galaxy's interstellar 
medium at the end of a star's lifetime in accordance with published yield prescriptions. 
For stars between $9-120$~M$_{\odot}$ we employ the yields for He, C, N, and O found in 
\citet[for Z~$= 10^{-5}$, 0.004, and 0.02]{meynet02,chiappini03b} and \citet[for Z~$=10^{-8}$]{hirschi07}, 
while yields for He, C, and N for $1-8$~M$_{\odot}$ stars are taken from \citet{karakas16}. 
We chose this yield combination because it proved to be the most capable of matching the observed trend in 
log(C/O) versus 12+log(O/H) for 12+log(O/H)~$\le 8.2$ for several object types (see Figure~11). 
Additionally, values for stellar main sequence lifetimes from \citet{schaller92} are employed. 
As a model computation progresses through a burst, stellar yields of H, He, C, N, O and 
numerous alpha elements are integrated over each time point. 
In the case of binary star contributions to the evolution of C, O, Si, S, Cl, Ar, and Fe through SNIa events, 
we employ the yields of \citet{nomoto97}, using both the W7 and W70 models to account for metallicity effects. 
At the conclusion of the burst the computation ceases, 
and the abundance ratios C/O, N/O, C/N, and O/H are reported.

With an aim of testing the claim of \citet{yin11} that the observed range in C/O among galaxies 
was primarily due to the differences in selective outflow of oxygen, 
we also include outflowing oxygen gas as a free parameter.
Relevant to the C, N, and O measurements of this work, selective oxygen outflows are possible if short-lived, 
massive stars dominate oxygen production and result in SNe-driven outflows, while C and N are predominantly 
produced at later times in low- and intermediate-mass stars.
In our models, a specified fraction of the total oxygen ($X_{\rm out}(O)$) produced by massive stars 
is subtracted off at each time point and accounted for in the gas mass.\footnotemark[9]
\footnotetext[9]{Note that $X_{\rm out}(O)$ is analogous 
to the $w_i\lambda$ parameter in equation~10 of \citet{yin11}.}

In order to match the observed details of each galaxy, the models were modulated by varying 
three free parameters, i.e., SFE in Gyr$^{-1}$, $X_{\rm out}(O)$, and $t_{burst}$ in Gyr.
As a first step, numerous models were run with various model inputs
to probe the sensitivity of C, N, and O abundances to each parameter.
In particular, through thorough testing we found that the C/O and O/H abundances are 
relatively insensitive to burst duration compared to N/O.
We then developed an informed model prescription that allowed us to produce reasonable models: 
we varied
(1) $\Delta t_{burst}$ until the observed N/O ratio was suitably matched,
(2) $X_{\rm out}(O)$ until we found agreement between the predicted and observed value of C/O, and
(3) the SFE parameter to bring the O/H abundance into agreement with the observed value. 
Small adjustments were then made to fine-tune the results. 
Each model was terminated at 13~Gyr, 
and, for consistency, the output values for the final time step of the burst were taken as the model result.  
Our goals were to match observed abundance ratio values (O/H, C/O, and N/O) to within 0.1~dex, 
while simultaneously matching the galaxy properties (log $M_\star$, log SFR, and $\mu_{gas}$) to within 0.3 to 0.4 dex.
More leniency was allowed in the case of galaxy properties because of the large uncertainties in those observations.

Our model fits are provided in Table~6. 
Object names are listed in column~1. 
In each of the six column pairs that follow, we list the observed-to-modeled (O/M) value ratio alongside the
model-predicted (Mod.) value for the parameter indicated to quantify the goodness of the model. 
Note that except for $\mu_{gas}$, the modeled values are expressed logarithmically, 
while O/M and $\mu_{gas}$ values are in linear form. 
For the discussion of the results, we have switched to expressing the parameter $X_{\rm out }(O)$ as its complement, 
$Y_{\rm{eff}}(O) = 1-X_{\rm out}(O)$, i.e., the effective yield fraction\footnotemark[10] of oxygen.
The next three columns list the values for our three variable parameters used in the most successful models to 
provide the best match to the observations, namely SFE, $Y_{\rm eff}(O)$, and $t_{burst}$. 
The matches between model and observation are mostly within our stated tolerances. 
The biggest exceptions are the SFRs, with the predicted values for J141851 and J132347 being nearly 
10 times higher than the observed level. 
No amount of tweaking of parameters could bring the SFR in line with the 
observations without worsening the agreement with several other observables. 
Other discrepancies are far less troublesome. 

\footnotetext[10]{Here we are defining {\it effective yield fraction} as the portion of the total 
stellar mass yield of an element at each time step that remains within the galaxy.}


\subsection{The Nature of the C/O and C/N Dispersion}
Figure~12 graphically displays the model results along with the observations 
for log(C/O) (left panels), 
log(N/O) (center panels), and 
log(C/N) (right panels) all versus 12+log(O/H). 
The observational data from Figure~10 pertaining to several object types are reproduced here
as black filled symbols (see the legend in the lower right panel). 
Model values in each case are shown as large diamonds and are color-coded
according to their model parameters: 
log SFE (top panels), 
$Y_{\rm{eff}}(O)$ (middle panels), and 
$t_{burst}$ (bottom panels). 
Note how the 10 model points nicely span the area occupied by our current objects 
(black diamonds) as well as metal-poor dwarf galaxies from other studies (circles and triangles).

We see in the left column of panels that greater values of C/O are associated with 
higher SFE (Fig.~12a), 
lower $Y_{\rm{eff}}(O)$ (Fig.~12b), and 
shorter $t_{burst}$ (Fig.~12c). 
From the center column of panels we see that objects with higher N/O values are associated with 
lower SFE values (Fig.~12d), 
higher $Y_{\rm{eff}}(O)$ (Fig.~12e), and 
longer $t_{burst}$ (Fig.~12f). 
In the right column, high C/N levels, like those of C/O, are associated with 
high SFE (Fig~12g), 
low $Y_{\rm{eff}}(O)$ (Fig.~12h), and 
short $t_{burst}$ (Fig.~12i). 

These trends suggest that SFE is indirectly related to both $Y_{\rm{eff}}(O)$ and $t_{burst}$. 
Apparently, efficient star formation is associated with greater oxygen outflow in these low mass systems (Fig.~12a and 12b).  
At the same time, more efficient star formation depletes a greater portion of interstellar gas by forming more stars at a higher rate, 
plus the greater number of stars per unit of gas drives more outflow of ambient gas. 
These two factors are likely reasons why the burst duration time is shorter (Fig.~12c)\footnotemark[11].
Additionally, when the duration time is shorter, less nitrogen from the slowly-evolving, lower-mass stars is ejected within that time (Fig.~12f). 
Thus, at the end of a burst, the ratio of N/O is lower in a system with high SFE (Fig.~12d). 
With N/O lower and C/O higher, C/N will behave like C/O, i.e., it is higher when the SFE is higher (Fig.~12g), 
$Y_{\rm{eff}}(O)$ is lower (Fig.~12h), and $T_{burst}$ is shorter (Fig.~12i). 

\footnotetext[11]{Our code does not account for stellar driven outflow, 
so we are unable to verify this idea computationally.}

\subsection{The C/O Trend in Metal-Poor, Dwarf Galaxies}
Carbon and Oxygen are thought to originate primarily from stars of different mass ranges; 
O is synthesized mostly in massive stars (MSs; $M > 10 M_\star$), 
while C is produced in both MSs and intermediate- mass stars. 
Theoretically, only primary (metallicity independent) nucleosynthetic processes 
are known to produce C so we would naively expect C/O to be constant in a closed-box model. 
Empirically, however, C/O increases proportionally with O/H for 12+log(O/H)~$>7.0$ (see Figure~11), 
suggesting some pseudo-secondary (metallicity dependent) C production is prominent
and/or galactic flows are modifying C/O. 
Certainly, some pseudo-secondary C enrichment results from low-mass asymptotic giant branch stars
\citep[e.g.,][]{kobayashi11} and the metallicity-dependent winds of MSs \citep[e.g.,][]{henry00},
but these effects are small in the metal-poor range by definition.

As suggested in B16, C/O may instead follow a bi-modal relationship,
where primary C production dominates in metal-poor dwarf galaxies, and
pseudo-secondary C production becomes important at some 12+log(O/H~)$>8$.
In this case, as discussed in Section~6.4, the C/O ratio in an individual metal-poor dwarf galaxy
depends most critically on its specific SFH and effective oxygen yield.
Then, higher total galaxy baryonic mass ($M_\star + M_{gas}$; see Tables 5 and 6)
will produce a stronger gravitational potential and likely 
cause a reduction of outflows both of fresh stellar ejecta as well as ambient interstellar gas. 
This is consistent with \citet{chisholm18b} who find that the observed outflows from low-mass galaxies remove
a larger fraction of their gas-phase metals compared to more massive galaxies.
Therefore, the more massive galaxies in our sample should exhibit longer burst durations 
(more fuel for star formation), higher effective oxygen yield fraction (less oxygen outflow), 
and lower C/O.

Indeed, in the top panel of Figure~13, we plot log(C/O) versus total baryonic mass, log($M_T$),
and find a decreasing relationship.
Further, this declining C/O trend with increasing mass may correspond to an increase in the 
retained O abundance, as indicated by the $Y_{\rm{eff}}(O)$ trend in the bottom panel of Figure~13.
We have fit the modeled data points in Figure~13 using a linear regression analysis (black line).
While the resulting trends are suggestive, with Pearson correlation coefficients of 
$r = -0.71$ for the C/O relationship and $r = 0.69$ for the oxygen effective yield fraction relationship,
the p-values are only 0.022 and 0.028, respectively.
Thus, these are only 2$\sigma$ trends at present, and are limited by the large uncertainty in the mass measurements
and the small number of modeled points.
Nonetheless, the range in mass of our sample, and the subsequent gravitationally
modulated $Y_{\rm{eff}}(O)$, could explain a significant portion of the observed scatter in C/O. \looseness=-2


\begin{figure}
\begin{centering}
   \includegraphics[scale=0.325,trim=0mm 0 0 0,clip]{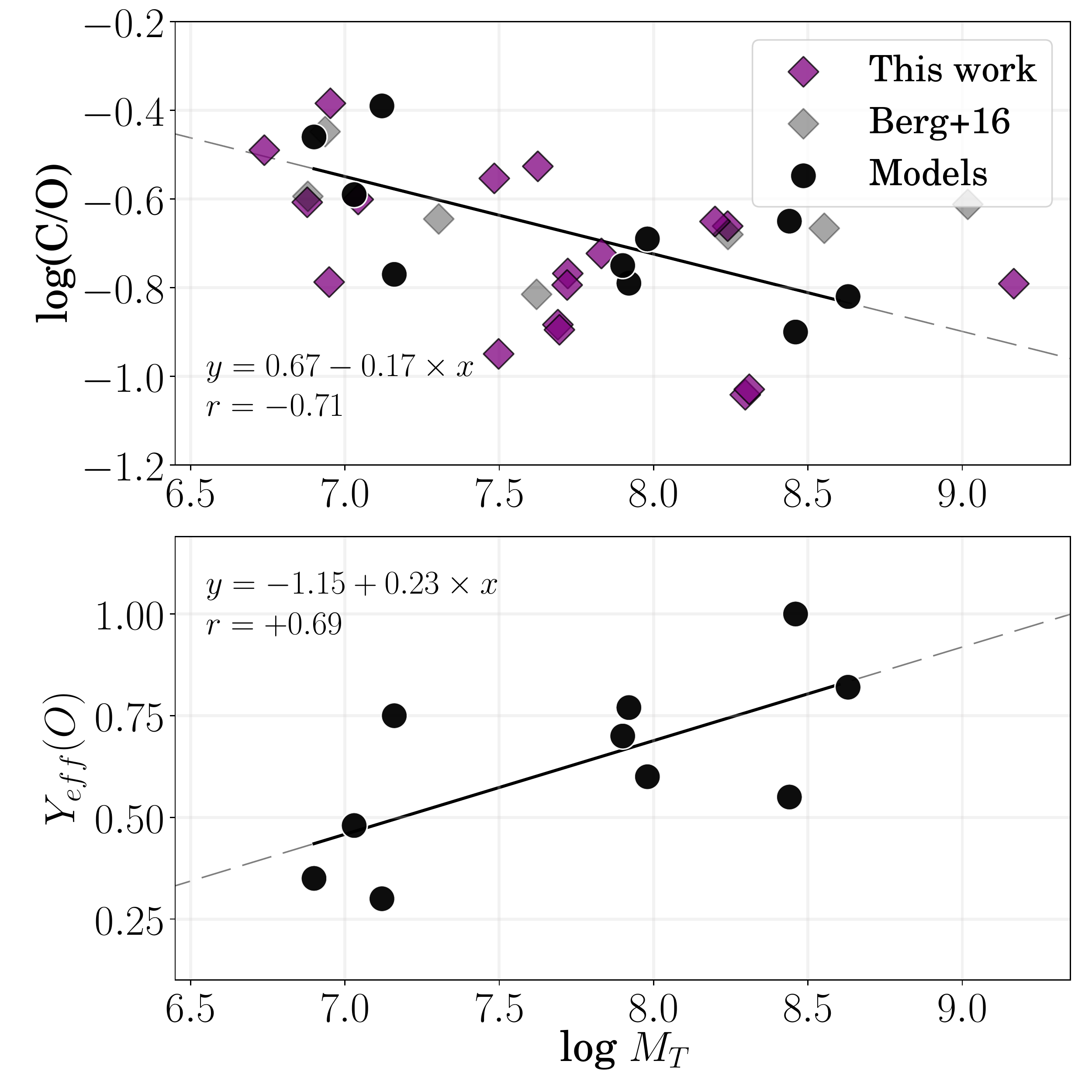} 
   \caption{ 
   \textit{Top:} Plot of log(C/O) and the log of the total baryonic mass, M$_T$,
   of each modeled galaxy (black points), as well as observations from 
   B16 (gray diamonds) and this work (purple diamonds). 
   \textit{Bottom:} Plot of $Y_{\rm{eff}}(O)$ versus the log of the total baryonic mass of each modeled galaxy, M$_T$.
   In both panels the model data points are fit with a linear regression analysis (black line), and 
   the corresponding Pearson's $r$ coefficient is reported.}
   \end{centering}
\label{fig13}
\end{figure}


\section{SUMMARY AND CONCLUSIONS} \label{sec:summary}

In this work, we present UV spectroscopy of 20 nearby low-metallicity high-ionization dwarf galaxies obtained 
using the Cosmic Origins Spectrograph on the Hubble Space Telescope. 
Building upon the study by B16, we build an expanded sample consisting of 40 local galaxies 
with the required detections of the UV O$^{+2}$ and C$^{+2}$ collisionally excited lines
and direct-method oxygen abundance measurements.
With these data we explore the relative abundance trends in high-ionization, low-metallicity (12+log(O/H) $< 8.0$) galaxies 
such as the large degree of scatter in nebular carbon measurements, their possible correlation with 
nitrogen abundance, and note the presence of very high ionization \ion{He}{2} and \ion{C}{4} emission lines. 

In order to measure the most accurate C/O abundances possible,
we produce new analytic functions of the carbon ionization correction factor (ICF).
To do so, we create a custom grid of {\sc cloudy} photoionization models, using 
Binary Population and Spectral Synthesis (BPASS) burst models as inputs, 
that are fine-tuned to match the properties of our sample.
Using our custom ICFs and measured C/O abundances, 
we confirm the flat trend in C/O versus O/H reported by B16 for local metal-poor galaxies, 
finding an average log(C/O) $= -0.71$ with a dispersion of $\sigma = 0.17$ dex. 
In contrast to earlier studies that found a linear relationship between C/O and O/H \citep[e.g.,][]{garnett95},
this work shows a large and real scatter in C/O over a large range in O/H.
Given the fact that UV emission-line spectra are increasingly being used to 
determine the physical properties of high-z galaxies, this result advocates for
more care in interpreting the distant universe; specifically,   
measurements of the UV C and O emission lines alone do not necessarily provide a good indicator of the O/H abundance.

The C/N ratio also appears to be constant at log(C/N) $= -0.75$, but with significant scatter ($\sigma = 0.20$ dex).
Given that N/O is known to follow a bi-modal relationship, where secondary nitrogen production
becomes important at oxygen abundances of 12+log(O/H)~$\gtrsim8.0$, this relatively flat
C/N trend suggests that carbon production may also experience a significant pseudo-secondary 
mechanism at moderate-to-high oxygen abundances. 
If true, then both C and N experience metallicity-dependent enrichment,
and the large scatter observed for C/N ratios is surprising.
This result could indicate that C and N production is dominated by stars of different 
masses on average such that C/N production is time sensitive. \looseness=-2

To better understand the observed abundance trends for our sample,
especially the broad variation in the observed C/O and C/N abundance ratios,
we have constructed individual heuristic chemical evolution models of 10 low metallicity star-forming dwarf galaxies.
By employing single-burst models in which the only variable parameters are the star formation efficiency, 
the burst duration, and the amount of newly synthesized oxygen which is lost from the galaxy due to outflow, 
we have closely matched the abundance ratios while approximating several global physical properties of each galaxy. 
Through this simple exercise, we find the C/O ratio to be very sensitive to the detailed star formation history,
where longer burst durations and lower star formation efficiencies correspond to low C/O ratios.
Additionally, we verify that variations in the effective yield fraction of oxygen can  
produce the large range in observed abundance ratios of C/O, N/O, and C/N, as originally suggested by \citet{yin11}.  \looseness=-2

Collectively, the characteristics of our group of models suggests that the total baryonic mass of a galaxy 
is the ultimate determinant for the effective oxygen yield and, therefore, 
the value of the relative abundance ratios involving CNO in our galaxy sample. 
Further work in modeling more of these galaxy types, especially those with detailed star formation histories,
will improve and solidify our understanding of the broad variation in the observed abundance ratios and 
what they tell us about the chemical evolution of the important CNO elements.   


\acknowledgements
DAB and DKE are grateful for partial support by the US National Science Foundation 
through the Faculty Early Career Development (CAREER) Program, grant AST-1255591.
Additionally, support for program \#14628 was provided by NASA through a grant from the Space Telescope Science Institute, 
which is operated by the Associations of Universities for Research in Astronomy, Incorporated, under NASA contract NAS5- 26555.
We also thank the referee for useful comments that helped make the analysis in this work 
more transparent.

Models in this work were computed using the computer cluster, Nemo, at 
The Leonard E Parker Center for Gravitation, Cosmology and Astrophysics is supported by NASA, 
the National Science Foundation, UW-Milwaukee College of Letters and Science, and UW-Milwaukee Graduate School. 
Any opinions, findings, and conclusions or recommendations expressed in this material are those of the author(s) 
and do not necessarily reflect the views of these organizations. 

Funding for the SDSS and SDSS-II has been provided by the Alfred P. Sloan Foundation, 
the Participating Institutions, the National Science Foundation, the U.S. Department of Energy, 
the National Aeronautics and Space Administration, the Japanese Monbukagakusho, 
the Max Planck Society, and the Higher Education Funding Council for England. 
The SDSS Web Site is http://www.sdss.org/.

The SDSS is managed by the Astrophysical Research Consortium for the Participating Institutions. 
The Participating Institutions are the American Museum of Natural History, Astrophysical Institute Potsdam, 
University of Basel, University of Cambridge, Case Western Reserve University, University of Chicago, 
Drexel University, Fermilab, the Institute for Advanced Study, the Japan Participation Group, 
Johns Hopkins University, the Joint Institute for Nuclear Astrophysics, 
the Kavli Institute for Particle Astrophysics and Cosmology, the Korean Scientist Group, 
the Chinese Academy of Sciences (LAMOST), Los Alamos National Laboratory, 
the Max-Planck-Institute for Astronomy (MPIA), the Max-Planck-Institute for Astrophysics (MPA), 
New Mexico State University, Ohio State University, University of Pittsburgh, University of Portsmouth, 
Princeton University, the United States Naval Observatory, and the University of Washington.

This research has made use of NASA's Astrophysics Data System Bibliographic Services and the 
NASA/IPAC Extragalactic Database (NED), which is operated by the Jet Propulsion Laboratory, 
California Institute of Technology, under contract with the National Aeronautics and Space Administration.


\bibliography{Dwarf_CO2.v9.resubmitted}{}

\clearpage


\appendix


\section{The Chemical Evolution Code}\label{A1}
Our numerical code is an expanded version of the one employed by \citet{henry00}. 
It is a one-zone chemical evolution program which follows the buildup of the elements 
H, He, C, N, O, Ne, Si, S, Cl, Ar, and Fe over time. 

We imagine an initially massless galaxy that accretes matter by way of infall of pristine, metal-free gas. 
The infall rate is given by:
\begin{equation}
f(t)_{in}=M_{T}\left\{\tau_{scale}\left[1-exp\left(-\frac{t_{now}}{\tau_{scale}}\right)\right]\right\}^{-1}exp\left(-\frac{t}{\tau_{scale}}\right) M_{\odot}/Gyr,
\end{equation}
where $t_{now}$ and $M_T$ are the current epoch in Gyr and total mass in M$_{\odot}$; the latter is given in column~5 of Table~5. 
This form of the infall rate is the one discussed and employed by \citet{timmes95} 
for their models of the Galactic disk. 
We adopted $\tau_{scale}=5$~Gyr, and $t_{now}$=13~Gyr for use in all of our models. 

As star formation commences, the galaxy mass M(t) becomes partitioned into interstellar gas of 
mass $g$ and stars of mass $s$ such that at time $t$
\begin{equation}
M(t)=g(t)+s(t),\label{mgs}
\end{equation}
each in units of M$_{\odot}$.
If $\psi(t)$, $e(t)$, and f(t)$_{out}$ are the rates of star formation, stellar ejection, and outflow, respectively, then:
\begin{equation}
\dot{s}=\psi(t)-e(t)
\end{equation}
and
\begin{equation}
\dot{g}=f(t)_{in}-\psi(t)+e(t)-f(t)_{out}.\label{g}
\end{equation}
All rates are in units of M$_{\odot}$/Gyr. $f(t)_{out}=\sum_x{e(t)_x\lambda_x}$, where $e_x(t)$ is
the rate at which newly formed element $x$ is ejected from stars, and $\lambda_x$ is the fraction of $e_x(t)$ 
which is subsequently lost to the circumgalactic medium through outflow. 
In our models, $\lambda_x=0$ for all elements except oxygen in which case $\lambda_O$ was used as a free parameter. 
The interstellar mass of element $x$ in the zone is $gz_x$, whose time
derivative is:
\begin{equation}
\dot{g} z_x + g \dot{z}_x = -z_x(t) \psi(t) + z_x(t)_{in} f(t)_{in} + e_x(t)(1-\lambda_x) ,
\end{equation}
where $z_x(t)$ and $z_x(t)_{in}$ are the mass fractions of $x$ in the
gas and in the infalling material, respectively. We assumed that $z(H)_{in}=0.76$, $z(He)_{in}=0.24$, 
and $z(x)_{in}=0.0$ for all other elements.  Solving for $\dot{z}_x(t)$ yields:
\begin{equation}
\dot{z}_x(t)=\{f(t)_{in}[z_x(t)_{in} - z_x(t)] -z_x(t)e(t)+ e_x(t)(1-\lambda_x) \}g^{-1}.
\end{equation}
The second term on the right hand side of eq.~A6 accounts for the injection of metals into the gas by stars, 
while the first and third terms account for effects of dilution due to infall and ejected stellar gas, respectively.

The rates of mass ejection $e(t)$ and ejection of element $x$, $e_x(t)$, are:
\begin{equation}
e(t)=\int_{m_{\tau_{m}}}^{m_{up}}[m-w(m)]\psi(t-\tau_m)\phi(m)dm\label{e}
\end{equation}
and
\begin{equation}
\begin{array}{lr}
e_x(t)=(1-C)\int_{m_{\tau_{m}}}^{m_{up}}\{[m-w(m)]z_{x}(t-\tau_{m})+mp_x(m,z_{t-
\tau_{m}})\}\psi(t-\tau_m)\phi(m) \ dm\\
\\
+ 
C\int^{M_{BM}}_{M_{Bm}}p_x(m,z_{t-\tau_m})m\phi(m)\int^{.5}_{\mu_{min}}24\mu^2
\psi(t-\tau_{m_2})d\mu\ dm_B. \label{ex}
\end{array}
\end{equation}
In Eqs.~\ref{e} and \ref{ex}, $m$ and $z$ are stellar mass and metallicity, 
respectively, and $m_{\tau_m}$ is the turn-off mass, i.e. the stellar mass whose 
main sequence lifetime corresponds to the current age of the system. This 
quantity was determined using results from \citet{schaller92}. 
$m_{up}$ is the upper stellar mass limit, taken to be 120~M$_{\odot}$, $w(m)$ is the remnant mass 
corresponding to ZAMS mass $m$ and taken from \citet{yoshii96}. 
$p_{x}(z)$ is the stellar yield, i.e. the mass fraction of a star of mass $m$ which is 
converted into element $x$ and ejected, and $\phi(m)$ is the initial mass function. 
Our choice of stellar yields is discussed below.  
In eq.~\ref{ex} the first integral gives the contributions to the ejecta of single stars, 
while ejected masses of C, O, Si, S, Cl, Ar, and Fe by SNIa through binary star 
formation are expressed by the second integral, where our formulation follows \citet{matteucci86}.
Assuming that the upper and lower limits for binary 
mass, M$_{BM}$ and M$_{Bm}$, are 16 and 3~M$_{\odot}$, respectively, 
eq.~\ref{ex} splits the contributions from this mass range between single and 
binary stars. The relative contributions are controlled by the parameter $C$ 
which we take to be equal to 0.05 when $3\le M \le 16\ M_{\odot}$ and zero otherwise. 
The variable $\mu$ is the ratio of the secondary star mass $m2$ to the total binary 
mass, $m_B$, where $\mu_{min}=max(1.5,m_{\tau_m},m_B-8)$.

The initial mass function $\phi(m)$ is the normalized \citet{salpeter55} relation
\begin{equation}
\phi(m)=\left[\frac{1-\alpha}{m_{up}^{(1-\alpha)}-m_{down}^{(1-\alpha)}}\right] 
m^{-(1+\alpha)},\label{psi}
\end{equation}
where $\alpha$=1.35 and $m_{down}=1$~M$_{\odot}$.
Finally, the star formation rate $\psi(t)$ is given by
\begin{equation}
\psi(t)=\nu  \left(\frac{g(t)}{M(t)}\right)^{1.4}M_{\odot}/Gyr\label{sfr}
\end{equation}
where $\nu$ is the star formation efficiency in Gyr$^{-1}$. The index of 1.4 in eq.~A10 is taken from \citet{kennicutt98}.

Nucleosynthetic products were presumed to be produced during, and expelled into the galaxy's 
interstellar medium at the end of, a star's lifetime in accordance with published yield prescriptions. 
For stars between 9-120~M$_{\odot}$ we employed the yields for He, C, N, and O found in 
\citet[for Z= 10$^{-5}$, 0.004, and 0.02]{meynet02,chiappini03} and \citet[for Z=10$^{-8}$]{hirschi07}, 
while yields for He, C, and N for 1-8~M$_{\odot}$ stars were taken from \citet{karakas16}. 
Additionally, values for stellar main sequence lifetimes from \citet{schaller92} were employed. 
As a model computation progressed through a burst, stellar yields of H, He, C, N, O 
and numerous alpha elements were integrated over each time point. In the case of binary star 
contributions to evolution of C, O, Si, S, Cl, Ar, and Fe through SNIa events, 
we employed the yields of \citet{nomoto97}, using both the W7 and W70 models 
to account for metallicity effects. 
At the conclusion of the burst the computation ceases, and the abundances of C/O, N/O, C/N, and O/H are reported.

Our calculations assumed a time step length of one million years, a value which is
less than the main sequence lifetime of a star with a mass equal to $m_{up}$, or 120~M$_{\odot}$. 
At each time point, the increment in $z_x$ was calculated by solving eq.~A6 and 
the required subordinate equations \ref{e} and \ref{ex}. 
This increment was then added to the current value and the program advanced to the next time step. 
Finally, the total metallicity at each point was taken as the sum of the mass 
fractions of all elements besides H and He. 
The unit of time is the Gyr, while the mass unit is the solar mass.


\section{Emission-Line Intensity and Abundance Tables}\label{A2}
Here we present the measured and derived quantities from the {\it HST}/COS rest-frame UV spectra of the sample presented in this work.
In Tables~$7-9$ we present the reddening-corrected emission-line intensities for our 20 nearby, compact dwarf galaxies.
The UV emission-line equivalent widths (EWs) of targets in our sample, and other studies referenced in this paper, 
with 3$\sigma$ UV CEL C/O detections are listed in Table~10.
Finally, we list the nebular gas conditions and relative abundance ratios derived from the rest-frame optical and UV spectra 
in Tables~$11-13$.


\begin{deluxetable*}{r r@{$\pm$}l r@{$\pm$}l r@{$\pm$}l r@{$\pm$}l r@{$\pm$}l r@{$\pm$}l r@{$\pm$}l}
\tabletypesize{\scriptsize}
\tablewidth{0pt}
\tablecaption{ Emission-Line Intensities for HST/COS Observations of Nearby Compact Dwarf Galaxies}
\tablewidth{0pt}
\tablehead{
\CH{}           			& \mc{J223831}         & \mc{J141851}       & \mc{J120202}         
					& \mc{J121402}         & \mc{J084236 }       & \mc{J171236}         & \mc{J113116} \\ 
\cline{2-15} 
\CH{Ion} & \multicolumn{14}{c}{$I(\lambda)/I$(\ion{C}{3}]) Measured from UV Spectra} }
\startdata 
C~\iv\ \W1548      		& 21.6 & 5.8   	& 19.6 & 3.4 	& \mc{\nodata} 	& 18.5 & 4.0 	& 53.6 & 14.6 	& 21.5 & 6.6 	& 62.2 & 13.7 	\\
C~\iv\ \W1550      		& 22.8 & 5.9 	& 13.0 & 3.0 	& 8.1 & 5.1 	& 9.5 & 3.7 	& 41.4 & 11.7 	& 22.1 & 6.7 	& 13.0 & 9.5 	\\
He~\ii\ \W1640     		& \mc{\nodata}	& 27.9 & 4.2 	& 7.3 & 3.6 	& \mc{\nodata}	& \mc{\nodata}	& 15.0 & 5.8 	& 19.2 & 12.2 	\\
O~\iii] \W1661     		& 13.7 & 4.9 	& 18.6 & 3.5 	& 5.6 & 6.4 	& 8.5 & 3.7 	& 11.2 & 6.1 	& 17.7 & 6.0 	& \mc{\nodata}	\\
O~\iii] \W1666     		& 25.3 & 5.5 	& 43.5 & 5.6 	& 34.2 & 7.3 	& 15.8 & 3.9 	& 47.7 & 13.4 	& 29.0 & 7.2 	& 33.1 & 12.2 	\\
N~\iii] \W1750     		& \mc{\nodata}	& \mc{\nodata}	& 9.8 & 22.3 	& \mc{\nodata}	& \mc{\nodata}	& \mc{\nodata}	& \mc{\nodata}	\\
Si~\iii] \W1883    		& 23.8 & 8.3    	& 17.7 & 4.9 	& 18.1 & 7.6 	& 16.6 & 6.9 	& \mc{\nodata}	& 36.7 & 13.4 	& \mc{\nodata}	\\
Si~\iii] \W1892    		& 7.3 & 7.9      	& 12.5 & 4.7 	& 8.9 & 7.4 	& \mc{\nodata}	& \mc{\nodata}	& 16.7 & 12.2 	& \mc{\nodata}	\\
C~\iii] \W1907     		& 61.4 &10.5   	& 59.6 & 10.3 	& 53.1 & 9.2 	& 38.6 & 7.7 	& 76.9 & 26.8 	& 53.5 & 15.0 	& 48.3 & 13.8 	\\
{[C~\iii] \W1909}  		& 38.6 & 9.0    	& 40.4 & 9.0 	& 46.9 & 8.8 	& 61.4 & 9.0 	& 23.1 & 19.1 	& 46.5 & 14.3 	& 51.7 & 14.1 \\
\hline \\
{} & \multicolumn{14}{c}{$I(\lambda)/I(\mbox{H}\beta)$ Measured from Optical Spectra} \\
\hline \vspace{-1ex} \\
{[Ne~\iii] \W3869} 		& 44.5 & 0.6 	& 8.8 & 0.1 	& 30.3 & 0.5 	& 39.9 & 0.6 	& 44.3 & 0.6 	& 53.9 & 0.8 	& 61.5 & 0.9 	\\
{[Ne~\iii] \W3968} 		& 22.0 & 0.3 	& \mc{\nodata}	& 22.9 & 0.4 	& 19.9 & 0.3 	& \mc{\nodata}	& \mc{\nodata}	& \mc{\nodata}       	\\
H\D\ \W4101        		& 27.0 & 0.4 	& 33.0 & 0.5 	& 25.4 & 0.4 	& 25.2 & 0.4 	& 26.9 & 0.4 	& 29.3 & 0.4 	& 28.5 & 0.4 	\\
H\G\ \W4340        		& 46.6 & 0.7 	& 50.4 & 0.7 	& 45.2 & 0.6 	& 45.8 & 0.6 	& 46.2 & 0.7 	& 63.9 & 0.9 	& 49.7 & 0.7 	\\
{[O~\iii] \W4363}  		& 13.5 & 0.2 	& 14.2 & 0.2 	& 9.7 & 0.1 	& 12.4 & 0.2 	& 14.5 & 0.2 	& 13.1 & 0.2 	& 14.8 & 0.2 	\\
He~\ii\ \W4686     		& 1.1 & 0.1 	& 2.0 & 0.1 	& 0.8 & 0.1 	& 0.5 & 0.5 	& 0.5 & 0.1 	& 1.6 & 0.1 	& 0.8 & 0.8 	\\
{[Ar~\iv] \W4711}  		& 1.8 & 0.1 	& 2.0 & 0.1 	& 1.2 & 0.1 	& 0.8 & 0.1 	& 2.0 & 0.1 	& 1.7 & 0.1 	& 2.1 & 0.1 	\\
{[Ar~\iv] \W4740}  		& 1.1 & 0.1 	& 1.5 & 0.1 	& \mc{\nodata}	& 0.7 & 0.1 	& 1.8 & 0.1 	& 1.1 & 0.1 	& 1.7 & 0.1 	\\
H\B\ \W4861        		& 100.0 & 1.4 	& 100.0 & 1.4 	& 100.0 & 1.4 	& 100.0 & 1.4 	& 100.0 & 1.4 	& 100.0 & 1.4 	& 100.0 & 1.4 	\\
{[O~\iii] \W4959}  		& 161.4 & 2.3 	& 160.1 & 2.3 	& 119.1 & 1.7 	& 168.1 & 2.4 	& 179.9 & 2.5 	& 180.2 & 2.5 	& 198.3 & 2.8 	\\
{[O~\iii] \W5007}  		& 481.2 & 6.8 	& 469.1 & 6.6 	& 369.6 & 5.2 	& 502.4 & 7.1 	& 529.3 & 7.5 	& 545.7 & 7.7 	& 595.2 & 8.4 	\\
O~\sc{i} \W6300    		& 1.2 & 0.1 	& 1.4 & 0.1 	& 1.2 & 0.1 	& 1.3 & 0.1 	& 1.8 & 0.1 	& 1.8 & 0.1 	& 1.2 & 0.1 	\\
{[S~\iii] \W6312}  		& 1.1 & 0.1 	& 1.0 & 0.1 	& 1.1 & 0.1 	& 1.2 & 0.1 	& 1.1 & 0.1 	& 1.5 & 0.1 	& 1.4 & 0.1 	\\
O~\sc{i} \W6363    		& \mc{\nodata}	& 0.4 & 0.1 	& 0.4 & 0.1 	& 0.5 & 0.1 	& 0.5 & 0.1 	& 0.6 & 0.1 	& 0.5 & 0.1 	\\
H\A\ \W6563        		& 275.8 & 3.9 	& 279.1 & 3.9 	& 278.5 & 3.9 	& 277.4 & 3.9 	& 277.1 & 3.9 	& 280.2 & 4.0 	& 278.0 & 3.9 	\\
{[N~\ii] \W6584}   		& 1.9 & 0.1 	& 2.1 & 0.1 	& 2.2 & 0.1 	& 1.6 & 0.1 	& 2.0 & 0.1 	& 3.3 & 0.1 	& 3.0 & 0.1 	\\
{[S~\ii] \W6717}   		& 4.5 & 0.1 	& 4.9 & 0.1 	& 4.8 & 0.1 	& 5.0 & 0.1 	& 5.4 & 0.1 	& 8.2 & 0.1 	& 6.0 & 0.1 	\\
{[S~\ii] \W6731}   		& 3.0 & 0.1 	& 3.6 & 0.1 	& 3.9 & 0.1 	& 3.7 & 0.1 	& 3.9 & 0.1 	& 5.5 & 0.1 	& 4.3 & 0.1	\\
{[Ar~\iii] \W7135} 		& 3.0 & 0.1 	& 3.2 & 0.1 	& 3.5 & 0.1 	& 3.8 & 0.1 	& 3.2 & 0.1 	& 5.0 & 0.1	& 5.1 & 0.2 	\\
{[O~\ii] \W7320}   		& 0.8 & 0.1 	& 0.9 & 0.1 	& 1.1 & 0.1 	& 1.0 & 0.1 	& 0.9 & 0.1 	& 1.2 & 0.1 	& 0.8 & 0.1	\\
{[O~\ii] \W7330}   		& 0.7 & 0.1 	& 0.6 & 0.1 	& 0.7 & 0.1 	& 0.7 & 0.1 	& 1.0 & 0.1 	& 1.2 & 0.1 	& 0.4 & 0.1 	\\
{[S~\iii] \W9069}  		& \mc{\nodata}	& \mc{\nodata}	& 6.9 & 0.1 	& 6.9 & 0.1 	& 5.8 & 0.1 	& 9.7 & 0.1 	& 10.4 & 0.1 	\\
{[S~\iii] \W9532}  		& \mc{\nodata} & \mc{\nodata}	& 13.0 & 0.2 	& \mc{\nodata}	& \mc{\nodata}	& \mc{\nodata}	& \mc{\nodata}	\vspace{0.5ex} \\
\hline \vspace{-1ex} \\
E(B$-$V)         			& 0.030 & 0.005 	& 0.140 & 0.004 	& 0.070 & 0.005 	& 0.010 & 0.005 	
					& 0.060 & 0.005 	& 0.080 & 0.004 	& 0.070 & 0.009	 \\
F$_{\rm C~III]}$ 	    	& \mc{50.9} 		& \mc{80.3} 		& \mc{47.2} 		& \mc{74.4}  
					& \mc{23.2}  		& \mc{42.6}  		& \mc{20.4}  		\\
F$_{H\beta}$     		& \mc{138.1} 		& \mc{267.2} 		& \mc{145.6} 		& \mc{188.7}  
					& \mc{124.5}  		& \mc{170.0}  		& \mc{90.1}  			
\enddata
\tablecomments{
The flux values for each object listed are the reddening corrected intensities ratios relative to \ion{C}{3}] \W\W1907,1909$\times100$
for the UV emission lines and H$\beta\times100$ for the optical emission lines.
The last three rows are the dust extinction and the raw fluxes for \ion{C}{3}] \W\W1907,1909 and H$\beta$, in units of 
$10^{-16}$ erg s$^{-1}$ cm$^{-2}$, measured from the HST and SDSS spectra, respectively.
Details of the spectral reduction and line measurements are given in Section~\ref{sec:observations}. }
\label{tbl6}
\end{deluxetable*}

\begin{deluxetable*}{r r@{$\pm$}l r@{$\pm$}l r@{$\pm$}l r@{$\pm$}l r@{$\pm$}l r@{$\pm$}l r@{$\pm$}l}
\tabletypesize{\scriptsize}
\tablewidth{0pt}
\setlength{\tabcolsep}{3pt}
\tablecaption{ Emission-Line Intensities for HST/COS Observations of Nearby Compact Dwarf Galaxies Cont.}
\tablewidth{0pt}
\tablehead{
\CH{}           			& \mc{J133126}         & \mc{J132853}       & \mc{J095430}         
					& \mc{J132347}         & \mc{J094718 }       & \mc{J150934}         & \mc{J100348} \\ 
\cline{2-15} 
\CH{Ion} & \multicolumn{14}{c}{$I(\lambda)/I$(\ion{C}{3}]) Measured from UV Spectra} }
\startdata 
C~\iv\ \W1548      		& \mc{\nodata}	& \mc{\nodata}	& 7.3 & 4.3 	& 49.0 & 8.8 	& 20.1 & 4.1 	& 17.3 & 4.8 	& 10.8 & 5.0 	\\
C~\iv\ \W1550      		& \mc{\nodata}	& \mc{\nodata}	& 21.1 & 4.8 	& 25.2 & 6.4 	& 10.2 & 3.9 	& 12.1 & 4.5 	& 9.9 & 4.9 	\\
He~\ii\ \W1640     		& 13.4 & 4.1 	& \mc{\nodata}	& 11.2 & 4.9 	& 24.7 & 5.4 	& \mc{\nodata}	& 12.9 & 5.4 	& \mc{\nodata}	\\
O~\iii] \W1661     		& 13.7 & 4.1 	& 20.0 & 9.8 	& 13.0 & 4.9 	& 28.4 & 5.7 	& 3.0 & 3.1 	& \mc{\nodata}	& \mc{\nodata}	\\
O~\iii] \W1666     		& 26.7 & 4.5 	& 52.8 & 14.9 	& 19.7 & 5.2 	& 67.7 & 10.5 	& 11.7 & 3.2 	& 34.0 & 6.9 	& 17.9 & 5.8 	\\
N~\iii] \W1750     		& 4.4 & 3.8 	& \mc{\nodata}	& \mc{\nodata}	& \mc{\nodata}	& \mc{\nodata}	& \mc{\nodata}	& \mc{\nodata}	\\
Si~\iii] \W1883    		& 21.6 & 6.2 	& \mc{\nodata}	& 22.5 & 7.8 	& 30.4 & 11.0 	& 16.8 & 5.3 	& 19.1 & 11.4 	& 12.9 & 17.1 	\\
Si~\iii] \W1892    		& \mc{\nodata} 	& \mc{\nodata}	& 21.8 & 7.8 	& 18.1 & 10.4 	& 12.8 & 5.2 	& 10.2 & 11.1 	& \mc{\nodata}	\\
C~\iii] \W1907     		& 46.4 & 7.1 	& 56.9 & 20.8 	& 63.9 & 10.0 	& 44.8 & 11.9 	& 34.8 & 5.8 	& 35.0 & 12.2 	& 42.8 & 19.7 	\\
{[C~\iii] \W1909}  		& 53.6 & 7.5 	& 43.1 & 19.0 	& 36.1 & 8.3 	& 55.2 & 12.8 	& 65.2 & 7.0 	& 65.0 & 14.9 	& 57.2 & 21.7 	\\
\hline \\
{} & \multicolumn{14}{c}{$I(\lambda)/I(\mbox{H}\beta)$ Measured from Optical Spectra} \\
\hline \vspace{-1ex} \\
{[Ne~\iii] \W3869} 		& 50.1 & 0.7 	& 47.1 & 0.7 	& 47.9 & 0.7 	& 60.1 & 0.9 	& 53.2 & 1.2 	& 52.3 & 0.7 	& \mc{\nodata}	\\
{[Ne~\iii] \W3968} 		& \mc{\nodata}	& 23.2 & 0.4 	& 23.9 & 0.4 	& 26.7 & 0.4 	& \mc{\nodata}	& 24.5 & 0.4 	& \mc{\nodata}	\\
H\D\ \W4101        		& 30.5 & 0.4 	& 26.5 & 0.4 	& 28.0 & 0.4 	& 25.4 & 0.4 	& 26.9 & 1.0 	& 24.9 & 0.4 	& 27.2 & 0.4 	\\
H\G\ \W4340        		& 48.8 & 0.7 	& 48.7 & 0.7 	& 46.7 & 0.7 	& 47.4 & 0.7 	& 48.8 & 0.7 	& 46.3 & 0.7 	& 47.9 & 0.7 	\\
{[O~\iii] \W4363}  		& 12.8 & 0.2 	& 11.5 & 0.2 	& 12.9 & 0.2 	& 20.1 & 0.3 	& 12.5 & 0.2 	& 15.4 & 0.2 	& 14.3 & 0.2 	\\
He~\ii\ \W4686     		& 1.3 & 0.1 	& 1.4 & 0.1 	& 0.8 & 0.8 	& 1.6 & 0.1 	& 1.5 & 0.1 	& 1.6 & 0.1 	& 0.9 & 0.1 	\\
{[Ar~\iv] \W4711}  		& 1.7 & 0.1 	& 1.5 & 0.1 	& 1.9 & 0.1 	& 4.4 & 0.1 	& 2.2 & 0.1 	& 2.8 & 0.1 	& 2.7 & 0.1 	\\
{[Ar~\iv] \W4740}  		& 1.1 & 0.1 	& 1.0 & 0.1 	& 1.3 & 0.1 	& 3.0 & 0.1 	& 0.9 & 0.1 	& 2.1 & 0.1 	& 1.5 & 0.1 	\\
H\B\ \W4861        		& 100.0 & 1.4 	& 100.0 & 1.4 	& 100.0 & 1.4 	& 100.0 & 1.4 	& 100.0 & 1.4 	& 100.0 & 1.4 	& 100.0 & 1.4 	\\
{[O~\iii] \W4959}  		& 174.7 & 2.5 	& 169.1 & 2.4 	& 174.6 & 2.5 	& 248.6 & 3.5 	& 186.5 & 2.6 	& 227.8 & 3.2 	& 223.3 & 3.2 	\\
{[O~\iii] \W5007}  		& 543.0 & 7.7 	& 506.2 & 7.2 	& 541.7 & 7.7 	& 742.1 & 10.5 	& 553.9 & 7.8 	& 677.0 & 9.6 	& 682.8 & 9.7 	\\
O~\sc{i} \W6300    		& 1.4 & 0.1 	& 2.5 & 0.1 	& 2.3 & 0.1 	& 0.6 & 0.1 	& 1.3 & 0.1 	& 1.5 & 0.1 	& 1.5 & 0.1 	\\
{[S~\iii] \W6312}  		& 1.4 & 0.1 	& 1.6 & 0.1 	& 1.6 & 0.1 	& 1.0 & 0.1 	& 2.1 & 0.1 	& 1.3 & 0.1 	& 1.6 & 0.1 	\\
O~\sc{i} \W6363    		& 0.5 & 0.1 	& 0.8 & 0.1 	& 0.6 & 0.1 	& 0.3 & 0.1 	& 0.5 & 0.1 	& 0.5 & 0.1 	& \mc{\nodata}	\\
H\A\ \W6563        		& 279.3 & 3.9 	& 279.3 & 4.0 	& 278.6 & 3.9 	& 277.7 & 3.9 	& 277.6 & 3.9	& 279.4 & 4.0 	& 279.8 & 4.0 	\\
{[N~\ii] \W6584}   		& 2.4 & 0.1 	& 3.3 & 0.1 	& 3.0 & 0.1 	& 1.0 & 0.1 	& 3.0 & 0.1 	& 2.6 & 0.1 	& 2.3 & 0.1 	\\
{[S~\ii] \W6717}   		& 6.1 & 0.1 	& 8.1 & 0.1 	& 8.6 & 0.1 	& 2.1 & 0.1 	& 7.5 & 0.1 	& 5.6 & 0.1 	& 6.6 & 0.1 	\\
{[S~\ii] \W6731}   		& 4.6 & 0.1 	& 6.6 & 0.1 	& 6.3 & 0.1 	& 2.2 & 0.1 	& 5.3 & 0.1 	& 4.3 & 0.1 	& 4.7 & 0.1 	\\
{[Ar~\iii] \W7135} 		& 4.4 & 0.1 	& 4.9 & 0.1 	& 5.3 & 0.1 	& 3.2 & 0.1 	& 6.1 & 0.2 	& 4.7 & 0.1 	& 6.0 & 0.1 	\\
{[O~\ii] \W7320}   		& 1.0 & 0.1 	& 1.5 & 0.1 	& 1.3 & 0.1 	& 0.5 & 0.1 	& 1.2 & 0.1 	& 1.0 & 0.1 	& 0.8 & 0.1 	\\
{[O~\ii] \W7330}   		& 0.9 & 0.1 	& 1.1 & 0.1 	& 0.9 & 0.1 	& 0.5 & 0.1 	& 0.9 & 0.1 	& 0.8 & 0.1 	& 0.7 & 0.1 	\\
{[S~\iii] \W9069}  		& 7.6 & 0.1 	& \mc{\nodata}	& 10.2 & 0.2 	& \mc{\nodata}	& 10.9 & 0.2 	& \mc{\nodata}	& \mc{\nodata}	\vspace{0.5ex} \\
\hline \vspace{-1ex} \\
E(B$-$V)         		& 0.080 & 0.005 		& 0.090 & 0.010 		& 0.130 & 0.010 		& 0.120 & 0.008 
				        & 0.100 & 0.009 		& 0.110 & 0.010 		& 0.040 & 0.012 		\\
F$_{\rm C~III]}$ 	    & \mc{100.9} 			& \mc{23.9} 			& \mc{39.9} 			& \mc{59.9}  
				        & \mc{55.0}  			& \mc{63.8}  			& \mc{45.5}  			\\
F$_{H\beta}$     	    & \mc{275.5} 			& \mc{140.8}			& \mc{122.3}			& \mc{169.5} 
				        & \mc{133.8}			& \mc{274.7} 			& \mc{155.4} 			
\enddata
\label{tbl7}
\end{deluxetable*}

\begin{deluxetable*}{r r@{$\pm$}l r@{$\pm$}l r@{$\pm$}l r@{$\pm$}l r@{$\pm$}l r@{$\pm$}l }
\tabletypesize{\scriptsize}
\tablewidth{0pt}
\setlength{\tabcolsep}{3pt}
\tablecaption{ Emission-Line Intensities for HST/COS Observations of Nearby Compact Dwarf Galaxies Cont.}
\tablewidth{0pt}
\tablehead{
\CH{}           			& \mc{J025346}         & \mc{J015809}       & \mc{J104654}         & \mc{J093006}	& \mc{J092055 } 	& \mc{{\it J084956} } \\ 
\cline{2-13} 
\CH{Ion} & \multicolumn{11}{c}{$I(\lambda)/I$(\ion{C}{3}]) Measured from UV Spectra} }
\startdata 
C~\iv\ \W1548      		& 10.7 & 4.8	& 20.0 & 9.4 	& \mc{\nodata}	& \mc{\nodata}	& 7.3 & 2.7 	& \mc{\nodata}	\\
C~\iv\ \W1550      		& 11.0 & 4.8 	& 31.2 & 10.3 	& 18.9 & 5.9 	& 21.7 & 7.3 	& 5.3 & 2.7 	& \mc{\nodata} 	\\
He~\ii\ \W1640     		& \mc{\nodata}	& \mc{\nodata}	& \mc{\nodata}	& \mc{\nodata}	& 12.2 & 5.5 	& \mc{\nodata}	\\
O~\iii] \W1661     		& \mc{\nodata}	& \mc{\nodata}	& \mc{\nodata}	& 10.6 & 5.2 	& \mc{\nodata}	& \mc{\nodata}	\\
O~\iii] \W1666     		& 16.1 & 4.1 	& 17.5 & 6.5 	& 24.8 & 5.9 	& \mc{\nodata}	& 18.2 & 5.5 	& 24.6 & 25.7 	\\
N~\iii] \W1750     		& \mc{\nodata}	& \mc{\nodata}	& \mc{\nodata}	& \mc{\nodata}	& \mc{\nodata}	& \mc{\nodata}	\\
Si~\iii] \W1883    		& 31.2 & 9.7 	& 20.2 & 13.0	& 16.1 & 7.9 	& 13.0 & 8.3 	& 33.9 & 7.0 	& \mc{\nodata} 	\\
Si~\iii] \W1892    		& \mc{\nodata}	& 14.5 & 12.8 	& \mc{\nodata}	& 18.7 & 8.4 	& 7.4 & 6.3 	& 21.9 & 42.0 	\\
C~\iii] \W1907     		& 53.8 & 11.2 	& 44.6 & 14.8 	& 58.4 & 9.9 	& 38.4 & 9.3 	& 50.0 & 7.7 	& 63.9 & 54.0 	\\
{[C~\iii] \W1909}  		& 46.2 & 10.7 	& 55.4 & 15.9 	& 41.6 & 8.9 	& 61.6 & 10.8 	& 50.0 & 7.7 	& 36.1 & 44.9   \\
\hline \\
{} & \multicolumn{12}{c}{$I(\lambda)/I(\mbox{H}\beta)$ Measured from Optical Spectra} \\
\hline \vspace{-1ex} \\
{[Ne~\iii] \W3869}	 	& 50.0 & 0.7 	& \mc{\nodata}	& 49.4 & 0.7 	& 50.3 & 1.5	& 51.8 & 0.7 	& 64.0 & 0.9	\\
{[Ne~\iii] \W3968} 		& 24.3 & 0.4 	& 30.9 & 0.5 	& 24.1 & 0.3 	& 30.6 & 1.4	& 34.2 & 0.5 	& 36.8 & 0.5	\\
H\D\ \W4101        		& 27.0 & 0.4 	& 25.6 & 0.4 	& 26.2 & 0.4 	& 27.5 & 0.4	& 29.7 & 0.4 	& 31.4 & 0.4	\\
H\G\ \W4340        		& 49.5 & 0.7 	& 47.8 & 0.7 	& 49.4 & 0.7 	& 47.2 & 0.7	& 53.7 & 0.8 	& 51.9 & 0.7	\\
{[O~\iii] \W4363}  		& 9.1 & 0.1 	& 13.2 & 0.2 	& 9.7 & 0.1 	& 7.1 & 0.2	    & 10.4 & 0.1 	& 7.8 & 0.1	\\
He~\ii\ \W4686     		& 0.4 & 0.1 	& 0.7 & 0.1 	& 1.0 & 0.1 	& 1.4 & 0.8	    & 1.2 & 0.1 	& 0.9 & 0.1 	\\
{[Ar~\iv] \W4711}  		& 1.1 & 0.1 	& 1.8 & 0.1 	& 1.3 & 0.1 	& 0.8 & 0.8	    & 1.2 & 0.1 	& 0.5 & 0.1	\\
{[Ar~\iv] \W4740}  		& \mc{\nodata}	& 1.1 & 0.1 	& 0.6 & 0.1 	& 0.6 & 0.8	    & 0.9 & 0.1 	& \mc{\nodata}	\\
H\B\ \W4861        		& 100.0 & 1.4 	& 100.0 & 1.4 	& 100.0 & 1.4 	& 100.0 & 1.4	& 100.0 & 1.4 	& 100.0 & 1.4	\\
{[O~\iii] \W4959}  		& 190.0 & 2.7 	& 204.2 & 2.9 	& 180.2 & 2.5 	& 165.9 & 2.4	& 187.7 & 2.7 	& 168.3 & 2.4	\\
{[O~\iii] \W5007}  		& 557.3 & 7.9 	& 625.6 & 8.9 	& 577.8 & 8.2 	& 496.8 & 7.1	& 560.4 & 7.9 	& 502.2 & 7.1	\\
O~\sc{i} \W6300    		& 1.3 & 0.1 	& 1.1 & 0.1 	& 2.3 & 0.1 	& 3.0 & 0.3	    & 3.4 & 0.1	    & 3.5 & 0.1	\\
{[S~\iii] \W6312}  		& 1.8 & 0.1 	& 1.7 & 0.1 	& 2.0 & 0.1 	& 1.9 & 0.3	    & \mc{\nodata}	& 1.8 & 0.1	\\
O~\sc{i} \W6363    		& 0.6 & 0.1 	& 0.6 & 0.1 	& 0.7 & 0.1 	& 1.0 & 0.3 	& 1.2 & 0.1 	& 1.1 & 0.1	\\
H\A\ \W6563        		& 280.5 & 4.0 	& 278.4 & 3.9 	& 281.8 & 4.0 	& 280.8 & 4.0	& 282.6 & 4.0 	& 283.2 & 4.0	\\
{[N~\ii] \W6584}   		& 4.1 & 0.1 	& 2.5 & 0.1 	& 4.9 & 0.1 	& 7.50 & 0.1	& 4.8 & 0.1 	& 6.9 & 0.1	\\
{[S~\ii] \W6717}   		& 8.8 & 0.1 	& 5.8 & 0.1 	& 11.4 & 0.2 	& 16.3 & 0.2	& 12.0 & 0.2 	& 15.6 & 0.2	\\
{[S~\ii] \W6731}   		& 6.3 & 0.1 	& 4.4 & 0.1 	& 8.6 & 0.1 	& 11.6 & 0.2	& 8.5 & 0.1 	& 11.6 & 0.2	\\
{[Ar~\iii] \W7135} 		& 7.3 & 0.1 	& 6.2 & 0.2 	& 7.3 & 0.2 	& 7.0 & 0.1	    & 5.3 & 0.1 	& 6.9 & 0.1	\\
{[O~\ii] \W7320}   		& 1.4 & 0.1 	& 1.1 & 0.1 	& 1.7 & 0.1 	& 2.1 & 0.1	    & 1.8 & 0.1 	& 2.3 & 0.1	\\
{[O~\ii] \W7330}   		& 1.1 & 0.1 	& 0.8 & 0.1 	& 1.3 & 0.1 	& 1.7 & 0.1 	& 1.5 & 0.1 	& 1.8 & 0.1	\\
{[S~\iii] \W9069}  		& 16.2 & 0.2 	& 11.3 & 0.2 	& 14.9 & 0.2 	& 15.4 & 0.2	& 11.8 & 0.2 	& \mc{\nodata}	 \vspace{0.5ex} \\
\hline \vspace{-1ex} \\
E(B$-$V)         			& 0.030 & 0.009	& 0.090 & 0.010	& 0.110 & 0.070	& 0.080 & 0.005 	& 0.160 & 0.009 	& 0.140 & 0.010 \\
F$_{\rm C~III]}$ 	    	& \mc{46.5} 		& \mc{26.8} 		& \mc{54.5} 		& \mc{58.9}  		& \mc{44.5}  		& \mc{44.3}   	\\
F$_{H\beta}$     		& \mc{183.5}		& \mc{83.1}		& \mc{196.5}		& \mc{34.70}		& \mc{267.9} 		& \mc{273.4} 				
\enddata
\label{tbl8}
\tablecomments{Note that J084956 is not a part of our 3$\sigma$ sample, but is included here for modeling purposes
(included to extend parameter space coverage). }
\end{deluxetable*}


\begin{deluxetable*}{ccccccccccc}
\tablecaption{UV Emission-Line Equivalent Widths for Nearby Galaxies}
\tablehead{
\CH{}				& 
\mc{\ion{C}{4} }			& 
\CH{\ion{He}{2} }		& 
\mc{\ion{O}{3}] }		& 
\mc{\ion{Si}{3}] }		& 
\mc{\ion{C}{3}]}			& 
\CH{}				\\
\CH{Target}			& 
\CH{\W1548}			& 
\CH{\W1550}			& 
\CH{\W1640}			& 
\CH{\W1661}			& 
\CH{\W1666}			& 
\CH{\W1883}			& 
\CH{\W1892}			& 
\CH{\W1907}			& 
\CH{\W1909}			& 
\CH{Ref.}				} 
\startdata
\multicolumn{11}{c}{From Newly Obtained UV Spectra} \\
\hline \vspace{-1ex} \\
{J223831}      		& 2.00	& 2.13	& \ND	& 1.46	& 2.65	& 3.58	& 1.10	& 9.31	& 5.86		& B18 	\\
{J141851}			& 1.11	& 1.41	& 3.36	& 2.26	& 5.31	& 3.09	& 2.22	& 10.95	& 7.46		& B18	\\
{J120202}     		& \ND	& 0.56	& 0.56	& 0.45	& 2.80	& 2.12	& 1.06	& 6.37	& 5.63		& B18	\\
{J121402}     		& 1.89	& 0.98	& \ND	& 1.00	& 1.87	& 3.27	& \ND	& 6.48	& 10.13		& B18	\\ 
{J084236}     		& 3.68	& 2.84	& \ND	& 0.73	& 3.10	& \ND	& \ND	& 7.51	& 2.28		& B18	\\
{J171236}       		& 1.93	& 1.99	& 1.49	& 1.78	& 2.91	& 6.02	& 2.73	& 8.65	& 7.43		& B18	\\ 
{J113116}			& 3.03	& 0.63	& 1.05	& \ND	& 1.79	& \ND	& \ND	& 3.10	& 3.30		& B18	\\
{J133126}      		& \ND	& \ND	& 1.24	& 1.30	& 2.55	& 2.94	& \ND	& 6.10	& 7.02		& B18	\\
{J132853}			& \ND	& \ND	& \ND	& 0.64	& 1.79	& \ND	& \ND	& 3.94	& 2.55		& B18	\\
{J095430}     		& 0.76	& 2.17	& 1.28	& 1.50	& 2.29	& 3.61	& 3.52	& 10.31	& 5.82		& B18	\\
{J132347}     		& 3.37	& 1.75	& 2.02	& 2.39	& 5.71	& 4.13	& 2.39	& 5.72	& 7.00		& B18	\\
{J094718}     		& \ND	& \ND	& \ND	& 0.33	& 1.49	& 3.84	& 2.79	& 7.16	& 13.27		& B18	\\
{J150934}       		& 1.20	& 0.84	& 0.92	& \ND	& 2.41	& 2.01	& 1.06	& 3.64	& 6.74		& B18	\\
{J100348}			& 0.93	& 0.86	& \ND	& \ND	& 1.54	& 1.59	& \ND	& 4.70	& 6.24		& B18	\\
{J025346}      		& \ND	& \ND	& \ND	& \ND	& 0.62	& 1.94	& \ND	& 3.66	& 3.16		& B18	\\
{J015809}			& \ND	& \ND	& \ND	& \ND	& 1.54	& 3.03	& 2.10	& 6.27	& 7.69		& B18 	\\
{J104654}     		& \ND	& 1.44	& \ND	& \ND	& 2.07	& 1.97	& \ND	& 7.10	& 5.04		& B18	\\
{J093006}    		& \ND	& \ND	& \ND	& 0.36	& \ND	& \ND	& \ND	& 1.73	& 2.85		& B18	\\
{J092055}     		& \ND	& 0.41	& 0.78	& \ND	& 1.19	& 3.14	& 0.67	& 4.33	& 5.51		& B18	\\ 
\hline \vspace{-1.5ex} \\
\multicolumn{11}{c}{From Archival Observations} \\
\hline \vspace{-1ex} \\
{SB2}			& 0.25    	& 0.22    	& 1.70	& 1.94	& 5.05	& \ND  	& \ND  	& \ND 	& 14.86		& S17	\\
{SB36}			& \ND	& \ND	& \ND	& 0.26	& 0.84	& \ND	& \ND	& \ND	& 4.98		& S17	\\
{SB82}			& 0.67	& 0.41	& 0.44	& 0.85	& 1.89	& \ND	& \ND	& \ND	&12.09		& S17	\\		
{SB179}			& \ND	& \ND	& \ND	& \ND	& 1.17	& \ND	& \ND	& \ND	& 8.71		& S17	\\
{SB182}			& 0.96	& 0.55 	& 0.94  	& 0.70	& 1.73	& \ND	& \ND	& \ND	& 13.35		& S17	\\	
{SB191}			& \ND	& \ND	& \ND	& 0.64	& 1.59	& \ND	& \ND	& \ND	& 11.33		& S17	\\	
{SBS0948+532}  	& \ND	& \ND	& \ND	& \ND	& \ND	& \ND	& \ND	& \ND	& 2			& PG17	\\	
{SBS1054+365}	& \ND	& \ND	& \ND	& \ND	& \ND	& \ND	& \ND	& \ND	& 2			& PG17	\\	
{SBS1319+579}	& \ND	& \ND	& \ND	& \ND	& \ND	& \ND	& \ND	& \ND	& 8			& PG17	\\	
{J082555}			& 1.06	& 0.76	& 0.78	& 1.16	& 1.74	& 2.98	& 3.26	& 7.15	& 9.34		& B16	\\
{J104457}			& 6.17	& 4.52	& 2.36	& 2.98	& 5.25	& 3.12	& 2.73	& 11.70	& 4.65		& B16	\\
{J120122}			& \ND	& \ND	& \ND	& \ND	& 1.72	& \ND	& \ND	& 7.82	& 4.10		& B16	\\
{J124159}			& \ND	& \ND	& \ND	& \ND	& 3.54	& 2.53	& 3.13	& 6.56	& 4.00		& B16	\\
{J122622}			& \ND	& \ND	& \ND	& \ND	& 2.55	& \ND 	& \ND	& 5.52	& 2.60		& B16	\\
{J122436}			& \ND	& \ND	& \ND	& 0.92	& 1.72	& \ND	& 1.81	& 4.72	& 4.12		& B16	\\
{J124827}			& 1.51	& 0.74	& 1.97	& \ND	& 1.60	& \ND	& \ND	& 5.02	& 2.49		& B16	
\enddata
\tablecomments{UV emission-line equivalent widths (EWs) of the nearby 3$\sigma$ UV CEL C/O targets used in this analysis. 
Units of EWs are given in \AA.
The last column gives the reference of the observations:
B18: this work; S17: \cite{senchyna17}; PG17: \cite{pena-guerrero17}; B16: \cite{berg16}.
Note that EWs are not available for the five additional C/O detections from B16 (not reported in their original studies).}
\label{tbl9}
\end{deluxetable*}


\begin{deluxetable*}{lccccccc}
\tabletypesize{\scriptsize}
\tablewidth{0pt} 
\setlength{\tabcolsep}{3pt}
\tablecaption{ Ionic and Total Abundance for HST/COS Compact Dwarf Galaxies }\label{tbl10}
\tablewidth{0pt}
\tablehead{
\CH{Property}                       	& \CH{J223831}      	& \CH{J141851}	& \CH{J120202}     		& \CH{J121402}      		& \CH{J084236}      	& \CH{J171236}   		& \CH{J113116}}
\startdata
\multicolumn{8}{c}{Properties Derived from Optical Spectra} \\
\hline \vspace{-1ex} \\
T$_e$ [O~\iii] (K)         		& 17870$\pm$240 	& 18730$\pm$240 	& 17230$\pm$250 		& 16690$\pm$170 		& 17627$\pm$230 	& 16440$\pm$190 		& 16790$\pm$170 	\\
T$_e$ [S~\iii] (K)$^a$      		& 16530$\pm$200 	& 18010$\pm$180 	& 16040$\pm$200		& 17030$\pm$140		& 17221$\pm$190	& 15440$\pm$160		& 14070$\pm$140	\\
T$_e$ [N~\ii] (K)$^a$   		& 15510$\pm$170 	& 16110$\pm$170 	& 15060$\pm$170 		& 14680$\pm$120 		& 15339$\pm$160 	& 14510$\pm$130 		& 14760$\pm$120 	\\
n$_e$ C~\iii] (cm$^{-3}$)  		& $<1913^b$         	& $<1750^b$ 		& $<14830^b$ 			& $<59600^b$ 			& 100$^b$        		& $<13580^b$ 			& $<26500^b$ 		\\
n$_e$ Si~\iii] (cm$^{-3}$)  	& 100$^b$ 		& $<3460^b$	 	& 100$^b$ 			&100$^b$ 			& \nodata         		& 100$^b$ 			& \nodata         		\\
n$_e$ [S~\ii] (cm$^{-3}$)  		& 100$^b$ 		& 60$\pm$40 		& 210$\pm$40			& 60$\pm$40 			& 40$\pm$40 		& 100$^b$ 			& 40$\pm$90 		\\
\\
O$^{+}$/H$^+$ (10$^5$)     	& 0.338$\pm$0.026 	& 0.312$\pm$0.018 	& 0.436$\pm$0.030 		& 0.526$\pm$0.037 		& 0.472$\pm$0.037	 & 0.718$\pm$0.041	 	& 0.375$\pm$0.039 	\\
O$^{+2}$/H$^+$ (10$^5$)    	& 3.528$\pm$0.198 	& 3.174$\pm$0.176 	& 3.826$\pm$0.219 		& 4.127$\pm$0.218 		& 3.644$\pm$0.202	 & 4.265$\pm$0.229 		& 4.069$\pm$0.213 	\\
12+log(O/H)                 		& 7.587$\pm$0.022 	& 7.542$\pm$0.022 	& 7.630$\pm$0.023 		& 7.668$\pm$0.021 		& 7.614$\pm$0.022	 & 7.698$\pm$0.020 		& 7.648$\pm$0.021 	\\
\\
N$^{+}$/H$^+$ (10$^7$)     	& 1.501$\pm$0.055 	& 1.502$\pm$0.055 	& 1.767$\pm$0.069 		& 1.431$\pm$0.083	 	& 1.573$\pm$0.077 	& 2.913$\pm$0.085 		& 2.599$\pm$0.130 	\\
log(N/O)                   			& $-1.353\pm$0.037	& $-1.317\pm$0.030 	& $-1.393\pm$0.034 		& $-1.566\pm$0.040 		& $-1.477\pm$0.040 	& $-1.392\pm$0.028 		& $-1.159\pm$0.050 	\\
\\
S$^{+}$/H$^+$ (10$^7$)     	& 0.745$\pm$0.017 	& 0.781$\pm$0.019 	& 0.934$\pm$0.025 		& 0.945$\pm$0.025 		& 0.930$\pm$0.024 	& 1.533$\pm$0.033 		& 1.100$\pm$0.030 	\\
S$^{+2}$/H$^+$ (10$^7$)    	& \nodata         		& 3.068$\pm$0.063 	& 4.181$\pm$0.070 		& 4.367$\pm$0.063 		& 3.628$\pm$0.116 	& 7.011$\pm$0.132 		& 8.824$\pm$0.130 	\\
S ICF                       			& \nodata         		& 2.652$\pm$0.155 	& 2.402$\pm$0.148 		& 2.238$\pm$0.134 		& 2.217$\pm$0.142 	& 1.908$\pm$0.087 		& 2.772$\pm$0.243 	\\
log(S/O)                   			& \nodata         		& $-1.533\pm$0.024	& $-1.540\pm$0.025 		& $-1.593\pm$0.024 		& $-1.610\pm$0.026 	& $-1.485\pm$0.023 		& $-1.208\pm$0.025 	\\
\\
Ne$^{+2}$/H$^{+}$ (10$^5$) 	& 0.745$\pm$0.028 	& 0.645$\pm$0.024 	& 0.558$\pm$0.027 		& 0.797$\pm$0.025 		& 0.768$\pm$0.029 	& 1.120$\pm$0.037 		& 1.208$\pm$0.037 	\\
Ne$^{+2}$/O$^{+2}$         		& 0.211$\pm$0.014 	& 0.203$\pm$0.014 	& 0.146$\pm$0.011 		& 0.193$\pm$0.012 		& 0.211$\pm$0.014 	& 0.263$\pm$0.016 		& 0.297$\pm$0.018 	\\
Ne ICF                      			& 1.096$\pm$0.083 	& 1.098$\pm$0.083 	& 1.114$\pm$0.086 		& 1.128$\pm$0.080 		& 1.130$\pm$0.084 	& 1.168$\pm$0.083 		& 1.092$\pm$0.078 	\\
log(Ne/O)                  			& $-0.636\pm$0.029 	& $-0.652\pm$0.029 	& $-0.790\pm$0.033 		& $-0.662\pm$0.026 		& $-0.624\pm$0.029 	& $-0.513\pm$0.027 		& $-0.489\pm$0.026 	\\
\\
Ar$^{+2}$/H$^{+}$ (10$^7$) 	& 0.934$\pm$0.037 	& 0.867$\pm$0.017 	& 1.126$\pm$0.028 		& 1.122$\pm$0.029 		& 0.922$\pm$0.033 	& 1.709$\pm$0.040 		& 2.076$\pm$0.062 	\\
Ar$^{+3}$/H$^{+}$ (10$^7$) 	& 0.760$\pm$0.159 	& 0.893$\pm$0.151 	& \nodata         			& 0.579$\pm$0.035 		& 1.237$\pm$0.264 	& 0.858$\pm$0.168 		& 1.273$\pm$0.323 	\\
Ar/O                        			& 0.005$\pm$0.001 	& 0.006$\pm$0.001 	& 0.003$\pm$0.001 		& 0.004$\pm$0.001 		& 0.006$\pm$0.001 	& 0.006$\pm$0.001 		& 0.008$\pm$0.002 	\\
Ar ICF                     			& 1.011$\pm$0.001 	& 1.011$\pm$0.001 	& 2.725$\pm$0.108 		& 1.014$\pm$0.001 		& 1.015$\pm$0.002 	& 1.021$\pm$0.002 		& 1.010$\pm$0.001 	\\
log(Ar/O)                  			& $-2.314\pm$0.096 	& $-2.251\pm$0.078	& $-2.096\pm$0.027	 	& $-2.379\pm$0.036 		& $-2.221\pm$0.097 	& $-2.212\pm$0.089 		& $-2.080\pm$0.113 	\\
\hline 
\multicolumn{8}{c}{Properties Derived from UV Spectra} \\
\hline \vspace{-1ex} \\
C$^{+2}$/O$^{+2}$          		& 0.221$\pm$0.053 	& 0.133$\pm$0.021 	& 0.161$\pm$0.038 		& 0.336$\pm$0.088 		& 0.116$\pm$0.035 	& 0.181$\pm$0.048 		& 0.161$\pm$0.061 	\\
C$^{+3}$/C$^{+2}$          		& 0.229$\pm$0.050 	& 0.161$\pm$0.030 	& 0.043$\pm$0.028  		& 0.154$\pm$0.049 		& 0.493$\pm$0.140 	& 0.243$\pm$0.064 		& 0.410$\pm$0.103 	\\
log U                       			& $-2.317\pm$0.053 	& $-2.416\pm$0.028 	& $-2.301\pm$0.025 		& $-2.409\pm$0.043 		& $-2.400\pm$0.047 	& $-2.477\pm$0.037 		& $-2.222\pm$0.084 	\\
C ICF                      			& 0.989$\pm$0.053 	& 0.987$\pm$0.029 	& 0.993$\pm$0.025 		& 0.963$\pm$0.043 		& 0.966$\pm$0.047 	& 0.942$\pm$0.037 		& 1.014$\pm$0.084 	\\
log(C/O)                   			& $-0.661\pm$0.095 	& $-0.883\pm$0.065 	& $-0.797\pm$0.093 		& $-0.490\pm$0.102 		& $-0.949\pm$0.114 	& $-0.768\pm$0.102 		& $-0.787\pm$0.142 	
\enddata
\tablecomments{
Ionic and total abundance calculations for our compact dwarf galaxy sample.
The T$_e$~[\ion{O}{3}] electron temperature, n$_e$~[\ion{S}{2}] density, and oxygen and nitrogen abundance are determined using the SDSS optical spectra. 
Due to the wavelength coverage of the SDSS optical spectra and the redshifts of our sample, 
[\ion{O}{2}] $\lambda3727$ was observed in only three of our galaxies (see Tables~$6-8$). 
Instead, O$^{+}$/H$^+$ ionic abundances waere determined from the [\ion{O}{2}] $\lambda\lambda7320,7330$ lines unless otherwise noted. 
C/O abundances were calculated using the C$^{+2}$/O$^{+2}$ ratio and the ICF relationship 
derived in Section 4.4 (see Table~3 and Figure~6). \\
$^{a}$ T$_e$[S~\iii], T$_e$[N~\iii] determined from T$_e$[O~\iii] using the theoretical \citet{garnett92} relationships. \\
$^{b}$ n$_e$[\ion{S}{2}], n$_e$\ion{C}{3}], and n$_e$\ion{Si}{3}] are set to 100 when the [\ion{S}{2}], \ion{C}{3}], or \ion{Si}{3}]
ratios exceed the low density theoretical limit and are given as upper limits when they are discrepantly large.  \\
$^{c}$ O$^{+}$/H$^+$ ionic was determined using [\ion{O}{2}] $\lambda3727$ line.}
\label{tbl10}
\end{deluxetable*}

\begin{deluxetable*}{lccccccc}
\tabletypesize{\scriptsize}
\tablewidth{0pt} 
\setlength{\tabcolsep}{3pt}
\tablecaption{Ionic and Total Abundance for HST/COS Compact Dwarf Galaxies - Cont.}
\tablewidth{0pt}
\tablehead{\CH{Property}           & \CH{J133126}	& \CH{J132853}	& \CH{J095430}	& \CH{J132347}	& \CH{J094718}	& \CH{J150934}	& \CH{J100348}}
\startdata 
\multicolumn{8}{c}{Properties Derived from Optical Spectra} \\
\hline \vspace{-1ex} \\
T$_e$ [O~\iii] (K)       		& 16330$\pm$170 	& 16030$\pm$230 	& 16390$\pm$240 	& 17540$\pm$220 	& 15990$\pm$260 	& 16040$\pm$170 	& 15420$\pm$190 	\\
T$_e$ [S~\iii]  (K)$^a$ 		& 17400$\pm$140 	& 15010$\pm$190 	& 15780$\pm$200 	& 16260$\pm$180 	& 17920$\pm$220 	& 15020$\pm$140 	& 14500$\pm$150 	\\
T$_e$ [N~\ii]  (K)$^a$  		& 14430$\pm$120 	& 14220$\pm$160 	& 14470$\pm$170 	& 15280$\pm$150	& 14190$\pm$120 	& 14220$\pm$120 	& 13790$\pm$130 	\\
n$_e$ C~\iii] (cm$^{-3}$) 		& $<31720^b$ 		& $<6520^b$  		& $100^b$        		& $<38170^b$ 		& $<76640^b$		& $<75840^b$		& $<42690^b$		\\
n$_e$ Si~\iii] (cm$^{-3}$) 		& \nodata         		& \nodata         		& $<15030^b$ 		& $100^b$    		& $<5720^b$    		& $100^b$    		& \nodata         		\\
n$_e$ [S~\ii] (cm$^{-3}$) 		& 90$\pm$30 		& 190$\pm$40 		& 70$\pm$30 		& 630$\pm$130	& 20$\pm$60 		& 110$\pm$30 		& 30$\pm$30 		\\
\\	
O$^{+}$/H$^+$ (10$^5$)     	& 0.604$\pm$0.031 	& 0.824$\pm$0.055 	& 0.699$\pm$0.051 	& 0.170$\pm$0.023 	& 0.776$\pm$0.062 & 0.590$\pm$0.021 & 0.603$\pm$0.056 \\
O$^{+2}$/H$^+$ (10$^5$)    	& 4.335$\pm$0.227 	& 4.514$\pm$0.258 	& 4.297$\pm$0.248 	& 3.667$\pm$0.202 	& 4.553$\pm$0.274 & 4.519$\pm$0.238 & 4.948$\pm$0.271 \\
12+log(O/H)                 		& 7.694$\pm$0.020 	& 7.727$\pm$0.021 	& 7.699$\pm$0.022 	& 7.584$\pm$0.023 	& 7.727$\pm$0.023 & 7.708$\pm$0.020 & 7.744$\pm$0.022 \\
\\
N$^{+}$/H$^+$ (10$^7$)     	& 2.175$\pm$0.068 	& 3.084$\pm$0.124 	& 2.689$\pm$0.115 	& 0.799$\pm$0.064 	& 2.793$\pm$0.128 	& 2.407$\pm$0.068 	& 2.223$\pm$0.102 \\
log(N/O)                   			& $-1.443\pm$0.026 	& $-1.427\pm$0.034 	& $-1.415\pm$0.037 	& $-1.328\pm$0.067 & $-1.444\pm$0.040 & $-1.390\pm$0.020 & $-1.433\pm$0.045 \\
\\
S$^{+}$/H$^+$ (10$^7$)     	& 1.217$\pm$0.026 	& 1.767$\pm$0.050 	& 1.669$\pm$0.047 	& 0.500$\pm$0.021 	& 1.474$\pm$0.044 	& 1.166$\pm$0.019 	& 1.377$\pm$0.035 	\\
S$^{+2}$/H$^+$ (10$^7$)    	& 4.627$\pm$0.066 	& \nodata         		& 7.156$\pm$0.145 	& \nodata         		& 6.348$\pm$0.146 	& \nodata         		& \nodata         		\\
S ICF                       			& 2.122$\pm$0.100 	& \nodata         		& 1.944$\pm$0.110 	& \nodata         		& 1.895$\pm$0.114 	& \nodata         		& \nodata         		\\
log(S/O)                   			& $-1.600\pm$0.023	& \nodata         		& $-1.464\pm$0.026 	& \nodata         		& $-1.556\pm$0.027 	& \nodata         		& \nodata         		\\
\\
Ne$^{+2}$/H$^{+}$ (10$^5$) 	& 1.059$\pm$0.032	& 1.047$\pm$0.044	& 1.004$\pm$0.043	& 1.056$\pm$0.038	& 1.190$\pm$0.086	& 1.161$\pm$0.037	& \nodata         		\\
Ne$^{+2}$/O$^{+2}$         		& 0.244$\pm$0.015 	& 0.232$\pm$0.016 	& 0.234$\pm$0.017 	& 0.288$\pm$0.019 	& 0.261$\pm$0.024 	& 0.257$\pm$0.016 	& \nodata         		\\
Ne ICF                      			& 1.139$\pm$0.080 	& 1.182$\pm$0.089 	& 1.163$\pm$0.089 	& 1.046$\pm$0.080 	& 1.171$\pm$0.094 	& 1.131$\pm$0.080 	& \nodata         		\\
log(Ne/O)                  			& $-0.555\pm$0.026 	& $-0.562\pm$0.031	& $-0.566\pm$0.031	& $-0.521\pm$0.029	& $-0.514\pm$0.041	& $-0.537\pm$0.027 	& \nodata         		\\
\\
Ar$^{+2}$/H$^{+}$ (10$^7$)    	& 1.255$\pm$0.023 	& 1.763$\pm$0.055	& 1.762$\pm$0.040 	& 1.007$\pm$0.041 	& 1.647$\pm$0.042 	& 1.674$\pm$0.039 	& 2.313$\pm$0.065 	\\
Ar$^{+3}$/H$^{+}$ (10$^7$) 	& 0.916$\pm$0.037 	& 0.889$\pm$0.307 	& 1.039$\pm$0.258 	& 1.995$\pm$0.180 	& 0.754$\pm$0.283 	& 1.785$\pm$0.143 	& 1.458$\pm$0.266 	\\
Ar/O			          		& 0.005$\pm$0.001 	& 0.006$\pm$0.002 	& 0.007$\pm$0.002 	& 0.008$\pm$0.001 	& 0.005$\pm$0.002 	& 0.008$\pm$0.001 	& 0.008$\pm$0.001 	\\
Ar ICF                     			& 1.016$\pm$0.001 	& 1.023$\pm$0.003 	& 1.020$\pm$0.003 	& 1.008$\pm$0.001 	& 1.021$\pm$0.003 	& 1.015$\pm$0.001 	& 1.014$\pm$0.002 	\\
log(Ar/O)                  			& $-2.293\pm$0.030 	& $-2.221\pm$0.153 	& $-2.177\pm$0.111 	& $-2.083\pm$0.049 	& $-2.269\pm$0.165 	& $-2.110\pm$0.043 	& $-2.112\pm$0.083 \\
\hline \vspace{-1.5ex} \\
\multicolumn{8}{c}{Properties Derived from UV Spectra} \\
\hline \vspace{-1ex} \\
C$^{+2}$/O$^{+2}$          		& 0.197$\pm$0.037 	& 0.098$\pm$0.030 	& 0.265$\pm$0.076 	& 0.082$\pm$0.015 	& 0.439$\pm$0.134 	& 0.152$\pm$0.033 	& 0.279$\pm$0.095 \\
C$^{+3}$/C$^{+2}$          		& \nodata         		& \nodata         		& 0.159$\pm$0.042 	& 0.387$\pm$0.077 	& 0.175$\pm$0.041 	& 0.167$\pm$0.044 	& 0.123$\pm$0.049 \\
log U                       			& $-2.408\pm$0.031 	& $-2.526\pm$0.033 	& $-2.469\pm$0.039 	& $-1.818\pm$0.097 	& $-2.479\pm$0.050 	& $-2.038\pm$0.024 	& $-2.268\pm$0.062 \\
C ICF                      			& 0.963$\pm$0.031 	& 0.926$\pm$0.033 	& 0.945$\pm$0.039 	& 1.137$\pm$0.099 	& 0.942$\pm$0.050 	& 1.064$\pm$0.026 	& 1.002$\pm$0.062 \\
log(C/O)                   			& $-0.723\pm$0.075 	& $-1.042\pm$0.116 	& $-0.601\pm$0.111	& $-1.029\pm$0.079	& $-0.384\pm$0.118 	& $-0.791\pm$0.087 	& $-0.553\pm$0.129 
\enddata
\label{tbl11}
\end{deluxetable*}

\begin{deluxetable*}{lcccccc}
\tabletypesize{\scriptsize}
\tablewidth{0pt} 
\setlength{\tabcolsep}{3pt}
\tablecaption{Ionic and Total Abundance for HST/COS Compact Dwarf Galaxies - Cont.}
\tablewidth{0pt}
\tablehead{\CH{Property}		& \CH{J025346}	& \CH{J015809}	& \CH{J104654}	& \CH{J093006}	& \CH{J092055}	& \CH{{\it J084956}}}
\startdata	
\multicolumn{7}{c}{Properties Derived from Optical Spectra}  \\
\hline \vspace{-1ex} \\
T$_e$ [O~\iii] (K)        		& 13790$\pm$190 	& 15490$\pm$210 	& 13950$\pm$150 	& 13100$\pm$170		& 14620$\pm$130 	& 13530$\pm$110 	 \\
T$_e$ [S~\iii] (K)$^a$       		& 12610$\pm$160 	& 15350$\pm$170 	& 13950$\pm$120 	& 13600$\pm$140		& 13830$\pm$110	& 12930$\pm$90	 \\
T$_e$ [N~\ii] (K)$^a$      		& 12660$\pm$130 	& 13840$\pm$140 	& 12760$\pm$110 	& 12200$\pm$120		& 13230$\pm$90 	& 12470$\pm$80	 \\
n$_e$ C~\iii] (cm$^{-3}$) 		& $<12310^b$ 		& $<36760^b$		& $<3310^b$    		& $<5800^b$			& $<21190^b$		& \nodata			\\
n$_e$ Si~\iii] (cm$^{-3}$) 		& \nodata         		& \nodata         		& \nodata         		& \nodata				& \nodata         		& \nodata			 \\
n$_e$ [S~\ii] (cm$^{-3}$) 		& 40$\pm$30 		& 100$\pm$60 		& 90$\pm$30 		& 28$\pm$22			& 20$\pm$20 		& 70$\pm$20		 \\
\\
O$^{+}$/H$^+$ (10$^5$)     	& 1.536$\pm$0.106 	& 0.725$\pm$0.095 	& 1.720$\pm$0.104 	& 2.839$\pm$0.167		& 1.664$\pm$0.069  & 2.640$\pm$0.100	 \\
O$^{+2}$/H$^+$ (10$^5$)    	& 6.606$\pm$0.393 	& 4.893$\pm$0.275 	& 6.403$\pm$0.354 	& 7.606$\pm$0.445		& 5.656$\pm$0.299 	& 6.954$\pm$0.362	 \\
12+log(O/H)                 		& 7.911$\pm$0.022 	& 7.750$\pm$0.022 	& 7.910$\pm$0.020 	& 8.019$\pm$0.020		& 7.865$\pm$0.018  & 7.982$\pm$0.010	 \\
\\
N$^{+}$/H$^+$ (10$^7$)     	& 4.772$\pm$0.174 	& 2.483$\pm$0.158 	& 5.700$\pm$0.154 	& 9.568$\pm$0.299		& 5.147$\pm$0.150 	& 8.407$\pm$0.193	 \\
log(N/O)                   			& $-1.508\pm$0.034 	& $-1.465\pm$0.063 	& $-1.480\pm$0.029 	& -1.472$\pm$0.029		& $-1.510\pm$0.022 	& $-1.497\pm$0.020	 \\
\\
S$^{+}$/H$^+$ (10$^7$)     	& 2.188$\pm$0.056	& 1.260$\pm$0.043 	& 2.885$\pm$0.061 	& 4.387$\pm$0.111		& 2.695$\pm$0.050	& 4.087$\pm$0.077 	 \\
S$^{+2}$/H$^+$ (10$^6$)    	& 1.660$\pm$0.031	& 0.828$\pm$0.025 	& 1.283$\pm$0.020	& 1.385$\pm$0.021		& 1.033$\pm$0.021	& \nodata			 \\
S ICF                       			& 1.635$\pm$0.073 	& 2.048$\pm$0.193 	& 1.544$\pm$0.055 	& 1.392$\pm$0.037		& 1.494$\pm$0.039	& \nodata			 \\
log(S/O)                   			& $-1.423\pm$0.025 	& $-1.459\pm$0.031 	& $-1.525\pm$0.022 	& $-1.614\pm$0.022		& $-1.575\pm$0.020	& \nodata			 \\
\\
Ne$^{+2}$/H$^{+}$ (10$^5$)	& 1.720$\pm$0.078	& \nodata         		& 1.642$\pm$0.061	& 2.043$\pm$0.103		& 1.489$\pm$0.047	& 2.343$\pm$0.070 	 \\
Ne$^{+2}$/O$^{+2}$ 		& 0.260$\pm$0.020 	& \nodata         		& 0.256$\pm$0.017	& 0.269$\pm$0.02		& 0.263$\pm$0.016  	& 0.337$\pm$0.020 	 \\
Ne ICF                     			& 1.233$\pm$0.096 	& \nodata         		& 1.269$\pm$0.091	& 1.373$\pm$0.102 		& 1.294$\pm$0.087 	& 1.380$\pm$0.090 	 \\
log(Ne/O)                 			& $-0.494\pm$0.032	& \nodata         		& $-0.488\pm$0.029	& $-0.433\pm$0.034 		& $-0.468\pm$0.027	& $-0.333\pm$0.026 	 \\
\\
Ar$^{+2}$/H$^{+}$ (10$^7$)    	& 3.623$\pm$0.071 	& 2.150$\pm$0.080	& 3.019$\pm$0.065 	& 3.019$\pm$0.050 		& 2.236$\pm$0.049	& 3.267$\pm$0.065 	 \\
Ar$^{+3}$/H$^{+}$ (10$^7$) 	& \nodata         		& 0.981$\pm$0.371 	& 0.701$\pm$0.135 	& \nodata				& 1.010$\pm$0.032 	& \nodata         		 \\
Ar/O			          		& 0.005$\pm$0.001 	& 0.006$\pm$0.002 	& 0.006$\pm$0.001 	& 0.004$\pm$0.001 		& 0.006$\pm$0.001  	& 0.005$\pm$0.001 	 \\
Ar ICF                     			& 1.973$\pm$0.079 	& 1.017$\pm$0.003 	& 1.043$\pm$0.006 	& 1.654$\pm$0.048 		& 1.050$\pm$0.006 	& 1.645$\pm$0.036 	 \\
log(Ar/O)                  			& $-1.966\pm0.027$ 	& $-2.186\pm0.167$ 	& $-2.218\pm0.087$ 	& $-2.183\pm$0.026 		& $-2.220\pm0.028$	& $-2.112\pm0.024$  \\
\hline \vspace{-1.5ex} \\
\multicolumn{7}{c}{Properties Derived from UV Spectra}  \\
\hline \vspace{-1ex} \\
C$^{+2}$/O$^{+2}$          		& 0.279$\pm$0.080 	& 0.285$\pm$0.111 	& 0.185$\pm$0.048 	& 0.162$\pm$0.041		& 0.264$\pm$0.083  & 0.181$\pm$0.191	 \\ 
C$^{+3}$/C$^{+2}$          		& 0.147$\pm$0.052 	& 0.306$\pm$0.097 	& 0.126$\pm$0.041 	& 0.158$\pm$0.057		& 0.079$\pm$0.025 	& 0.186$\pm$0.155	 \\
log U                       			& $-2.544\pm$0.054 	& $-2.108\pm$0.024 	& $-2.582\pm$0.054	& -$2.764\pm$0.047		& $-2.632\pm$0.046 	& $-2.765\pm0.020$	 \\
C ICF                      			& 0.884$\pm$0.054 	& 1.044$\pm$0.025 	& 0.869$\pm$0.054 	& 0.784$\pm$0.047		& 0.848$\pm$0.046	& 0.783$\pm$0.045	 \\
log(C/O)                   			& $-0.607\pm$0.111 	& $-0.526\pm$0.143 	& $-0.794\pm$0.102 	& $-0.895\pm$0.101		& $-0.651\pm$0.121	& $-0.848\pm0.313$	 
\enddata
\tablecomments{
Note that J084956 is not a part of our 3$\sigma$ sample, but is included here for modeling purposes. }
\label{tbl12}
\end{deluxetable*}


\end{document}